\documentclass[fleqn,usenatbib,useAMS]{mnras}

\usepackage{graphicx}	
\usepackage{amsmath}	
\usepackage{amssymb}	
\usepackage{multicol}        
\usepackage{bm}		
\usepackage{pdflscape}	
\usepackage{xcolor}
\usepackage{subfigure}
\usepackage{threeparttable}
\usepackage{soul}
\usepackage{comment}  

\usepackage[T1]{fontenc}
\usepackage{ae,aecompl}

\usepackage{newtxtext,newtxmath}

\title[Multiphase Outflows in (U)LIRGs]{Properties of the Multiphase Outflows in Local (Ultra)luminous Infrared Galaxies}

\author[A Fluetsch et al.]{A. Fluetsch$^{1,2}$, R. Maiolino$^{1,2}$, S. Carniani$^{3}$, S. Arribas$^{4}$, F. Belfiore$^{5, 6}$, E. Bellocchi$^{7}$,  \newauthor S. Cazzoli$^{8}$, C. Cicone$^{9}$, G. Cresci$^{6}$, A. C. Fabian$^{10}$, R. Gallagher$^{1,2}$, W. Ishibashi$^{11}$, \newauthor F. Mannucci$^{6}$, A. Marconi$^{6,12}$, M. Perna$^{4,6}$, E. Sturm$^{13}$, G. Venturi$^{14}$ \\
 \\
$^{1}$University of Cambridge, Cavendish Laboratory, Cambridge CB3 0HE, UK\\
$^{2}$University of Cambridge, Kavli Institute for Cosmology, Cambridge CB3 0HE, UK\\
$^{3}$Scuola Normale Superiore, Piazza dei Cavalieri 7, I-56126 Pisa, Italy \\
$^{4}$Centro de Astrobiolog\'ia (CSIC-INTA), Departamento de Astrof\'isica, Cra. de Ajalvir Km.~4, 28850, Torrej\'on de Ardoz, Madrid, Spain \\
$^{5}$European Southern Observatory, Karl-Schwarzschild-Strasse 2, D-85748 Garching, Germany \\
$^{6}$INAF -- Osservatorio Astrofisico di Arcetri, Largo Enrico Fermi 5, I-50125 Firenze, Italy \\
$^{7}$ Centro de Astrobiolog\'ia (CSIC-INTA), ESAC Campus, 28692 Villanueva de
la Ca\~nada, Madrid, Spain \\
$^{8}$Instituto de Astrof\'isica de Andaluc\'ia (CSIC), Apdo. 3004, 18008, Granada, Spain \\
$^{9}$Institute of Theoretical Astrophysics, University of Oslo, PO Box 1029, Blindern 0315, Oslo, Norway \\
$^{10}$Institute of Astronomy, University of Cambridge, Madingley Road, Cambridge CB3 0HA, UK \\
$^{11}$Physik-Institut, Universitaet Zurich, Winterthurerstrasse 190, CH-8057 Zurich, Switzerland \\
$^{12}$Dipartimento di Fisica e Astronomia, Universit\`a di Firenze, Via G. Sansone 1, I-50019 Sesto Fiorentino (Firenze), Italy \\
$^{13}$Max-Planck-Institut fuer extraterrestrische Physik, Postfach 1312, D-85741, Garching, Germany \\
$^{14}$Instituto de Astrof\'isica, Facultad de F\'isica, Pontificia Universidad Cat\'olica de Chile, Avda. Vic\~una Mackenna 4860, 8970117 Macul, Santiago, Chile \\
}

\date{}

\pubyear{2015}

\begin{document}
\label{firstpage}
\pagerange{\pageref{firstpage}--\pageref{lastpage}}
\maketitle

\begin{abstract}
Galactic outflows are known to consist of several gas phases, however, the connection between these phases has been investigated little and only in a few objects. In this paper, we analyse MUSE/VLT data of 26 local (U)LIRGs and study their ionized and neutral atomic phases. We also include objects from the literature to obtain a sample of 31 galaxies with spatially resolved multi-phase outflow information. We find that the ionized phase of the outflows has on average an electron density three times higher than the disc ($n_{\rm e, disc}$ $\sim$ 145~cm$^{-3}$ vs $n_{\rm e, outflow}$ $\sim$ 500~cm$^{-3}$), suggesting that cloud compression in the outflow is more important than cloud dissipation. We find that the difference in extinction between outflow and disc correlates with the outflow gas mass. Together with the analysis of the outflow velocities, this suggests that at least some of the outflows are associated with the ejection of dusty clouds from the disc. This may support models where radiation pressure on dust contributes to driving galactic outflows. The presence of dust in outflows is relevant for potential formation of molecules inside them. We combine our data with millimetre data to investigate the molecular phase. We find that the molecular phase accounts for more than 60~$\%$ of the total mass outflow rate in most objects and this fraction is higher in AGN-dominated systems. The neutral atomic phase contributes of the order of 10~$\%$, while the ionized phase is negligible. The ionized-to-molecular mass outflow rate declines slightly with AGN luminosity, although with a large scatter.
\end{abstract}




\newpage
\section{Introduction}

Massive outflows of gas, driven either by active galactic nuclei (AGN) or star formation, can inject energy and momentum into the interstellar medium (ISM) and thereby affect the evolution of galaxies. 
These feedback effects are potentially capable of suppressing or even shutting down star formation in galaxies \citep[e.g.][]{Fabian2012, Zubovas2012}. AGN feedback can also explain the tight correlation between the mass of the central supermassive black hole (SMBH) and the velocity dispersion $\sigma$, the mass or luminosity of the host galaxy \citep[e.g.][]{Kormendy2013, Mcconnell2013}.  
In simulations of galaxy evolution, AGN feedback is needed to reproduce the lack of massive star forming galaxies in the local Universe \citep[e.g.][]{Dave2011, Vogelsberger2014, Beckmann2017}, 
whereas feedback from supernovae and stellar winds is invoked to explain the shape of the stellar mass function at the low stellar mass end \citep[e.g.][]{Puchwein2013}.

In recent years, galactic winds and outflows have been observed in several different gas phases, including in the hot highly-ionized \citep[using X-ray spectroscopy,][]{Tombesi2013,Nardini2015} the warm ionized \citep[e.g.][]{Bellocchi2013, Cazzoli2018}, the neutral atomic phase through absorption in the sodium doublet \citep[e.g.][]{Perna2017, Roberts-Borsani2019}, [CII] emission \citep[e.g][]{Maiolino2012, Bischetti2019a} or in HI absorption \citep{Morganti2016} and the molecular gas phase \citep[e.g.][]{Sturm2011, Cicone2014, Fluetsch2019}. These measurements trace outflows at different spatial scales (from sub-pc in the X-ray up to 10--20 kpc in the cold molecular and ionized phases) and in different physical conditions (e.g. they are sensitive to different temperatures and densities). The aim of most of these works was to study the physical mechanisms that drive outflows and to quantify whether some of them can indeed quench star formation in galaxies as suggested by models of galaxy evolution \citep[see e.g.][]{Dave2011, Somerville2015}. 

The vast majority of these studies, however, focus on one single outflow phase. Only a small number of works have studied several phases in a single galaxy and found that galactic outflows are multi-phase and can be as complex as the normal ISM. For instance, in the Seyfert galaxy IC 5063, a remarkable kinematic similarity between the ionized, neutral atomic and the molecular phase was found \citep{Morganti2015, Oosterloo2017}, suggesting a link between the different phases. Similarly, HE 1353-1917 \citep{Husemann2019} and IRAS F08572+3915 \citep{Herrera-Camus2020} also display co-spatiallity between the molecular and the ionized outflow phase. In other galaxies, such as Mrk 231 \citep{Feruglio2015}, the interplay between gas phases appears to be much more complex and there might have been several feedback episodes. In a sample of Seyfert I galaxies, it was found that the spatial correlation between neutral atomic outflows traced by the sodium doublet and the ionized outflow varies greatly from galaxy to galaxy, but overall, the different phases seem to be unrelated \citep{Rupke2017}. In addition, some works also complemented such multi-phase studies with measurements of the hot, highly-ionized phased probed by X-ray observations \citep{Veilleux2017a, Bischetti2019, Sirressi2019}.

Despite these efforts to detect multiphase outflows, the connection (if there is any) between the different outflow phases and therefore the physical mechanism driving them remains mostly unknown \citep{Cicone2018}. One major open question is the relative contribution of different phases to the total mass or energy budget of the outflow. It is crucial to know the \textit{total} outflow mass and energy as only this will enable a thorough comparison to simulations of AGN or star formation-driven feedback \citep[e.g.][]{Biernacki2018, Nelson2019} and allow us to distinguish between different driving mechanisms (i.e. energy-driven, momentum-driven or radiation-pressure driven) \citep[see e.g. ][]{Fabian2012, Costa2014}.

Several observational studies have suggested that the molecular phase is dominant with a mass that is a factor of 10-100 larger than that in the ionized phase \citep{Carniani2015, Fluetsch2019, RamosAlmeida2019}, although the uncertainties in both phases are significant. \cite{Fiore2017} confirmed with disjoint samples that the molecular outflow mass rate is significantly higher than the ionized outflow mass rate, but suggested that at high AGN luminosity, the ionized phase becomes more prominent and almost comparable in outflow rate at luminosities above $\sim$~10$^{47}$~erg~s$^{-1}$. By only selecting objects with measurements of both gas phases instead of disjoint samples, \cite{Fluetsch2019} were unable to reproduce the increasing contribution of the ionized outflow phase at high AGN luminosities. However, most galaxies in \cite{Fluetsch2019} have AGN luminosities below 10$^{46}$~erg~s$^{-1}$ and might therefore not properly test the regime discussed in \cite{Fiore2017}.

In this work, we investigate multiphase outflows in a sample of 26 local ($z$ $<$ 0.2) (ultra)luminous infrared galaxies ((U)LIRGs).  ULIRGs, formally defined as a class of galaxies with an infrared luminosity >10$^{12}$ $L_{\odot}$ (ULIRGs) (or >10$^{11}$ $L_{\odot}$ for LIRGs), are galaxies with a strong AGN, starburst activity or a combination of the two. Usually, (U)LIRGs are (advanced) mergers and have large amounts of gas and dust \citep{Lonsdale2006, Kartaltepe2010}. This makes them suitable laboratories to study feedback as they are likely to exhibit fast and massive outflows \citep[see e.g.][]{Rodriguez-Zaurin2013, Rose2017} and resemble massive active galaxies at higher redshift, where we expect feedback processes to be enhanced. Indeed, (U)LIRGs have shown the clearest examples of outflows in the ionized \citep[e.g.][]{Rupke2013, Arribas2014}, neutral atomic \citep[e.g.][]{Cazzoli2016} and molecular phase \citep[e.g.][]{Cicone2014, Pereira-Santaella2018, Fluetsch2019}. These studies have found that outflow velocities in ULIRGs are high in all three phases, usually between several hundred to a few 1000~km~s$^{-1}$ \citep{Westmoquette2012, Gonzalez-Alfonso2017, Perna2021, Perna2020}. The momentum fluxes (several $L/c$) and kinetic power especially in the molecular phase indicate that outflows in these objects could significantly affect their hosts \citep{Cicone2014, Gonzalez-Alfonso2017}.
Furthermore, we supplement our sample with multiphase outflows from the literature, increasing our total sample size to 31 galaxies with multiphase outflows. This makes our work the largest spatially resolved multiphase outflow study to date.

The goal of this study is twofold. First, we aim to characterise the multiphase outflows in our sample of local (U)LIRGs in terms of morphology, kinematics and outflow properties (e.g. electron density). Second, we shed light on the connection between the different gas phases as well as their relative contribution, both on a spatially resolved basis and in terms of global properties such as mass outflow rate or kinetic power of the outflow.

Throughout this work, a \textit{H$_{\rm 0}$} = 70~km s$^{-1}$ Mpc$^{-1}$, $\Omega_{\rm M}$ = 0.27 and $\Omega_{\Lambda}$ = 0.73 cosmology is adopted.

\section{Methods}

\subsection{Sample and Observations}

Our sample consists of 26 local ($z$ < 0.2) (U)LIRGs. These galaxies host a strong AGN, significant star formation or a combination of the two. According to their BPT diagrams, 14 galaxies are star-forming, 8 AGN-dominated (2 of which are type I AGN) and 4 are classified as composite. The objects span several order of magnitude in AGN luminosity, from $\lesssim$ 10$^{42}$ erg s$^{-1}$ to $\sim$10$^{46}$ erg s$^{-1}$. Out of the 26 galaxies, 2 galaxies (Mrk463 and IRAS 23389+0300) have a radio excess according to \citet{Drake2004}, which is defined as log($S_{\nu}$(60 $\mu$m)/($S_{\nu}$(4.8~GHz)) < 1.8. Unfortunately, as discussed later on, both these galaxies only show evidence of ionized outflows, so they cannot be used for the multi-phase analysis.

These galaxies were observed as part of several different ESO Multi Unit Spectroscopic Explorer (MUSE) observing programmes at the Very Large Telescope. 11 galaxies were observed under programme 101.B.0368 (PI: Maiolino). Four galaxies are from the programme 102.B.0617 (PI: Fluetsch). These two programmes targeted (U)LIRGs for which the presence of an outflow, either ionized or neutral atomic has been established by previous studies \citep{Bellocchi2013, Arribas2014,Cazzoli2016}. The remaining targets are taken from a variety of different programmes: three targets from 095.B-0049, two targets from each 094.B-0733 and 096.B-0230, and one target from each of the following observing programmes 60.A.9315, 097.B.0165, 097.B-0313 and 094.B-0733. These targets were selected among (U)LIRGs for which there is already a molecular outflow detected \citep{Fluetsch2019} and by searching MUSE data in the ESO archive. Due to these heterogeneous selection criteria our sample is not necessarily representative of the whole (U)LIRG population. Therefore it should not be used, for instance, to infer statistics about the occurence of outflows among the (U)LIRG population.


\subsection{Data analysis}

The MUSE data cubes are used in their reduced form after running the MUSE pipeline. This produces a data cube of about 300 $\times$ 300 spaxels with a spatial sampling of 0.2'' $\times$ 0.2''. The wavelength coverage of MUSE cubes ranges from 4650~\AA\: to 9300~\AA\: and, at the redshift of the targets, covers all the main optical emission lines of interest (from H$\beta$ to [SII]$\lambda\lambda$6716,6731). It achieves a resolution between 1750 (at 4650~\AA) and 3750 (at 9300~\AA). The field of view (1' $\times$ 1') covers between $\sim$10~kpc to $\sim$170~kpc of the galaxy depending on its redshift. The average seeing for the observations is about 1''. The typical observation time for the targets is approximately 40 mins. The details of the observations are listed in Table \ref{tab:obs} in the Appendix.

In a first step, the data cubes are binned using Voronoi tessellation to achieve an average signal-to-noise ratio (SNR) of 30--40 (this varies from galaxy to galaxy) per wavelength channel in each bin in the r-band (5405-6982~\AA). In this step, all spaxels with a SNR < 3 in the r-band are discarded to avoid noisy spectra in the Voronoi bins. 

The stellar continuum is modelled using the MILES stellar library templates, which include about 1000 stars, which range across a large number of atmospheric parameters (such as ages and metallicities) \citep{Blazquez2006}. These templates cover the wavelength range from 3525--7500~\AA, which includes all nebular lines which are of interest for our analysis. The fitting is performed using an adapted Penalized Pixel-Fitting (pPXF) routine \citep{Cappellari2004, Cappellari2017} and allowing an additive Legendre polynomial to correct the shape of the the template continuum. The main emission lines (H$\beta$, [OIII]$\lambda \lambda$4959,5007, [OI]$\lambda \lambda$6300,6366, H$\alpha$, [NII]$\lambda \lambda$6548,6584 and [SII]$\lambda \lambda$6717,6732) are fitted simultaneously with the stellar continuum to allow optimal recovery of the total flux of emission lines, in particular the Balmer lines \citep[see e.g.][]{Sarzi2006}. We fit with one, two or three Gaussian emission line components. During this step, we mask five strong sky emission lines (5577.4~\AA, 5889.9~\AA, 6157.7~\AA, 6300.3~\AA\: and 6363.7~\AA) as well as the Na~ID absorption lines (at 5889.9~\AA\: and 5895.9~\AA) at the target's redshift since this might not exclusively originate from stars in these galaxies, but also from absorption by neutral atomic gas in the ISM (see Section \ref{sec:neutral_atomic}). As the best fit we select the model with the lowest $\chi_{\rm red}^{2}$.
Next we subtract the stellar continuum from the total spectra in each spaxel, scaling the fit from bin to each spaxel according to their r-band flux. We performed visual checks of the fits for a sub-sample of spaxels to verify that the stellar continuum subtraction works well.

\subsection{Ionized gas}

The resulting continuum-subtracted data cube is fitted on a spaxel-by-spaxel basis using the Python library lmfit \citep{Newville2014}. Lmfit provides non-linear optimisation and is built on the Levenberg-Marquardt algorithm \mbox{\citep{Levenberg1944, Marquardt1963}}. We fit the same emission lines as in the simultaneous fitting with one or two Gaussian components. For Seyfert 1 galaxies (two in our sample: IRAS 00509+1225 and IRAS 23389+0300), we allow a third, very broad component for the Balmer lines to account for emission from the AGN broad line region (BLR). 
However, the deblending of the very broad component (from the BLR) and the outflow component results in significant uncertainties in the characterization of the latter. Thus, we have decided to discard the information obtained from the outflow of these two Seyfert 1 galaxies. For galaxies with just two components, the width ($\sigma$) of the narrow component has to be smaller than 200~km~s$^{-1}$ and the broad component has to have a width of at least 125~km~s$^{-1}$ and has to be larger than the narrow component. For each component (narrow, broad and very broad if needed), we tie velocity centroid and dispersion, while the amplitudes are allowed to vary freely. Exceptions are the amplitudes of [OIII]$\lambda \lambda$4959,5007; [OI]$\lambda \lambda$6300,6366; and [NII]$\lambda \lambda$6548,6584, for which the ratio of the doublets is given by the Einstein coefficients. The ratio of the two emission lines of the [SII] doublet is allowed to vary between 0.44 and 1.5 \citep{Osterbrock2006}. For each spaxel we employ the reduced $\chi^2$ \citep{Andrae2010}, $\chi^2_{\rm red}$, to determine whether a model with one, two or three components is preferred. We then select the model with the lowest $\chi_{\rm red}^{2}$ value. If based on the $\chi^2_{\rm red}$ metric, we find two or more components, we classify the broad component as a signature of an ionized outflow only if it is kinematically clearly distinct from the narrow component. In addition, we require the broad component to be strong enough to appear in the integrated spectrum. For some of the galaxies in our sample additional higher spectral resolution and higher sensitivity data were obtained with X-shooter (programmes 0103.B-0478(A) and 097.B-0918(A)). This data confirms the decomposition of the various components, although limited to the central region (an extensive description of the X-shooter data will be part of a separate paper).
Finally, to get the intrinsic fluxes for all emission lines, we use the Balmer decrement and we apply a \cite{Calzetti2000} attenuation curve for a galactic diffuse ISM ($R_{\rm V}$ = 4.05) to correct for dust reddening. This should account for the mixing between dust and gas emitting regions in star forming galaxies.

\subsection{Neutral atomic gas}
\label{sec:neutral_atomic}
To study the neutral atomic gas, we analyse the Na~I doublet (at 5889.9~\AA\: and 5895.9~\AA) profile in the data cube obtained by dividing the original data cube by the continuum fit \citep[for details see][]{Rupke2005}. The Na~ID line probes neutral gas (i.e. HI), thanks to the low ionization potential of the associated transition, which is 5.14~eV. One caveat of the sodium absorption technique is that the detection of outflowing gas requires a strong background of stellar continuum light and hence outflows outside the projected stellar disc or far from the galaxy's bright centre might be missed. Residual stellar contribution (even after running pPXF) to the Na~ID as well as contamination from Na~ID emission and from the nearby He~I$\lambda$5876 emission line may further complicate the measurement.

As the sodium absorption feature is rather weak, we apply Voronoi binning to this data cube. We require a typical SNR of $>$ 10--15 for all targets on the sodium absorption feature in the stellar-continuum subtracted cube. We then fit the sodium absorption profile in each Voronoi bin.

Our fitting approach is based on a model of partially overlapping Na atoms \citep[see][]{Rupke2005} with the analytical form:
\begin{equation}
    I(\lambda) = 1 - C_{\mathrm{f}} + C_{\mathrm{f}} \times e^{-\tau_{\mathrm{B}}(\lambda) - \tau_{\mathrm{R}}(\lambda)}
\label{eq:atomic_overlap}
\end{equation}
with $C_{\rm f}$ being the covering fraction and $\tau_{\rm B}$ and $\tau_{\rm R}$ being the optical depths of the blue and the red Na~I lines (5889.9~\AA\: and 5895.9~\AA). The optical depth of a line, $\tau (\lambda)$ is given by a Gaussian:
\begin{equation}
    \tau (\lambda) = \tau_{0}e^{-(\lambda - \lambda_{0} + \Delta \lambda_{\rm offset})^{2}/((\lambda_{0}+\Delta \lambda_{\rm offset})b_{\rm D}/c)^{2}},
\label{eq:tau}
\end{equation}
where $\tau_{0}$ and $\lambda_{0}$ are the central optical depth and wavelength of each line component and $b_{\rm D}$ is the Doppler linewidth. $\Delta \lambda_{\rm offset}$, the wavelength offset, is linked to the velocity offset: $\Delta \lambda_{\rm offset}$ = $\Delta$v$\lambda_{\rm 0}$/c with c being the speed of light. The doublet is fit with one, two or three absorption components and up to one sodium emission component. We allow for simultaneous fitting of the emission and absorption components. This simultaneous fitting can lead to degeneracy. In order to minimize that, we tie the width of the Na~ID emission component to the He~I emission line width. The ratio of the central optical depths of the sodium doublet is fixed to $\tau_{0,B}$/$\tau_{0,R}$ = 2 for all absorption components which is given by the ratio of the Einstein parameters for the two transitions \citep{Morton1991}. We require the covering fractions to be between 0 and 1. Simultaneously we also fit the He~I emission line (5876 \AA). This is important as the sodium absorption profile sometimes extends bluewards and its true amplitude might be underestimated if He~$\lambda$5876 is not considered. To fit several components, we multiply the different $I_{\rm i}(\lambda)$'s, e.g. for two components $I(\lambda)$ = $I_{\rm 1}(\lambda)$ $I_{\rm 2}(\lambda)$, where $I_{\rm 1}(\lambda)$ and $I_{\rm 2}(\lambda)$ are of the form given in equation \ref{eq:atomic_overlap} with different covering fractions and optical depths for each component.

As in the emission line fitting, we use the lmfit library \citep{Newville2014}. The best fit is determined using the Bayesian information criterion (BIC). The BIC selects the best fit while taking into account the complexity of the model \citep[i.e. models with more parameters are penalized,][]{Liddle2007}. 

We use the 50th percentile, $v_{\rm 50}$, of the whole sodium absorption profile, as a measure of outflow velocity. Similar to \cite{Rupke2005}, we classify a component as outflowing if it has a velocity shift of $\Delta v$ < --50~km~s$^{-1}$ relative to the systemic velocity. We adopt this conservative assumption to account for possible errors in the fitting. Redshifted sodium emission may also be additional evidence of receding, outflowing gas \citep[see e.g.][]{Rupke2015, Roberts-Borsani2019}.

The advantage of this model is that the optical depth can be related to the sodium column density, \textit{N}(\rm Na~I), via:

\begin{equation}
    N(\mathrm{Na\: I}) = \frac{\tau_{\mathrm{0}}b_{\mathrm{D}}}{1.497\times 10^{-15} \lambda_{\mathrm{0}}f}.
\end{equation}
$\lambda_{0}$ = 5897.55~\AA\: and $f$ = 0.318 are the vacuum wavelength and the oscillator strength, respectively. $b_{\rm D}$ is the Doppler parameter, which is related to the velocity dispersion via $b_{\rm D}$ = $\sqrt{2}\sigma$.

The hydrogen column density, \textit{N}(\rm H), is then given by:
\begin{equation}
    N(\mathrm{H}) = N(\mathrm{Na\: I})(1-y)^{-1}10^{-(a+b)},
\end{equation}
where $y$ is the ionization fraction, $a$ the galaxy's Na abundance and $b$ the depletion onto dust. We assume the following values: $y$ = 0.9, $a$ = --5.69 and $b$ = --0.95 \citep{Savage1996, Rupke2005}. The hydrogen column density can then be used to calculate the outflow mass (see Section \ref{sec:calc_of_prop}). Typical errors for the covering fraction and the velocity are around 20~$\%$ and 20~km~s$^{-1}$, respectively \citep{Rupke2005b}, and we therefore conservatively assume an error of 0.5~dex on the neutral atomic mass outflow rate.

\subsection{Ancillary properties}
\label{sec:ancill}

In this paper, we compare the neutral atomic, the ionized and the molecular outflow phases. The properties of the molecular outflows (velocity, radius and mass of the outflow as well as derived properties) are taken from \cite{Fluetsch2019}. This paper uses the same cosmology, the same definition of mass outflow rate and the same convention for the measurement of outflow radius and velocity as we do here and hence allows direct comparison.

The AGN luminosity, $L_{\rm AGN}$, is provided in \cite{Fluetsch2019}, who used the absorption-corrected X-ray luminosity in the range from 2--10~keV and the bolometric correction by \cite{Marconi2004} to determine the AGN bolometric luminosity, $L_{\rm AGN}$, in most cases. If no X-ray data is available, then the AGN bolometric luminosity is inferred from the extinction-corrected [OIII] luminosity using the relation given in \cite{Heckman2004}. 

We calculate the AGN contribution, $\alpha_{\rm bol}$ = $L_{\rm AGN}$/$L_{\rm bol}$, where $L_{\rm bol}$ is the bolometric luminosity of the galaxy, which for ULIRGs is given by $L_{\rm bol}$ $\approx$ 1.15 $L_{\rm IR}$ \citep{Veilleux2009}.

The star formation rates (SFRs) from \cite{Fluetsch2019} are used if possible. The calculation is based on the total infrared luminosity (8--1000~$\mu$m), taking into account the AGN fraction ($\alpha_{\rm bol}$) \citep{Sturm2011}. Otherwise, we calculate the SFR ourselves using the same prescription. There are a few exceptions: for HE 1351-1917, zC400528, 3C 298 we use the SFRs provided in the respective papers (see Table \ref{tab:sample}) based on spectral energy density (SED) fitting due to the lack of reliable IR data.
\newline The optical classification in star forming, LI(N)ER and Seyfert galaxies is based on the galaxy's BPT diagrams, which will be presented in a forthcoming paper. An overview of the sample including ancillary data of the galaxies can be found in Table \ref{tab:sample}.

\subsection{Calculation of Outflow Properties}
\label{sec:calc_of_prop}
In order to perform a comparison to previous outflow studies, we calculate the outflow properties in the integrated spectrum. To calculate the integrated spectrum we sum all spaxels with a r-band SNR > 3. In our analysis, we use the broad H$\alpha$ component to calculate outflow properties. Some studies have used [OIII] instead, which gives lower outflow rate values \citep{Carniani2015}, likely because it does not properly account for the lower ionization phases of the outflow. The ionized outflow mass can be inferred from the H$\alpha$ luminosity in the outflow (i.e. broad component), $L_{\rm H\alpha, OF}$, and the electron density, $n_{\rm e}$, as follows \citep{Carniani2016}:

\begin{equation}
    M_{\mathrm{OF,ion}} [\mathrm{M}_{\odot}] = 6.1\times10^{8}\left(\frac{L_{\mathrm{H\alpha, OF}}}{10^{44}~\mathrm{erg s^{-1}}}\right)\left(\frac{n_{\mathrm{e}}}{500~\mathrm{cm^{-3}}}\right)^{-1}.
    \label{eq:mass_outflow}
\end{equation}
We then obtain the mass outflow rate at radius $r_{\rm OF, ion}$ by dividing the outflowing gas mass by the dynamical time-scale, $\tau_{\rm OF}$:
\begin{equation}
    \dot{M}_{\rm OF, ion} = \frac{M_{\rm OF, ion}}{\tau_{\mathrm{OF, ion}}} = \frac{M_{\rm OF, ion}v_{\rm OF, ion}}{r_{\mathrm{OF, ion}}},
    \label{eq:mass_outflow_rate}
\end{equation}
where $v_{\rm OF, ion}$ and $r_{\rm OF, ion}$ are the velocity and extent of the ionized outflow, respectively. We define the outflow velocity $v_{\rm OF}$ = $\Delta v$ + FWHM$_{\rm broad}$/2 to be consistent with previous works \citep[e.g.][]{Rupke2005,Fluetsch2019}. 

The radius of the ionized outflow is defined as the radius which encompasses 50~$\%$ of the total flux of the broad H$\alpha$ component as observed in the data. This definition is used to have a comparable definition as to the one we used for the molecular outflow phase \citep{Fluetsch2019}. By directly determining the electron density in the outflow (see Section \ref{sec:density_extinction}), we can drastically reduce the main uncertainty of this computation. Hence, the typical uncertainty on the ionized mass outflow rate is estimated to be $\sim$ 0.3~dex, mainly due to projection effects which affect the velocity and radius measurements as well as uncertainties in determining the electron density.
The kinetic power of the outflow is given by 0.5$v_{\rm OF}^{2}\dot{M}_{\rm OF}$. The properties (mass, radius and velocity) of the ionized outflows in our sample and the extended sample can be found in Table \ref{tab:ion_of_prop}.

The mass outflow rate of the neutral atomic outflow can be computed using equation \ref{eq:mass_outflow_rate} and using the neutral outflow velocity and radius and mass. Using the hydrogen column density obtained in Section \ref{sec:neutral_atomic}, the neutral mass outflow rate is given by 

\begin{equation}\label{eq:neut_outflow_rate}
\begin{split}
    \dot{M}_{\rm OF,neu} [\mathrm{M}_{\odot}/\mathrm{yr}] &= 11.5 \sum_{i=1}^{N} \left(\frac{C_{\Omega}}{0.4} C_{\mathrm{f}}\right)  \left(\frac{r}{10~\mathrm{kpc}}\right)\\& \times \left(\frac{N \mathrm{(H)}}{10^{20}\: \mathrm{cm^{-2}}}\right)\left(\frac{\Delta v}{200\: \mathrm{km~s^{-1}}}\right),
\end{split}
\end{equation}
where we sum over all outflowing components, from $i$ = 1 to N.
$C_{\rm f}$ and $C_{\rm \Omega}$ are the local (see eq. \ref{eq:atomic_overlap}) and large-scale covering factor, respectively. Following \cite{Rupke2005}, we assume $C_{\rm \Omega}$ = 0.4 for LIRGs and $C_{\rm \Omega}$ = 0.8 for ULIRGs. The local covering factor is determined through the fitting of the sodium doublet. $N$(H) is the hydrogen column density along the line of sight and $r$ is the radius. The parameters for the fitting of the Na~I~D profile are shown in Table \ref{tab:neu_of_prop}.

Similar to the ionized outflow, we also calculate the properties of the neutral atomic outflow in the integrated spectrum. The spatial extent of the neutral outflow, $r_{\rm OF, neu}$,
is calculated as the radius that encompasses 50~$\%$ of the total flux of all outflowing sodium components similar to what was done for the ionized outflow radius. For the Seyfert 1 galaxies (2 objects) we do not fit the sodium absorption doublet because of the degeneracy caused by the strong quasar continuum (see Section \ref{sec:map_outflows}). 

\subsection{Extended sample}
\label{sec:extended_sample}
To study the relation between different outflow phases, we include galaxies studied in the literature with outflows detected in several phases. 
We refer to these galaxies combined with our galaxies with multiphase outflows as the \textit{extended sample}. In total, the extended sample consists of 31 galaxies with outflows studied in multiple phases.
Our extended sample includes the multiphase outflows studied by \cite{Rupke2013}, \cite{Rupke2017}, \cite{Fluetsch2019} and \cite{Husemann2019}. We also add three targets with redshifts $z$ > 1, which all have measurements of outflows in several phases: zC400528 \citep{Herrera-Camus2019}, XID2028 \citep{Cresci2015a,Brusa2018} and 3C 298 \citep{Vayner2017}. 

The outflow measurements in these galaxies are homogenised to allow a fair comparison to our MUSE sample. In particular, to re-calculate the properties of the ionized outflows, we use the same
\begin{itemize}
    \item formula to calculate mass outflow rate (eq. \ref{eq:mass_outflow_rate}), i.e.  without the factor of three as used in some other works,
    \item electron density ($n_{\rm e}$ = 500~cm$^{-3}$, average electron density of outflows, see below)
    \item definition of outflow velocity if possible ($v_{\rm OF}$ = $\Delta v$ + FWHM/2)
    \item attenuation curve to correct for extinction \citep{Calzetti2000} with $R_{\rm V}$ = 4.05
    \item tracer of outflowing gas, i.e. H$\alpha$ instead of [OIII] (see Section \ref{sec:calc_of_prop})
    \item conversion from H$\alpha$ luminosity into outflow mass, i.e. equation \ref{eq:mass_outflow}.
\end{itemize}
For the calculation of neutral atomic outflow properties, we adopt the same
\begin{itemize}
    \item formula to calculate mass outflow rate (eq. \ref{eq:mass_outflow_rate}), i.e.  without the factor of three to be consistent with the calculation of mass outflow rates in other phases,
    \item values for the ionization fraction, galaxy's sodium abundance and the depletion onto dust (see Section \ref{sec:neutral_atomic}).
\end{itemize}
Finally, to ensure that molecular outflow rates and energetics are calculated consistently across the sample, we assume the same
\begin{itemize}
    \item formula to calculate mass outflow rate (eq. \ref{eq:mass_outflow_rate}), i.e.  without the factor of three as used in some other works
    \item same conversion factor ($\alpha_{\rm CO}$ = 0.8 M$_{\odot}$/(K km s$^{-1}$ pc$^{2}$)$^{-1}$
    \item definition of outflow velocity if possible ($v_{\rm OF}$ = $\Delta v$ + FWHM/2).
\end{itemize}
\section{Results}

\subsection{Prevalence of outflows}

Out of the 26 galaxies analysed in our MUSE sample, 13 galaxies show clear signs of an ionized outflow and 10 show evidence of neutral atomic outflowing gas, 8 galaxies have outflows in both phases. The number of neutral outflows may be larger than this number as any sodium absorption in Seyfert I galaxies
(see Section \ref{sec:calc_of_prop}) might be overshadowed by broad He~I lines. However, we found in other objects a more extended Na~ID outflow could partially absorb the He~I emission line (see e.g. Fig. \ref{fig:IRAS14378}).

In our study, we find that most galaxies with outflows have outflow signatures in both phases, but there are a few exceptions. For instance, IRAS 23128-5919  \citep[also  studied in][]{Maiolino2017} has a clear ionized outflow, but no detectable neutral atomic gas at high velocity. 
However, one should be aware, that the discrepancy might be due to sensitivity in the different phases rather than due to the physical properties of the outflow. For each galaxy we present maps of the ionized and neutral atomic phases and the integrated spectra in \mbox{\ref{sec:map_outflows}}.

\subsection{Characterisation of outflows}

\subsubsection{Electron density}
\label{sec:density_extinction}
We compare the electron density of the outflowing gas and the disc. This not only informs our understanding of the physical conditions of the ionized ISM, but it is also important to calculate the mass and kinetic energy of the ionized outflow (which are both inversely proportional to the electron density $n_{\rm e}$, see Section \ref{sec:calc_of_prop}). Indeed, the electron density has been one of the main uncertainties in these calculations \citep{Harrison2018}.

\begin{figure}
\centering
\includegraphics[width=\columnwidth]{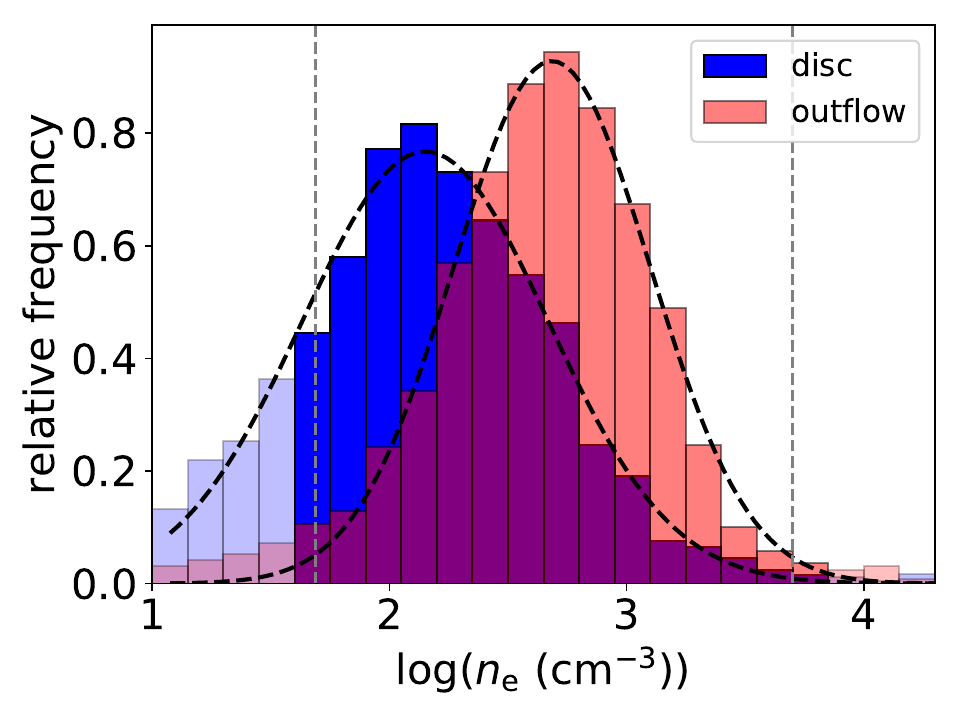}
\caption{In blue and red, the distribution of the electron densities ($n_{\rm e}$) of the narrow (disc) and broad (outflow) components of all galaxies with ionized outflows are shown, respectively. The dashed lines are the Gaussian fit to the distributions. Densities outside the interval 50--5000~cm
$^{-3}$ (outside the grey dashed lines) should be considered as lower/upper limits since the [SII] doublet ratio is little sensitive to the electron density outside this range.}
\label{fig:density_all}
\end{figure}

We estimate the electron density based on the [SII]$\lambda$6717/[SII]$\lambda$6731 ratio, assuming a temperature of $T_{\rm e}$ = 10$^{4}$~K \citep{Osterbrock2006}. The density is calculated for each spaxel in a galaxy using the prescription in \cite{Sanders2016}:

\begin{eqnarray}
n_{\rm e} = \frac{cR-ab}{a-R},
\end{eqnarray}
where $R$ = [SII]$\lambda$6717/[SII]$\lambda$6731 and $a$ = 0.4315, $b$ = 2107, $c$ = 627.1. The [SII] ratio is sensitive to density in the range 50 < $n_{\rm e}$ < 5000 cm$^{-3}$ \citep{Osterbrock2006}. Below and above these limits, the [SII] ratio varies little and is not suitable to accurately determine densities.

In Fig. \ref{fig:density_all} we show the electron density distribution of all spaxels of the narrow and broad components which trace dynamically quiescent gas in the disc and gas in the outflow, respectively. The blue and red distribution show the electron density values for the narrow and broad components, respectively. For each [SII] component we require a SNR of at least 2 for both sulphur lines. We observed very little change in these relative distributions even if we increase the S/N threshold to 3 (or even 5) for each of the sulphur lines (although the statistics becomes (much) poorer). As mentioned above, densities outside the interval 50--5000~cm$^{-3}$ should be considered as lower/upper limits at the corresponding boundaries, as the [SII] doublet ratio is little sensitive to the electron density outside this interval. To account for the fact that for nearby galaxies we have more spaxels and therefore the distribution might be biased towards them, we weigh spaxels of each galaxy differently. We do that in such a manner that each galaxy contributes equally to the final histogram regardless of the number of spaxels in the broad and narrow components. This is done by weighing each spaxel by the inverse of the total number of spaxels in the galaxy to which this spaxel belongs. The distributions in Fig. \ref{fig:density_all} are wide because we add up several galaxies with different densities, but there is also a considerable range of densities in each individual galaxy as well as the uncertainty associated to individual measurements.
We fit these distributions with Gaussians (shown as black dashed lines in Fig. \ref{fig:density_all}) and obtain the following centroids: $n_{\rm e,broad}$ = 490~cm$^{-3}$ and $n_{\rm e,narrow}$ = 145~cm$^{-3}$, and dispersions $\sigma_{\rm broad}$ = 0.41~dex and $\sigma_{\rm narrow}$ = 0.52~dex. If the values outside of the range 50--5000~cm$^{-3}$ are ignored, the density of the outflow remains the same and the density of the narrow component is $n_{\rm e,narrow}$ = 190~cm$^{-3}$. The density of the outflowing gas is on average a factor of three higher than the density of the disc. This trend remains even if we do not weigh spaxels differently as described above. The trend is also robust if we consider spaxels where both components are present (i.e. we include only parts of the disc where we also have an outflowing component). The density of the broad component also does not change if we apply more stringent criteria for the identification of the outflowing gas, such as we require a minimum blueshift of --100~km~s$^{-1}$ and a width of $\sigma_{\rm broad}$ >150~km~s$^{-1}$ for the broad component to ensure we only trace outflowing gas with the broad component. The finding is equally clear both in galaxies with higher AGN fraction ($\alpha_{\rm bol}$>0.5) and with lower AGN fraction ($\alpha_{\rm bol}$<0.5). The galaxies with stronger AGN show slightly higher densities ($\sim$ 0.1--0.2~dex higher) for both the narrow and the broad components. In Appendix \ref{fig:density_agn_vs_sf_host}, we show that AGN host galaxies have on average a higher disc electron density than star forming galaxies. In a forthcoming paper, we will show the electron densities in several individual galaxies.

We note that recent studies, based on the investigation of auroral lines have shown that the [SII] doublet tends to underestimate the gas density in galactic outflows even by 1--2 orders of magnitude \citep{Holt2011, Rose2017, Davies2020}. However, this would make our result that outflows are denser than their galactic discs even stronger.

In addition, we observe the trend that outflows are denser than the disc in all individual galaxies apart from IRAS 21453-3511 (discussed below) and IRAS 15115+0208 (although in this galaxy the number of spaxels with sufficient SNR (i.e. > 3) is not high enough to definitely confirm this).

This result qualitatively agrees with several previous studies \citep{Holt2011, Arribas2014, Villar2014, Perna2017, Rose2017, Mingozzi2019, Davies2020}, who all found outflowing gas to be denser than gas in the disc. The simplest explanation for dense gas in the outflow is that the expelled gas is compressed \citep[e.g.][]{Bourne2015, Decataldo2019}. Our average values are similar to the ones found by \cite{Arribas2014}, who found densities of $n_{\rm e, broad}$ = 460 $\pm$ 70 cm$^{-3}$ and $n_{\rm e, narrow}$ = 300 $\pm$ 30 cm$^{-3}$ also in a sample of (U)LIRGs. \cite{Villar2014} and \cite{Perna2017} find a higher electron density than this work of $\gtrsim$ 1000 cm$^{-3}$ in the outflowing gas, likely because they focus on more extreme objects, considering only AGN hosts or higher redshift targets.



As mentioned above, although this trend seems robust over a range of galaxy properties, there are few rare individual objects, where the density in the disc is similar or higher than in the outflow. For instance, in IRAS 21453-3511, the density of the disc and outflow are very similar. 
In this galaxy the electron densities for the two components are $n_{\rm e,broad}$ = 275~cm$^{-3}$ and $n_{\rm e,narrow}$ = 288~cm$^{-3}$. Another isolated case is NGC 6810, where some parts of the outflow are less dense than the disc (see Venturi et al. in prep.).
We cannot yet explain why some galaxies show this trend, but it seems to be unique to sources that are classified as star forming galaxies according to their [SII]-BPT diagrams. We can only speculate that possibly some star formation driven outflows do not lead to enhanced densities in the outflowing gas due to a different mechanism at play.

\begin{figure}
\centering
\includegraphics[width=\columnwidth]{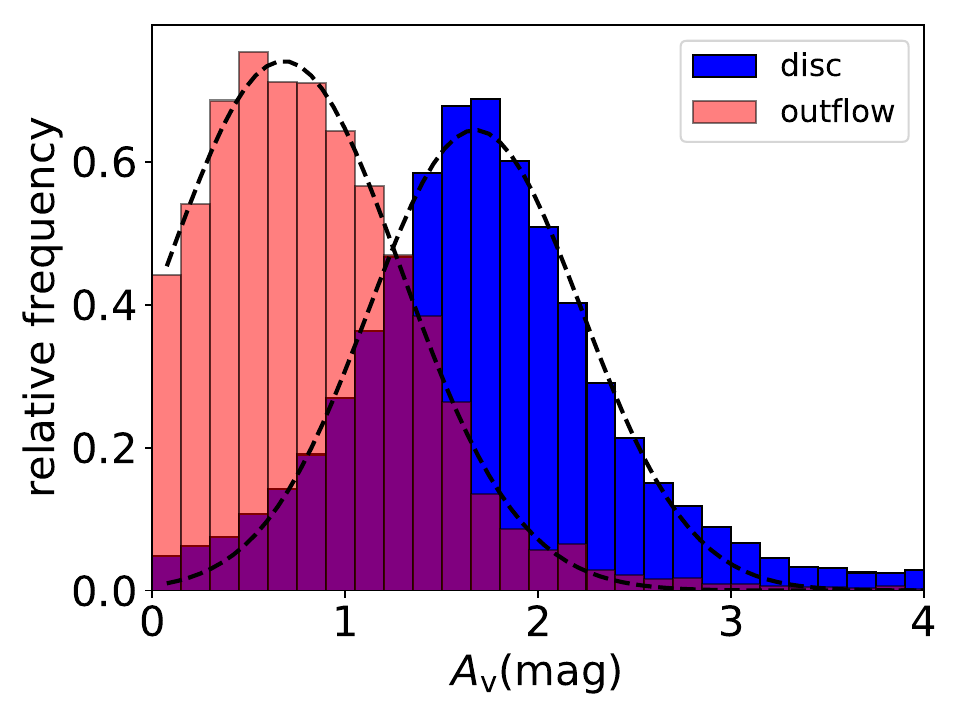}
\caption{Distribution of the values of visual extinction of the narrow (disc, in blue) and broad components (outflow, in red). The centroids of the Gaussian to these distributions are $A_{\rm V}$ = 1.67~mag and $A_{\rm V}$ = 0.68~mag for the narrow and broad components, respectively.}
\label{fig:extinction_all}
\end{figure}

\subsubsection{Dust extinction}

We compare the dust extinction, $A_{\rm V}$, of the narrow and broad components. We estimate the extinction using the Balmer decrement H$\alpha$/H$\beta$ and assuming the \cite{Calzetti2000} attenuation law with $R_{\rm V}$ = 4.05. The intrinsic flux ratio of H$\alpha$/H$\beta$ is little dependent on density and temperature and any significant deviation in the observed ratio can thus by attributed to interstellar dust extinction. We assume an intrinsic H$\alpha$/H$\beta$ = 2.86 for $T$ = 10$^{4}$~K \citep{Osterbrock2006}. Furthermore, we require the SNR on H$\beta$ to be $>$3. As in any external galaxy, the observed Balmer decrement (or colour excess) does not trace a dusty screen but dust in the ISM mixed with the ionized regions emitting H$\alpha$ and H$\beta$.

The resulting extinction distributions are shown in Fig. \ref{fig:extinction_all} with the narrow component shown in blue and the broad component in red. The narrow component has a visual extinction more than twice as large as the broad component. The centres of the Gaussian fit to the distributions are $A_{\rm V, broad}$ = 0.68~mag and $A_{\rm V, narrow}$ = 1.67~mag. The median extinction values for different galaxies range from $\sim$1 to $\sim$ 4~mag in the narrow component and from $\sim$0.4~mag to $\sim$2.7~mag in the broad component. The difference in extinction between narrow and broad components, $\Delta A_{\rm V}$, varies considerably across the sample and this will be discussed in more detail below.
\newline Some works \citep{Rose2017, Mingozzi2019} confirm the finding that outflows are less affected by extinction than the disc, others find the opposite result \citep{Holt2011, Villar2014}. It should be noted that the studies have used different samples. \citet{Villar2014} studies type II quasars (with various levels of radio luminosities), while \citet{Holt2011} investigates a type I radio-loud quasar. Also \citet{Mingozzi2019} focusses on Seyfert 2 galaxies and \citet{Rose2017} studies ULIRGs with a mix of type I and type II AGN. These different results regarding the relative dust extinction can be explained by a combination of two effects. First, the near side of the outflow is generally less obscured than the disc, while the far side may be extinct by the disc to such a degree that it cannot be detected at optical wavelengths in many galaxies. This seems to be the main explanation for galaxies with lower extinction in the outflow than in the disc.
Second, as there are a large number of objects with lower extinction in the disc as e.g. in \cite{Villar2014}, it seems likely that at least some outflows are very dusty, and dustier than the disc. 
This has indeed been proposed to explain very high UV luminosities as a consequence of dust scattering in the outflows of the starburst galaxies NGC 253 and M82 \citep[e.g.][]{Hoopes2005}. Radiation pressure-driven models of feedback also predict that dusty gas should preferentially be expelled by powerful outflows \citep{Murray2005, Ishibashi2016}. Unequivocal observational evidence of dust in most or all outflows would also help us greatly to understand molecular outflows. The presence of dust in the outflow is a requisite in many models to explain molecule formation in the outflow \citep{Richings2018}.  
To quantify these effects, one would determine the extinction for the approaching and receding outflow part separately. This could be achieved by separate fitting of the red and blue components of the Balmer emission lines \citep[see][]{Venturi2018}. Such an analysis suffers from additional fitting degeneracy and will therefore not be discussed in more detail in this paper.

\begin{figure}
\centering
\includegraphics[width=\columnwidth]{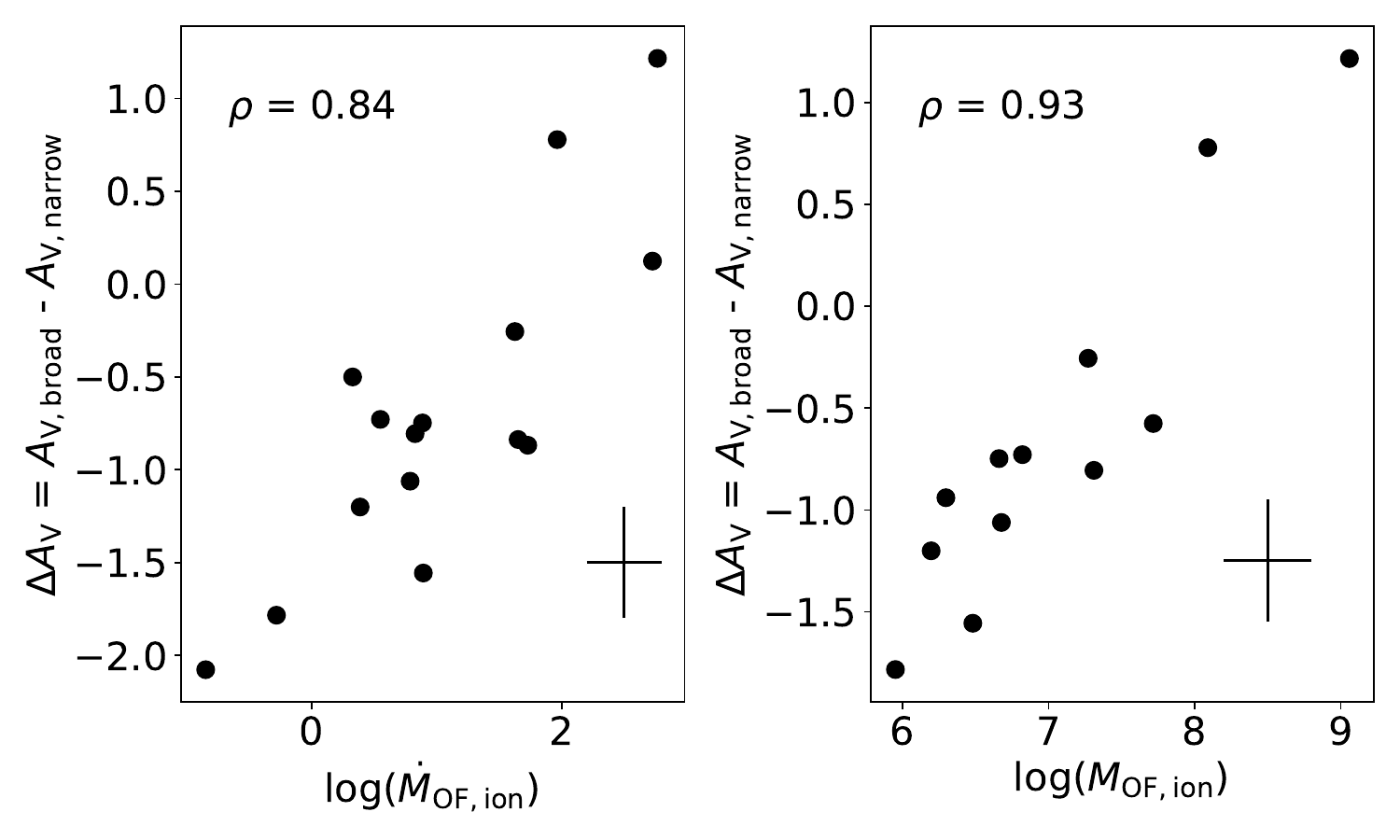}
\caption{Difference in extinction between the narrow and broad components, $\Delta A_{\rm V}$ = $A_{\rm V, broad}$ - $A_{\rm V, narrow}$, as a function of ionized mass outflow rate (left) and ionized outflowing gas mass (right). The typical error is given by the black cross.}
\label{fig:extinction_delta_correlations}
\end{figure}

\begin{figure}
\centering
\includegraphics[width=\columnwidth]{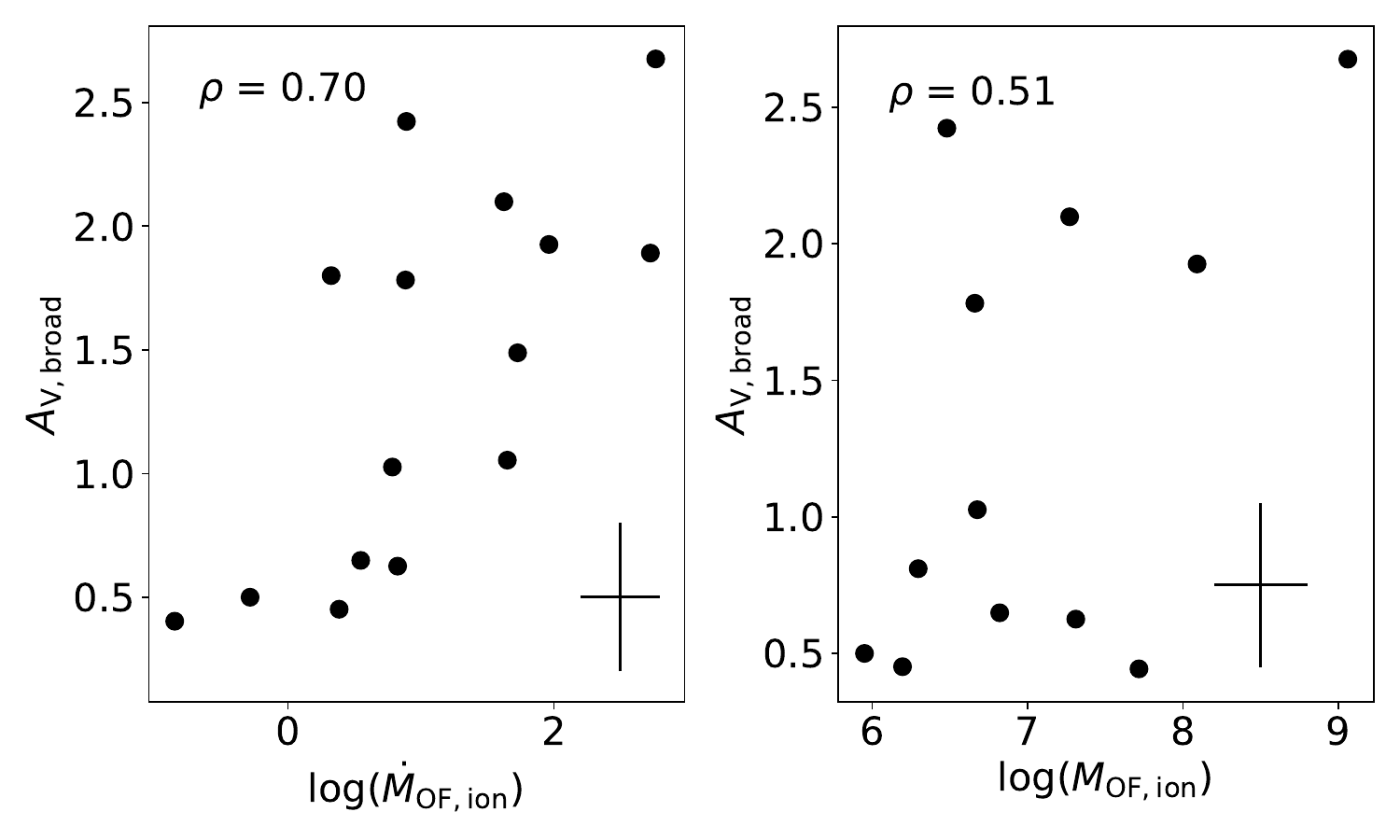}
\caption{Extinction of the broad component, $A_{\rm V, broad}$, as a function of ionized mass outflow rate (left) and ionized gas mass (right). The typical error is given by the black cross.}
\label{fig:extinction_broad_correlations}
\end{figure}

To evaluate the driver of different extinction values in the broad and narrow components, we look at how the galaxy's median $\Delta A_{\rm V}$ = $A_{\rm V, broad}$ - $A_{\rm V, narrow}$ scales with integrated outflow or galaxy properties. There is no clear correlation between $\Delta A_{\rm V}$ and $\alpha_{\rm bol}$, SFR, the neutral atomic or the molecular outflow mass or mass rate. However, the ionized mass outflow rate ($\dot{M}_{\rm OF, ion}$) correlates with $\Delta A_{\rm V}$ (with a Pearson correlation coefficient of $\rho$ = 0.84) as shown in Fig. \ref{fig:extinction_delta_correlations} on the left. This correlation appears to be driven by the mass of the ionized outflow (right-hand side of Fig. \ref{fig:extinction_delta_correlations}), as this leads to an even tighter correlation with $\Delta A_{\rm V}$ (correlation coefficient $\rho$ = 0.93).
Intuitively one might assume that this trend is mainly driven by $A_{\rm V, broad}$, which is proportional to the outflow mass. Higher $A_{\rm V, broad}$ gives a higher intrinsic H$\alpha$ luminosity for the same observed H$\alpha$ flux. This
H$\alpha$ luminosity in turn is proportional to the outflow mass (see equation \ref{eq:mass_outflow}).
In order to test this hypothesis, we plot the extinction of the broad component ($A_{\rm V, broad}$) as a function of the ionized mass outflow rate and ionized outflow mass in Fig. \ref{fig:extinction_broad_correlations}.
The trends in \ref{fig:extinction_broad_correlations} are much weaker than in Fig. \ref{fig:extinction_delta_correlations}, hence indicating that the latter are not purely driven by a potential co-variance between the two axes through H$\alpha_{\rm broad}$. Part of this might be explained by errors in measuring both the visual extinction and the outflow mass (rate). To quantify the effect of errors leading to a spuriously higher correlation, we run a Monte Carlo simulation, assuming errors of 0.3 dex on outflow mass (rate) and 0.3 dex on $A_{\rm V}$ and there is only a 2~$\%$ chance that the tighter correlation as seen here with $\Delta A_{\rm V}$ than with $A_{\rm V, broad}$ is due to errors. Hence, it seems unlikely, that errors are solely responsible for the tighter correlation in Fig. \ref{fig:extinction_delta_correlations}.
Instead it seems the ionized outflow mass does drive the \textit{difference} in extinction between broad and narrow components. In other words, the most massive ionized outflows have comparatively more dust in the outflow than in the disc. In these outflows, mainly dusty gas might be expelled, enhancing the extinction in the outflow while simultaneously lowering the extinction in the disc. This could be further indication that dust plays an important role in outflows, especially for the most massive of them. {In particular this finding would support models according to which galactic outflows may be due to radiation pressure on dusty clouds \citep{Fabian2012, Ishibashi2018}.}

\subsection{Comparison of outflow phases}
\label{sec:comp_phases}

\begin{figure}
\centering
\includegraphics[width=\columnwidth]{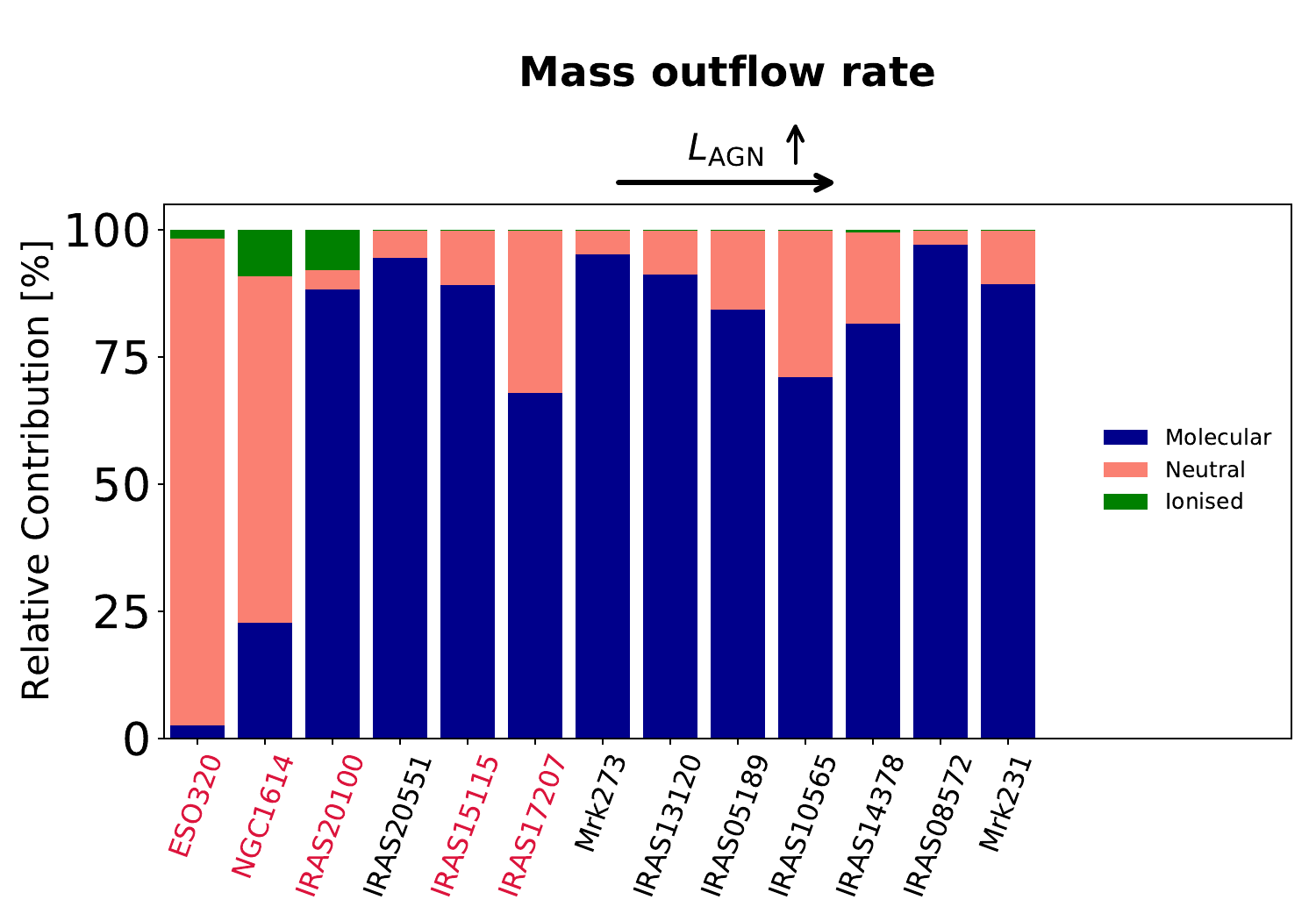}
\caption{Relative contribution of the molecular (blue), ionized (green) and neutral atomic (red) phases to the total mass outflow rate. The galaxies whose names are in red are star forming galaxies. The galaxies are sorted by increasing AGN luminosity from left to right.}
\label{fig:three_phase_comp}
\end{figure}
\begin{figure}
\centering
\includegraphics[width=\columnwidth]{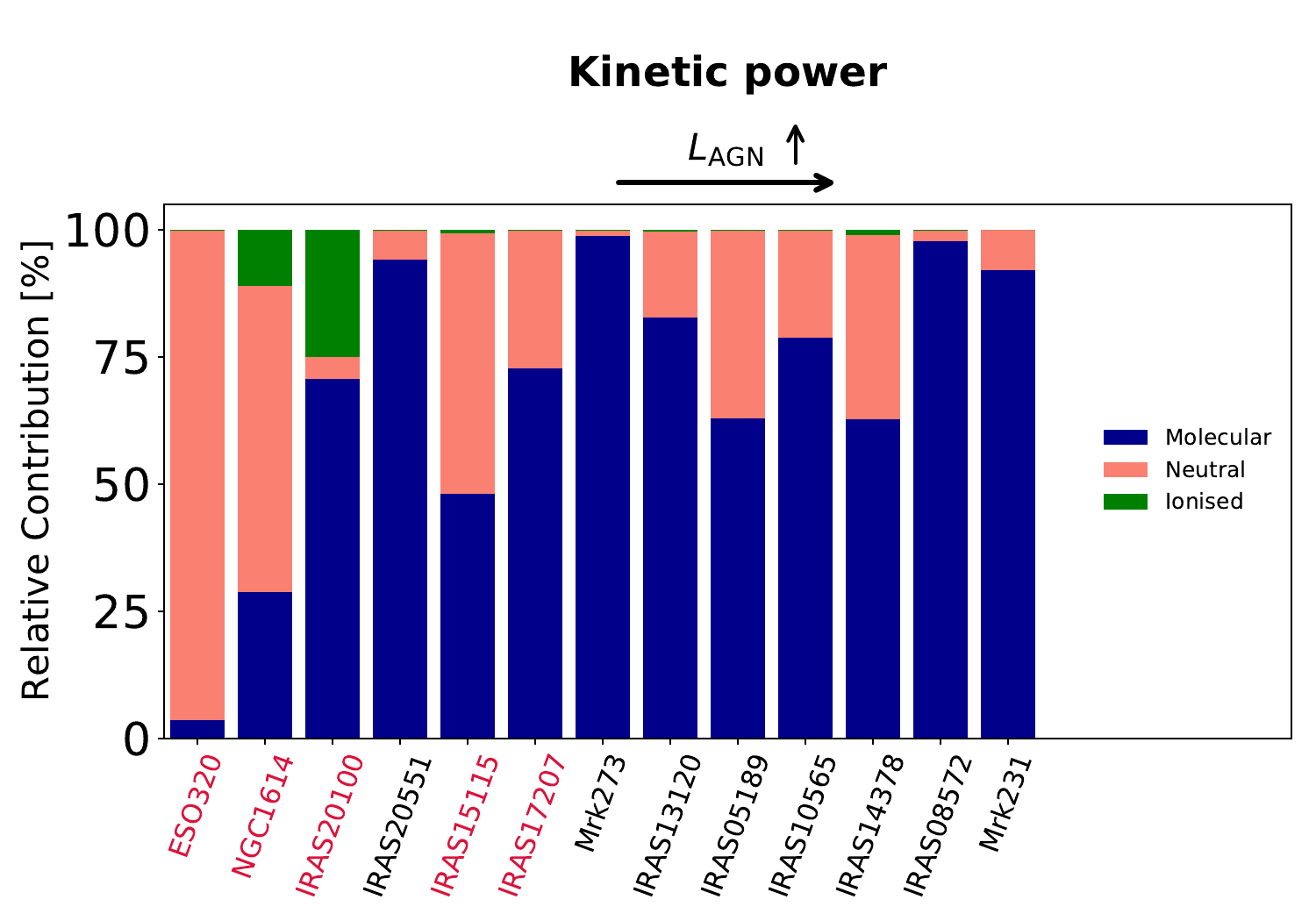}
\caption{Relative contribution of the molecular (blue), ionized (green) and neutral atomic (red) phases to the total kinetic power of the outflow. The galaxies whose names are in red are star forming galaxies.}
\label{fig:three_phase_comp_power}
\end{figure}

In this subsection we compare the ionized and neutral atomic outflows to the molecular outflow properties. 

For the following analysis, we also add the extended sample described in Section \ref{sec:extended_sample}. which includes 31 galaxies with at least two outflow phases and 13 galaxies with all three outflow phases (excluding galaxies which have only an upper limit in at least one phase). 

We first test how much each phase (molecular, ionized and neutral atomic) contributes to the total outflow rate for galaxies with measurements of all outflow phases. This is important to allow for a fair comparison to predictions of models and simulations of galaxy evolution, which look at the total impact of the outflow, while most observations focus on one single phase instead and thus might underestimate outflow properties. The outflow masses were homogenised according to the procedure outlined in Section \ref{sec:extended_sample}.
The relative contributions of different phases to the total outflow rate are shown in Fig. \ref{fig:three_phase_comp}, where the molecular contribution is shown in blue, the neutral atomic and the ionized in red and green, respectively. The galaxies are sorted by AGN luminosity which increases from left to right. The molecular mass outflow rate amounts to more than half the total mass outflow rate in 11 out of the 13 galaxies. The two galaxies where this is not the case are both classified as star forming galaxies. Some star forming galaxies, whose names are shown in crimson in Fig. \ref{fig:three_phase_comp}, display larger mass outflow rate contributions from the ionized and the neutral atomic phase than AGN and LI(N)ER galaxies. The ionized phase is negligible (<5~$\%$ of the total outflow rate) in AGN and LI(N)ER galaxies and is only larger than that in two star forming objects (NGC1614 and IRAS 20100-4156). If the gas densities in the ionized outflow are even larger than estimated through the [SII] doublet, as suggested by the results based on the auroral lines discussed in Section \ref{sec:density_extinction}, then this would make our results even stronger. In other words, the fraction of the outflow rate associated with the ionized component would be even lower.

A similar picture emerges if we study the relative contribution of different phases to the total kinetic power. The kinetic power is given by $P_{\rm OF}$ = $\dot{M}_{\rm OF} v_{\rm OF}^{2}$/2. The associated distribution is shown in Fig. \ref{fig:three_phase_comp_power}, where the relative contribution of the molecular (blue), neutral atomic (red) and ionized phases (green) are displayed. Star forming galaxies have significant contributions from the ionized and neutral atomic phases, ranging from $\sim$40~$\%$ to $>$95~$\%$ of the total kinetic power.
In AGN and LI(N)ER galaxies, the molecular component is more dominant than in star forming galaxies. There is, however, significant variation from galaxy to galaxy. Studying the relative contribution of different phases to the outflow rate and the kinetic power as a function of SFR or $\alpha_{\rm bol}$ revealed no clear trends (see e.g. Section \ref{fig:ion_molec_SFR}).

\begin{figure*}
\centering
\includegraphics[width=\textwidth]{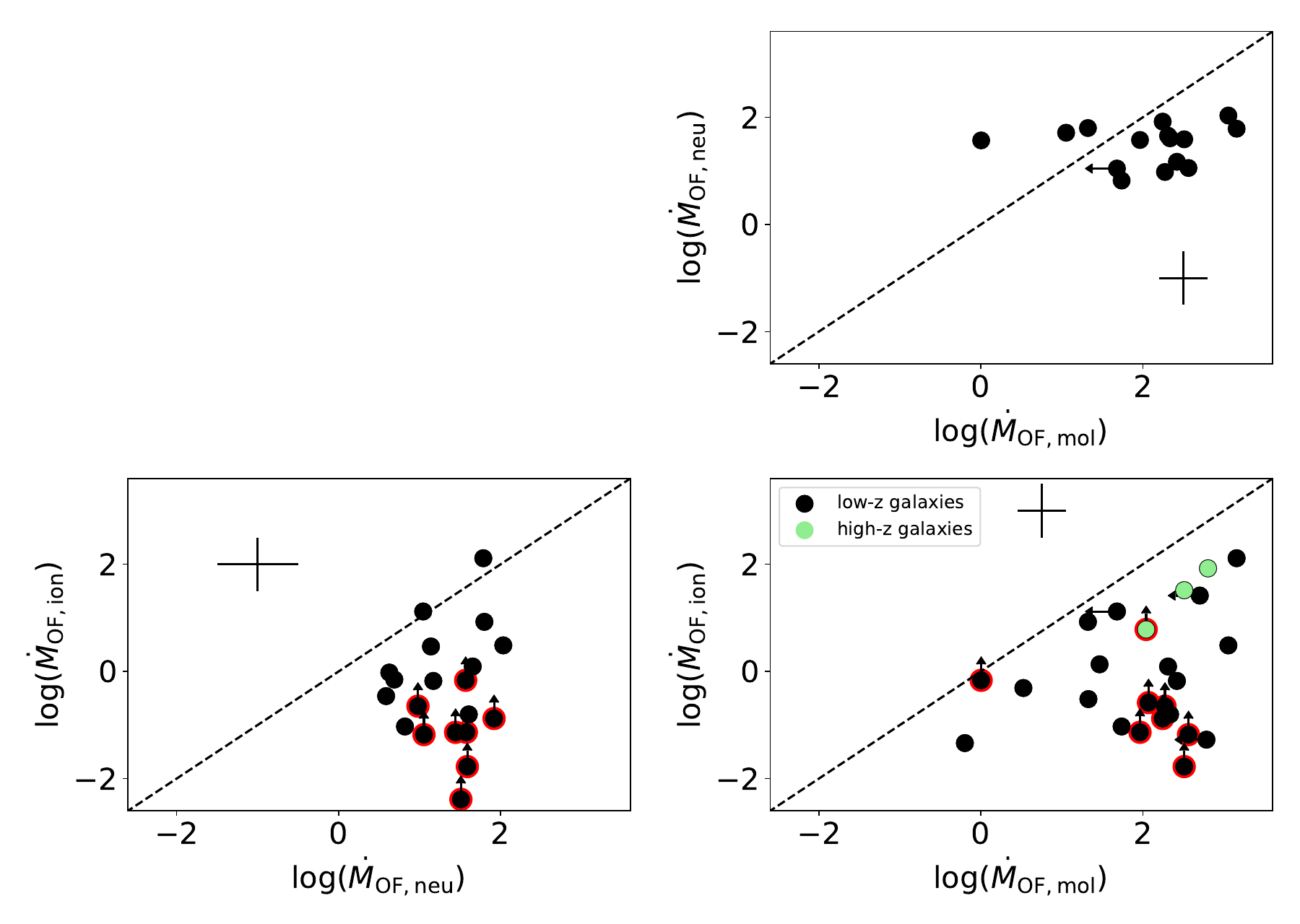}
\caption{Comparison between the outflow rates in the ionized, molecular and neutral gas phases. The black symbols represent the low-redshift ($z$<0.2) sources while the green symbols are the high-redshift ($z$>1) targets. The dashed black line in all plots shows the 1:1 correlation. Typical errors are given by the black crosses. The data points with red edges have an uncertain or no extinction correction in the ionized outflow phase and should be treated as lower limits.}
\label{fig:multiphase_comp}
\end{figure*}

Next we investigate the mass outflow rate of the three different phases, ionized, neutral atomic and molecular for galaxies which have measurements (or upper limits) for at least two phases. This comparison can be seen in Fig. \ref{fig:multiphase_comp}, where we plot the different mass outflow rates against each other The samples in the three plots are slightly different. To increase the statistics, in each plot we show the galaxies with measurements in both of the phases plotted. We distinguish between high-redshift sources ($z$ > 1, green) and low-redshift sources ($z$ < 0.2, black). The data points with a red edge have no reliable extinction correction for the ionized phase and thus should be treated as lower limits.
These figures clearly show that the ionized mass outflow rate is much smaller (up to 2--3 orders of magnitude smaller) than the neutral atomic or molecular mass outflow rate. A major caveat is that some ionized outflow rates might be underestimated because of an uncertain extinction correction. But even if these points are ignored, the ionized outflow rates are significantly lower (cf. Fig. \ref{fig:three_phase_comp}, which only includes galaxies with proper extinction correction). This confirms previous findings which estimate the ionized phase to contribute only minimally \citep{Rupke2013, Carniani2015, RamosAlmeida2019}.
The figure on the top right highlights that outflow rates in the molecular and neutral atomic phase are similar (usually within 1 dex). This picture might be slightly skewed, however, as we include upper limits in the molecular and ionized, but not in the atomic neutral outflow phase. However, the relative contribution of the ionized outflow rate to the total outflow rate varies from galaxy to galaxy (from $\leq$ 1~$\%$ to 10's of $\%$) and a simple prescription as in \cite{Fluetsch2019} might not be fully appropriate to determine the total outflow budget.  If electron densities from auroral or trans-auroral lines are a better estimate than the densities based on the [SII] doublet (as suggested by \cite{Davies2020}, see Section \ref{sec:density_extinction}), then the contribution of the ionized phase would drop even further by additional 1--2 orders of magnitude and hence it would become negligible also in star forming galaxies.

\begin{figure}
\centering
\includegraphics[width=\columnwidth]{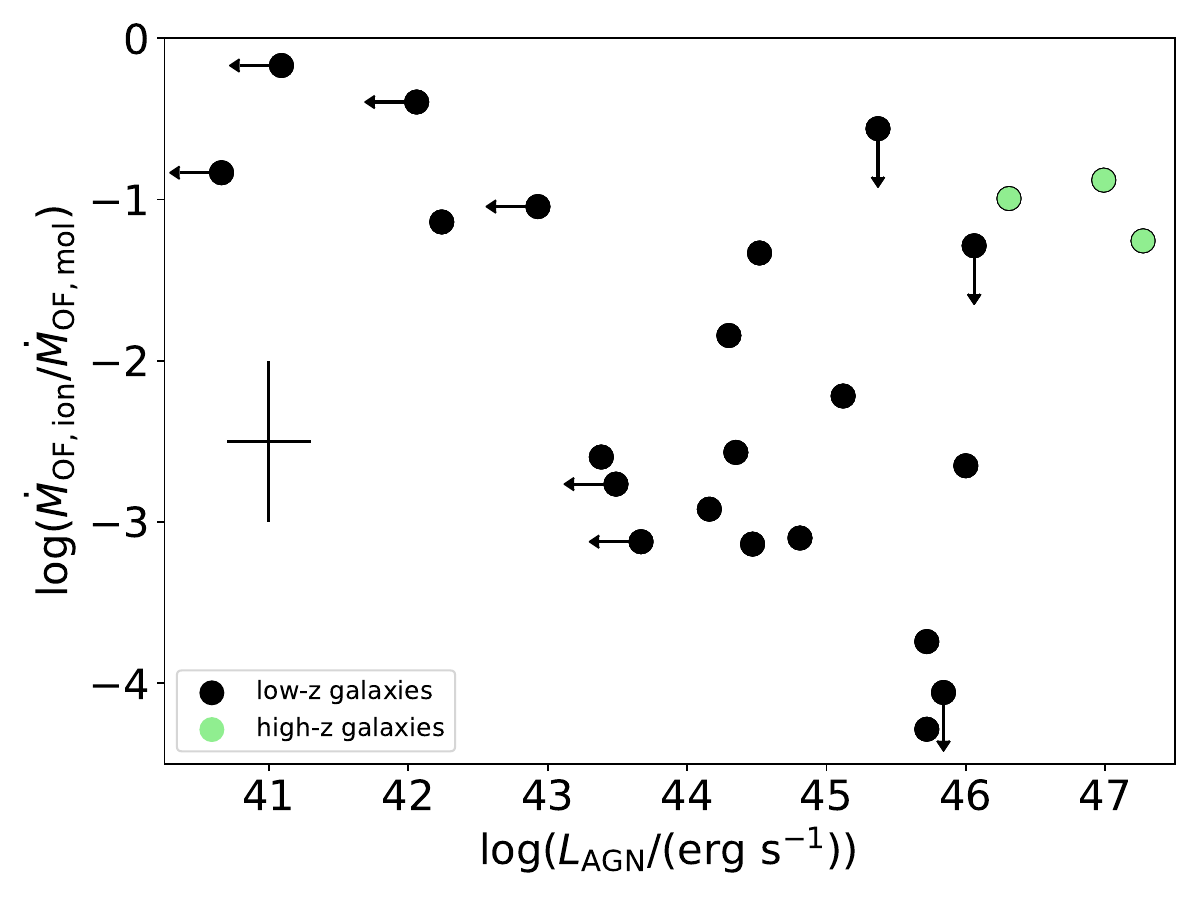}
\caption{Ratio of ionized and molecular mass outflow rate as a function of AGN luminosity. The green data points represent the high-redshift sources and the black data points the low-redshift targets.}
\label{fig:ion_molec_LAGN}
\end{figure}

Another open question is also how the relative contribution of ionized and molecular gas to the total outflow rate changes with increasing AGN luminosity. Previous studies suggested that the ionized gas contribution to the total mass outflow rate becomes increasingly important with higher AGN luminosity and is similar to the molecular outflow rate above AGN luminosities of 10$^{46}$ erg~s$^{-1}$ \citep{Fiore2017, Bischetti2019}. However, these studies worked with disjoint samples, with a limited number of sources or by restricting the AGN luminosity range. We study this trend in Fig. \ref{fig:ion_molec_LAGN}, where we plot the ratio $\dot{M}_{\rm OF,ion}/\dot{M}_{\rm OF,mol}$ as a function of the AGN luminosity, $L_{\rm AGN}$. The green data points represent galaxies above redshift 1, whereas low-redshift objects are shown as black data points. Contrary to some previous studies, we do not find that the ionized phase becomes more prominent at higher AGN luminosities. Instead, there is no clear trend. The contribution of the ionized mass outflow rate might actually decrease at higher $L_{\rm AGN}$ as initially found by \cite{Fluetsch2019}. The three high-redshift sources all seem to have large ionized outflow rates, which are only slightly smaller to their molecular ones. They are also the only objects with $L_{\rm AGN}$ > 10$^{46}$ erg s$^{-1}$ and are consistent with the results by \cite{Fiore2017}.
However, given the small sample, it is impossible to establish whether higher redshifts targets indeed have a fundamentally different outflow composition or whether objects with higher AGN luminosities have comparatively higher ionized outflow rates. An additional factor possibly contributing to the deviation of high-$z$ galaxies is that they suffer from observational limitations, such as reduced sensitivity to low-surface brightness and high-velocity gas.

\subsection{Spatial and velocity comparison of outflow phases}

MUSE has the capability to resolve the gas kinematics with excellent spatial resolution and we therefore also attempt to compare the different outflow phases on a spatially resolved basis. The maps of all 13 galaxies (with outflows) are presented in Section \ref{sec:map_outflows}. What is striking is that galaxies with outflows in the ionized and in the neutral atomic phase display similar morphology and kinematic structures. For instance, in IRAS 13120-5453, the velocity gradient across the field of view is very similar for the two phases. In addition, the two phases are co-spatial for large parts of the galaxy, displaying similar blueshifts. This seems to be contradicting an earlier study by \citet{Bae2018}, who finds almost no correlation between the two phases. However, their work is based on SDSS single fibre (3'') observations that, depending on the distance of the galaxy, provide integrated emission of most of the galaxy. This can blend different velocity components. In other cases, only the central regions is sampled (hence providing only a partial view of large-scale outflows). \citet{Concas2019} also finds no multiphase outflows in SF galaxies, but does detect multiphase winds in AGN hosts. They also use SDSS single fibre information (hence affected by the same issues as \citet{Bae2018}). In addition, they use the stacking technique which may wash away differences in individual galaxies. However, apart from the technical issues associated with the two works above, a possible explanation is that neutral outflows are mostly driven by star formation feedback while black hole accretion gives rise to ionized outflows, as suggested by \citet{Concas2019}. The study by \citet{Rupke2017} supports this explanation to some degree as in some of their type-I AGN with large star formations rates (e.g. IRAS 05189-2524) they also find a correlation between different phases. Similarly, also in the local type-II AGN IRAS 08572+3915 (with an SFR $\sim$ 70 M$_{\odot}$/yr) spatial coincidence between different phases (ionised, neutral atomic and even the cold and warm molecular phases) is found \citep{Herrera-Camus2020}. Hence, in many (U)LIRGs where we find an active AGN as well as star formation, the properties of the atomic neutral and ionized outflows appear to be linked. This is possibly because the AGN and SF driving mechanisms team up to drive (stronger) multi-phase outflows, as suggested by some models \citep{Koudmani2019}.
\begin{figure}
\centering
\includegraphics[width=\columnwidth]{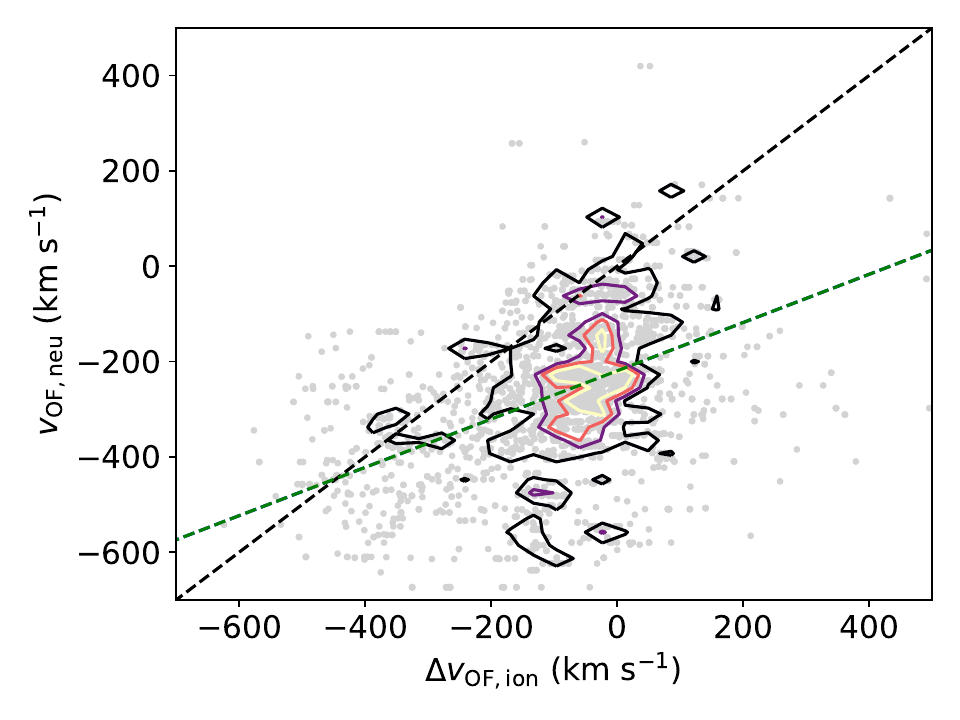}
\caption{Neutral outflow velocity ($v_{\rm 50}$) versus ionized outflow blueshift for all spaxels of all galaxies. The coloured isodensity contours correspond to 10, 30, 50 and 70 points in that bin (the parameter space has been divided into 50 bins). The dashed black line indicates the 1:1 correlation, the green dashed line is the best fit to the data points.}
\label{fig:all_velocity}
\end{figure}
In order to quantify this better, in Fig. \ref{fig:all_velocity} we compare the outflow velocity of the neutral phase with that of the ionized phase for each spaxel, for which we find outflows in both phases. The dashed black line indicates the 1:1 relation and the dashed green line is the best fit to the data points. We note that the properties of the neutral atomic phase are inferred from binned data, whereas the ionized lines can be fit in the individual spaxel. In order to account for that, we took the average of all ionized outflow velocities for each bin of neutral outflow velocity. Each grey data point represents one spaxels in one galaxy with outflows in both phases. In general, most data points lie below the black lines, i.e. the neutral outflows seem to have a larger blueshift than in the ionized gas. We remind the reader that the outflow velocity of the neutral phase is measured using $v_{\rm 50}$, the 50th percentile of the whole sodium absorption profile, whereas for the ionized outflow phase, the velocity is the blueshift of the broad component (see Section \ref{sec:calc_of_prop}). It is not possible to use the same definition for the outflow velocity due to the different spectral properties of the two tracers.
Thus, we do not expect to see a perfect 1:1 correlation in this plot, but the correlation (although with large scatter) suggests that the two phases are somehow connected.

While the ionized and the neutral atomic phases are often co-spaitial, there is no clear trend regarding their relative sizes. In some galaxies the neutral atomic outflow has a larger radius than the ionized one (e.g. IRAS 14378-3641) while in other galaxies we see the opposite trend (e.g. IRAS 13229-2934). Especially because of the Voronoi binning, which is necessary to characterise the Na~ID outflow, the need to have a strong stellar continuum for Na~ID detection, different SNRs in both phases, different tracers and also because of the faintness of outflow features, we are unable to draw unbiased conclusions about the relative sizes of outflows in different phases.



\section{Summary and conclusions}

This study investigates the properties of multiphase outflows in local (U)LIRGs. We used MUSE observations of 26 (U)LIRGs, which include both AGN hosts and starburst-dominated objects, to study the ionized gas phase including estimates of the electron density and visual extinction of the ionized gas. In addition, we investigate the neutral atomic gas phase traced by the Na~I doublet. To study the connection between different outflow phases, we supplemented our sample with outflow studies from the literature. This brings our sample of galaxies with outflows detected in at least two phases to 31 objects, out of which 13 galaxies have outflows in all three phases, molecular, neutral atomic and ionized. This makes it the largest resolved study of multiphase outflows to date.
Our main findings are:

\begin{itemize}
    \item Out of the 26 galaxies observed with MUSE/VLT, 12 galaxies show clear signs of ionized outflows, 10 galaxies have neutral atomic outflows and 8 galaxies have outflow in both phases simultaneously. 
    \item Outflowing ionized gas has on average an electron density a factor of three higher than the disc ($<n_{\rm e, disc}>$ = 145~cm$^{-3}$ and $<n_{\rm e, outflow}>$ = 490~cm$^{-3}$) (see Fig. \ref{fig:density_all}). This is also the case individually for nearly all galaxies. This finding could indicate that cloud compression is more important than cloud dissipation in the outflow. However, some star forming galaxies show similar densities in the disc and the outflow (e.g. IRAS 21453-3511).
    \item The average visual extinction is twice as high in the narrow component than in the broad component (see Fig. \ref{fig:extinction_all}). This is interpreted as the approaching side of the outflow (the one most easily detected) being less obscured than the disc. The most massive ionized outflows, however, show higher visual extinction in the outflow, 
    possibly because of their high dust content (inside the outflow).
    \item The difference in extinction between the broad and narrow components increases with higher ionized outflow mass, which could be explained by mainly dusty material being swept up by the outflow. These findings may support scenarios according to which galactic outflows are also driven by radiation pressure on dusty clouds.
    \item The molecular mass outflow rate accounts for the majority of the total outflow rate, amounting from 60 to 95 per cent in 11 out of 13 objects. The atomic phase contributes less than the molecular phase, while the ionized phase is negligible in most objects studied here. Star forming galaxies have less molecular gas in the outflow and the molecular phase is not always the dominant phase in these galaxies.
    \item In many galaxies, the ionized and neutral atomic phases are co-spatial and show a similar kinematic structure (e.g. correlated velocities), hinting at the possibility that the two phases are linked. This is contrary to some earlier studies which found no link between different outflow phases.
    \item The relative contribution of the ionized and molecular phase varies little with increasing AGN luminosity, but if anything, the ratio of ionized to molecular mass outflow rate declines slightly with higher AGN luminosity (with a large scatter).
\end{itemize}

\section*{Acknowledgements}
This work was based on data obtained at ESO's Very Large Telescope through programmes 101.B.0368, 102.B.0617 , 095.B-0049, 094.B-0733, 096.B-0230, 60.A.9315, 097.B.0165, 097.B-0313 and 094.B-0733. The authors are grateful to Asa Bluck and Sara Ellison for their useful comments and discussion. RM and AF acknowledge ERC Advanced Grant 695671 "QUENCH" and support by the Science and Technology Facilities Council (STFC). SA acknowledges support from the Spanish Ministerio de Ciencia e Innovaci\'on through grant ESP2017-83197-P. S Cazzoli acknowledges financial support from the State Agency for Research of the Spanish MCIU through the "Center of Excellence Severo Ochoa" award to the Instituto de Astrof\'isica de Andaluc\'ia (SEV-2017-0709). AM, GC and FM acknowledge the grant PRIN 2017PH3WAT$\_$003 from the Italian Ministry for University and Research. MP is supported by the Programa Atracci\'on de Talento de la Comunidad de Madrid
via grant 2018-T2/TIC-11715. GV acknowledges support from ANID programs FONDECYT Postdoctorado 3200802 and Basal-CATA AFB-170002. EB acknowledges the support from Comunidad de Madrid through the Atracci\'on de Talento grant 2017-T1$/$TIC-5213.
This research has made use of the NASA/ IPAC Infrared Science Archive, which is operated by the Jet Propulsion Laboratory, California Institute of Technology, under contract with the National Aeronautics and Space Administration. This publication makes use of data products from the Two Micron All Sky Survey, which is a joint project of the University of Massachusetts and the Infrared Processing and Analysis Center/California Institute of Technology, funded by the National Aeronautics and Space Administration and the National Science Foundation.

\section*{Data availability}
This work is based on MUSE data which can be found in the ESO archive (\href{http://archive.eso.org}{http://archive.eso.org}; programme IDs 101.B.0368, 102.B.0617 , 095.B-0049, 094.B-0733, 096.B-0230, 60.A.9315, 097.B.0165, 097.B-0313 and 094.B-0733). Further, derived data underlying this article will be shared on reasonable request to the corresponding author. 




\bibliographystyle{mnras}
\bibliography{bibliography_corr1.bib} 


\appendix

\section{Electron density}

\subsection{Density of AGN and star forming hosts}

\begin{figure}
\centering
\includegraphics[width=\columnwidth]{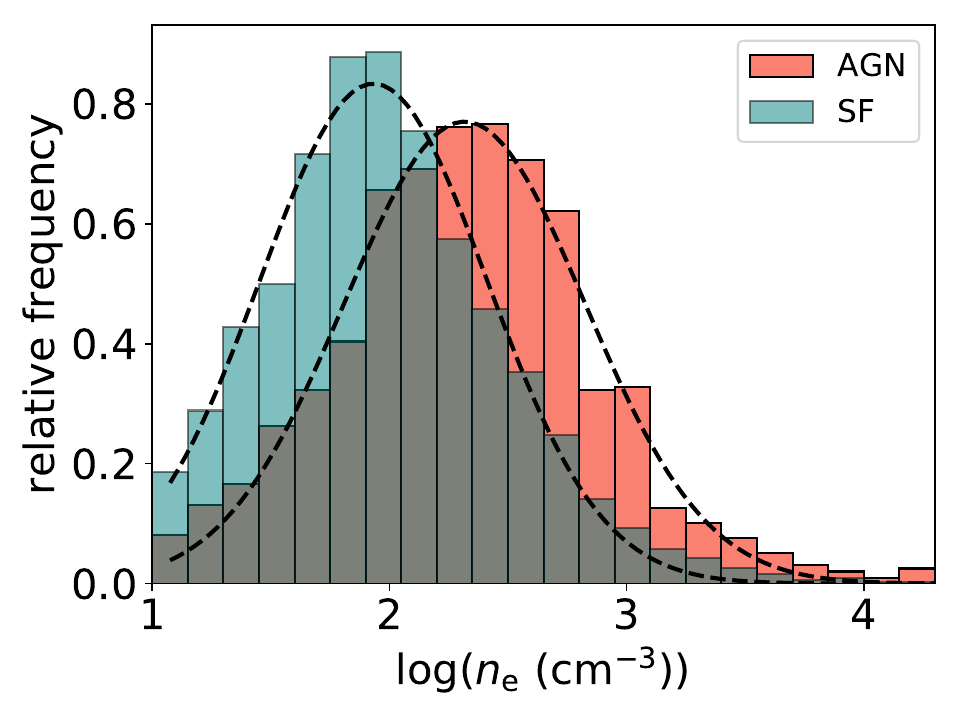}
\caption{In teal and light red, the distributions of the electron densities ($n_{\rm e}$) of the narrow components for star forming and AGN host galaxies are shown, respectively. The dashed lines are the Gaussian fit to the distributions.}
\label{fig:density_agn_vs_sf_host}
\end{figure}

Fig. \ref{fig:density_agn_vs_sf_host} shows the electron density in the disc for star forming (teal) and AGN host galaxies (light red). The centroids for the two distributions are $n_{\rm e, narrow, SF}$ = 85.6~cm$^{-3}$ and $n_{\rm e, narrow, AGN}$ = 207.4~cm$^{-3}$.

\section{Relative importance of ionized and molecular phase as function of star formation}

\begin{figure}
\centering
\includegraphics[width=\columnwidth]{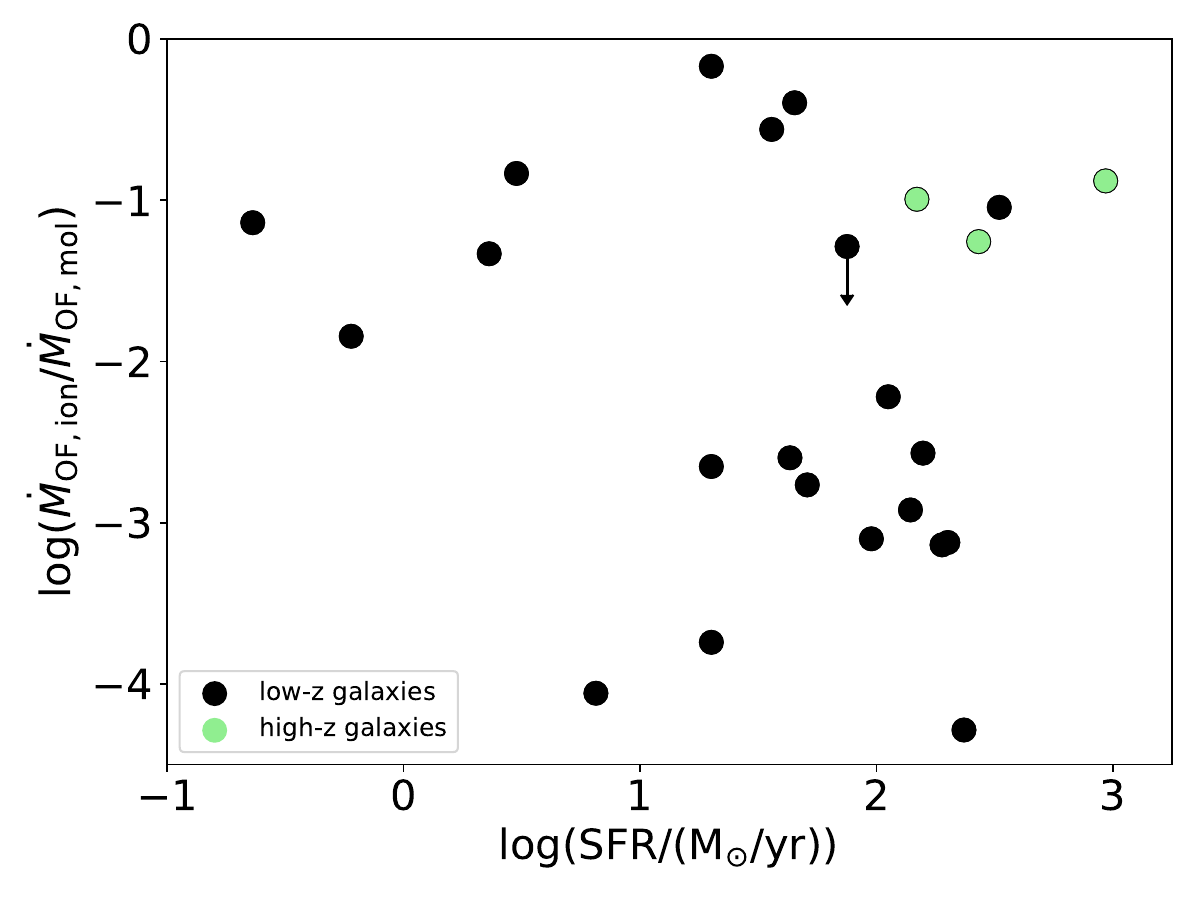}
\caption{Ratio of ionized to molecular mass outflow rate as a function of SFR. The green data points represent high-redshift galaxies and the black data points low-redshift galaxies.}
\label{fig:ion_molec_SFR}
\end{figure}

The ratio of ionized to molecular mass outflow rate does not show a trend with increasing SFR as shown in Fig. \ref{fig:ion_molec_SFR}.

\section{Target list}
\begin{table*}
\centering
\small
\caption{Details of the observations}
\label{tab:obs}
\begin{tabular}{ c c c c } 
	\hline
	Galaxy & exposure time & seeing & phyiscal FOV \\
	 & [min] & [arcsec] & [kpc]\\
	 (1) & (2) & (3) & (4) \\
	\hline
	\multicolumn{4}{c}{\textbf{MUSE sample}} \\
	\hline
  	IRAS 00509+1225 & 25 & 1.08 & 71$\times$71 \\
   	IRAS 01159-4443 & 53 & 0.89 & 28$\times$28 \\
   	IRAS 01341-3735 & 53 & 1.50 & 21$\times$21 \\
  	IRAS 06259-4708N & 80 & 1.05 & 46$\times$46 \\
  	IRAS 10257-4339 & 60 & 0.63 & 12$\times$12 \\
 	IRAS 10409-4556 & 53 & 1.13 & 26$\times$26 \\
 	IRAS 12043-3140 & 53 & 0.82 & 28$\times$28 \\
 	IRAS 13120-5453 & 33 & 0.82 & 37$\times$37 \\
 	IRAS 13156+0435N & 25 & 1.20 & 124$\times$125 \\
 	IRAS 13156+0435S & 25 & 1.20 & 124$\times$124 \\ 	
  	IRAS 13229-2934 & 25 & 1.08 & 17$\times$17 \\
        Mrk 463$^{*}$ & 25 & 1.02 & 59$\times$59 \\
  	IRAS 14378-3651 & 30 & 0.72 & 78$\times$78 \\
  	IRAS 14544-4255 & 25 & 0.72 & 19$\times$19 \\
  	IRAS 15115+0208 & 33 & 0.60 & 107$\times$107 \\
  	IRAS 17207-0014 & 25 & 0.57 & 51$\times$51 \\
  	IRAS 18093-5744 & 13 & 0.97 & 21$\times$21 \\
  	IRAS 19542+1110 & 25 & 0.99 & 75$\times$75 \\
  	IRAS 20100-4156 & 53 & 0.55 & 140$\times$140 \\
  	IRAS 20551-4250 & 41 & 0.63 & 51$\times$51 \\
  	IRAS 21130-4446 & 25 & 0.71 & 104$\times$104 \\
  	IRAS 21453-3511 & 53 & 1.21 & 20$\times$20 \\
 	IRAS 22491-1808 & 25 & 0.54 & 87$\times$87 \\
 	IRAS 23060+0505 & 33 & 0.93 & 178$\times$178 \\
    IRAS 23128-5919 & 53 & 2.03 & 53$\times$53 \\
 	IRAS 23389+0300$^{*}$ & 25 & 1.11 & 154$\times$154 \\
	\hline
\end{tabular}
\newpage
\begin{tablenotes}
\item Columns (1): Galaxy name, (2) exposure time in minutes, (3): seeing in arcsec and (4): physical field of view in kpc, $^{*}$: these galaxies have a radio excess (see text).
\end{tablenotes}
\end{table*}

\begin{table*}
\centering
\small
\caption{Galaxy properties of the sample}
\label{tab:sample}
\begin{tabular}{ c c c c c c c c c c c } 
	\hline
	Galaxy & \textit{z} & \textit{D$_{L}$} & class & log($L_{\rm AGN}$) & $\alpha_{\rm bol}$ = $L_{\rm AGN}$/$L_{\rm bol}$ & SFR & io & no & mo & References \\ 
	 & & [Mpc] & & [erg s$^{-1}$] & & [M$_{\odot}$/yr] & & & & \\
	 (1) & (2) & (3) & (4) & (5) & (6) & (7) & (8) & (9) & (10) & (11) \\
	\hline
	\multicolumn{11}{c}{\textbf{MUSE sample}} \\
	\hline
  	IRAS 00509+1225 & 0.0608 & 272 & AGN (type I) & 45.37 & 0.59 & 36 & \checkmark & \checkmark$^{***}$ & (\checkmark) & (a) \\
  	IRAS 01159-4443 & 0.022903 & 100 & SF & & & 27 & & & & \\
  	IRAS 01341-3735 & 0.01731 & 75 & SF & $\leq$42.11 & $\leq$0.0025 & 11 & & & & (b) \\
  	IRAS 06259-4708N & 0.038790 & 171 & comp. & $\leq$42.10 & $\leq$0.00037 & 147 & & & & (c) \\
 	IRAS 10257-4339 & 0.009354 & 40 & SF & $\leq$41.88 & $\leq$0.00038 & 36 & & \checkmark & \checkmark & (d) \\
 	IRAS 10409-4556 & 0.021011 & 91 & SF & & & 17 & & & & \\
 	IRAS 12043-3140 & 0.023203 & 101 & comp. & & & 22 & & & & \\
  	IRAS 13120-5453 & 0.030761 & 139 & AGN & 44.35 & 0.028 & 157 & \checkmark & \checkmark & \checkmark & (e) \\
 	IRAS 13156+0435N & 0.1130 & 525 & SF & & & 129$^{**}$ & & & & \\
 	IRAS 13156+0435S & 0.1130 & 525 & AGN & 45.5 & & 129$^{**}$ & & & & (f) \\
  	IRAS 13229-2934 & 0.01369 & 59 & AGN & 44.65 & 0.54 & 8.5 & \checkmark & & & (g) \\
    Mrk 463 & 0.050382 & 190 & AGN & 44.74 & 0.19 & 53 & \checkmark & & & (h) \\
  	IRAS 14378-3651 & 0.067637 & 314 & comp. & 45.12 & 0.21 & 112 & \checkmark & \checkmark & \checkmark & (e) \\
  	IRAS 14544-4255 & 0.015728 & 68 & SF & $\leq$41.94 & $\leq$0.0017 & 12 & & & & (b) \\
  	IRAS 15115+0208 & 0.095482 & 452 & SF & $\leq$43.49 & $\leq$0.014 & 50.9 & \checkmark & \checkmark & \checkmark & (e) \\
  	IRAS 17207-0014 & 0.04281 & 189 & SF & $\leq$43.67 & $\leq0.0042$ & 200 & & \checkmark & \checkmark & (i) \\
  	IRAS 18093-5744 & 0.017345 & 75 & SF & $\leq$41.66 & $\leq$0.00038 & 27 & & & & (d) \\
  	IRAS 19542+1110 & 0.064955 & 292 & SF & $\leq$43.76 & $\leq$0.0093 & 138 & \checkmark & \checkmark & & (j) \\
  	IRAS 20100-4156 & 0.12958 & 608 & SF & $\leq$42.93 & $\leq$0.00038 & 330 & \checkmark & \checkmark & \checkmark & (e) \\
   	IRAS 20551-4250 & 0.0430 & 190 & comp. & 43.39  & 0.0047 & 43 & \checkmark & \checkmark & \checkmark & (i) \\
  	IRAS 21130-4446 & 0.092554 & 424 & SF & $\leq$44.54 & $\leq$0.057 & 129 & & & & (k) \\
  	IRAS 21453-3511 & 0.016151 & 70 & AGN & 44.25 & 0.16 & 20 & \checkmark & \checkmark & & (b) \\
 	IRAS 22491-1808 & 0.0760 & 344 & SF & $\leq$41.83 & $\leq$0.00010 & 160 & & & \checkmark & (l) \\
  	IRAS 23060+0505 & 0.17301 & 861 & AGN & 46.06 & 0.35 & 75 & \checkmark & & (\checkmark) & (e) \\
  	IRAS 23128-5919 & 0.0448 & 198 & SF & $\leq$42.93 & $\leq$0.0019 & 112 & \checkmark & & & (l) \\
 	IRAS 23389+0300 & 0.145 & 687 & AGN (type I) & 45.63 & 0.49 & 100 & \checkmark & & & (m) \\

	\hline
 	\multicolumn{11}{c}{\textbf{Extended sample}} \\
 	\hline
 	NGC 253 & 0.00081 & 3.47 & SF & $\leq$40.66 & $\leq$0.0004 & 3 & \checkmark & & \checkmark & (e) \\
 	NGC 1433 & 0.00359 & 15.4 & AGN & 42.24 & 0.20 & 0.23 & \checkmark & & \checkmark & (e) \\
 	NGC 1614 & 0.01594 & 69.1 & SF & $\leq$42.07 & $\leq$0.0006 & 45 & \checkmark & \checkmark & \checkmark & (e) \\
 	IRAS 05189-2524 & 0.04256 & 188 & AGN (type I) & 44.47 & 0.05 & 189 & \checkmark & \checkmark & \checkmark & (e) \\
 	IRAS 07599+5506 & 0.1483 & 704 & AGN (type I) & & & 95 & \checkmark & \checkmark & & \\
 	IRAS 08572+3915 & 0.05835 & 261 & AGN & 45.72 & 0.86 & 20 & \checkmark & \checkmark & \checkmark & (e) \\
 	zC400528 & 2.387 & 19282 & AGN & 46.31$^{*}$ & & 148 & \checkmark & & \checkmark & (n) \\
 	XID 2028 & 1.5930 & 11750 & AGN & 47.21$^{*}$ & & 270 & \checkmark & & \checkmark & (o) \\
 	IRAS 10565+2448 & 0.0431 & 191 & AGN & 44.81 & 0.170 & 95 & \checkmark & \checkmark & \checkmark & (e) \\
 	ESO 320-G030 & 0.01078 & 46.6 & SF & $\leq$41.09 & $\leq$0.0001 & 20 & \checkmark & \checkmark & \checkmark & (e) \\
 	Mrk 231 & 0.04217 & 186 & AGN (type 1) & 45.72 & 0.340 & 234 & \checkmark & \checkmark & \checkmark & (e) \\
 	IRAS 13218+0552 & 0.2051 & 1008 & AGN (type I) & & & 80 & \checkmark & \checkmark & & \\
 	IRAS F13342+3932 & 0.17931 & 868 & AGN (type I) & & & 95 & \checkmark & \checkmark & & \\
 	Mrk 273 & 0.03778 & 167 & AGN & 44.16 & 0.080 & 139 & \checkmark & \checkmark & \checkmark & (e) \\
 	HE 1353-1917 & 0.03502 & 154 & AGN & 44.23$^{*}$ & & 2.3 & \checkmark & & \checkmark & (p) \\
 	SDSS J1356+1026 & 0.12297 & 575 & AGN & 46 & 0.43 & 20 & \checkmark & & \checkmark & (e) \\
 	3C 298 & 1.43812 & 10357 & AGN (type I) & 46.99 & 0.276 & 930 & \checkmark & & \checkmark & (q) \\
 	VV 705: NW & 0.04019 & 177 & comp. & & & 120 & \checkmark & \checkmark & & \\
 	Mrk 876 & 0.129 & 605 & AGN (type I) & 45.84 & 0.93 & 6.5 & \checkmark & & (\checkmark) & (e) \\
 	IC 5063 & 0.01135 & 49 & AGN & 44.30 & 0.9 & 0.6 & \checkmark & & \checkmark & (e) \\
	\hline
\end{tabular}
\newpage
\begin{tablenotes}
\item Columns: (1): Galaxy name, (2): redshift, (3): luminosity distance, (4): optical class, star forming (SF), AGN-dominated (AGN) or composite (comp.), (5): AGN luminosity, $^{*}$: no estimate of total bolometric luminosity, some galaxies have no available X-ray, IR or [OIII] data to calculate $L_{\rm AGN}$, (6): AGN contribution, (7): star formation rate, $^{**}$: SFR of both galaxies, (8)-(10): evidence of outflow in the ionized (io), neutral atomic (no) and molecular (mo) phase, $^{***}$: neutral outflow estimate from \citet{Rupke2017}, (\checkmark) means there exist an estimate for the upper limit of the outflow mass rate (11): references for AGN luminosity: (a): \cite{Veilleux2009}, (b): \cite{Alonso-Herrero2011}, (c): \cite{Iwasawa2011}, (d): \cite{Pereira-Santaella2011}, (e): \cite{Fluetsch2019} and references therein, (f): this work, derived from [OIII] luminosity, (g): \cite{Burtscher2015}, (h): \cite{Yamada2018}, (i): \cite{Nardini2010}, (j): \cite{Laha2018}, (k): \cite{Farrah2003}, (l): \cite{Franceschini2003}, (m): \cite{Imanishi2014}, (n): \cite{Alonso-Herrero2011}, (o): \cite{Brusa2015b}, (p): \cite{Husemann2019}, (q): \cite{Runnoe2012}
\end{tablenotes}
\end{table*}

\begin{table*}
\centering
\small
\caption{Ionized outflow properties}
\label{tab:ion_of_prop}
\begin{tabular}{ c c c c c c } 
	\hline
	Galaxy & $M_{\rm OF, ion}$ & $r_{\rm OF, ion}$ & $v_{\rm OF, ion}$ & $n_{\rm e}$ & log($L_{\rm H\alpha}$/(erg s$^{-1}$)\\
	 & [M$_{\odot}$] & [kpc] & [km~s$^{-1}$] & [cm$^{-3}$] \\
	 (1) & (2) & (3) & (4) & (5) & (6) \\
	\hline
	\multicolumn{6}{c}{\textbf{MUSE sample}} \\
	\hline
 	IRAS 13120-5453 & 6.47 & 0.3 & 780 & 500 & 41.30 \\
  	IRAS 13229-2934 & 6.19 & 0.43 & 410 & 520 & 41.01 \\
  	Mrk 463$^{*}$ & 6.32 & 1.2 & 790 & 540 & 41.14 \\
  	IRAS 14378-3651 & 6.65 & 1.1 & 740 & 500 & 41.48 \\
  	IRAS 15115+0208 & 5.95 & 1.2 & 300 & 250 & 40.77 \\
  	IRAS 17207-0014 & 5.76 & 1.1 & 610 & 250 & 41.89 \\
  	IRAS 19542+1110 & 6.67 & 0.87 & 670 & 1030 & 41.49 \\
  	IRAS 20100-4156 & 9.06 & 3.2 & 900 & 470 & 43.87 \\
  	IRAS 20551-4250 & 6.82 & 1.8 & 440 & 170 & 41.64 \\
  	IRAS 21453-3511 & 6.29 & 1.2 & 300 & 270 & 41.11 \\
  	IRAS 23060+0505 & 8.08 & 2.1 & 1090 & 720 & 42.90 \\
  	IRAS 23128-5919 & 7.30 & 2.6 & 520 & 500 & 42.13 \\
  	\hline
 	\multicolumn{6}{c}{\textbf{Extended sample}} \\
 	\hline
 	NGC 253 & 6.60 & 0.66 & 200 & 700--2000 & n/a \\
 	NGC 1433 & 4.50 & 0.11 & 370 & n/a & 39.33 \\
 	NGC 1614 & 7.98 & 1.7 & 350 & n/a & n/a \\
 	IRAS 05189-2524 & 6.60 & 3 & 300 & $^{**}$ & 41.42 \\
 	IRAS 07599+5506 & 5.72 & 8 & 160 & $^{**}$ & 40.54 \\
 	IRAS 08572+3915 & 5.62 & 2 & 760 & n/a & n/a \\
 	zC400528 & 8.48 & 3 & 800 & n/a & 43.30 \\
 	XID 2028 & 8.11 & 13 & 1500 & 1000-3000 & 42.93$^{a}$\\
 	IRAS 10565+2448 & 6.58 & 5 & 240 & n/a & n/a \\ 	
 	ESO 320-G030 & 6.70 & 0.31 & 100 & n/a & n/a \\
 	Mrk 231 & 5.67 & 3 & 260 & n/a & 39.91 \\
 	IRAS 13218+0552 & 7.50 & 12 & 320 & $^{**}$ & 42.32 \\
 	IRAS F13342+3932 & 7.98 & 10.7 & 200 & $^{**}$ & 42.81 \\
 	Mrk 273 & 6.77 & 4 & 370 & n/a & n/a \\
 	HE 1353-1917 & 6.14 & 0.2 & 480 & 214/1500$^{***}$ & 40.97 \\
 	SDSS J1356+1026 & 7.53 & 12 & 230 & $\lesssim$ 850 & 42.35 \\
 	3C 298$^{*}$ & 8.73 & 4 & 1550 & 270 & 43.55 \\
 	VV 705: NW & 6.36 & 3 & 230 & n/a & n/a \\
	Mrk 876 & 6.71 & 8 & 200 & $^{**}$ & 41.53 \\
 	IC 5063$^{*}$ & 5.70 & 0.5 & 800 & 770$^{****}$ & 40.50 \\
	\hline
\end{tabular}
\newpage
\begin{tablenotes}
\item Columns: (1): Galaxy name, (2) ionized gas mass of the outflow, (3): radius of the ionized outflow in kpc, (4): ionized outflow velocity, (5): electron density from our work or the relevant paper (see caption of Table \ref{tab:outflow_sample}) and (6) H$\alpha$ luminosity of the broad (outflowing) component.
$^{*}$: these galaxies have a radio excess (see text).
$^{**}$: no numerical values provided in \citet{Rupke2017}, but density maps are shown in the paper.
$^{***}$: for north-east and south-west regions, see \citet{Husemann2019} for details
$^{****}$: value at the centre
$^{a}$: converted from H$\beta$ luminosity.
\end{tablenotes}
\end{table*}

\begin{table*}
\centering
\small
\caption{Neutral outflow properties}
\label{tab:neu_of_prop}
\begin{tabular}{ c c c c c c } 
	\hline
	Galaxy & $\Delta v$ & $b_{\rm D}$ & $\tau$ & $C_{\rm f}$ & $r_{\rm OF, neu}$\\
		 & [km~s$^{-1}$] & [km~s$^{-1}$] & & & [kpc] \\
	 (1) & (2) & (3) & (4) & (5) & (6) \\
	\hline
	\multicolumn{6}{c}{\textbf{MUSE sample}} \\
	\hline
  	IRAS 10257-4339  & -72 & 97 & 2.8 & 0.093 & 1.5 \\
  	 & -378 & 153 & 4.0 & 0.091 & \\
 	IRAS 13120-5453 & -51 & 222 & 0.40 & 0.642 & 3.1  \\
 	 & -477 & 520 & 0.15 & 1.000 & \\
  	IRAS 14378-3651 & -119 & 291 & 0.50 & 0.478 & 2.5  \\
  	 & -861 & 352 & 0.19 & 0.216 & \\
  	IRAS 15115+0208 & -179 & 156 & 0.10 & 0.999 & 5.1  \\
  	IRAS 17207-0014 & -276 & 296 & 0.55 & 0.690 & 3.0  \\
  	IRAS 19542+1110 & -473 & 263 & 0.53 & 0.417 & 0.6  \\
  	IRAS 20100-4156 & -123 & 121 & 4.2 & 0.091 & 1.6 \\
  	 & -379 & 168 & 2.2 & 0.624 & \\
  	IRAS 20551-4250 & -92 & 30 & 0.02 & 0.903 & 2.9 \\
  	 & -524 & 139 & 1.8 & 0.043 & \\
  	IRAS 21453-3511 & -109 & 100 & 0.18 & 0.278 & 3.0  \\
  	 & -307 & 219 & 0.04 & 0.988 & \\
	\hline
\end{tabular}
\newpage
\begin{tablenotes}
\item Parameters of the outflowing components in the Na~I~D fitting. If there are several outflowing components, the parameters for each component are on different rows. Columns: (1): Galaxy name, (2) velocity offset in km~s$^{-1}$ (3): Doppler linewidth (4): optical depth, (5): local covering factor and (6) neutral atomic outflow radius.
\end{tablenotes}
\end{table*}

\begin{table*}
\centering
\small
\caption{Outflow properties of the sample}
\label{tab:outflow_sample}
\begin{tabular}{c c c c c} 
	\hline
	Galaxy & $\dot{M}_{\rm OF, ion}$ & $\dot{M}_{\rm OF, neu}$ & $\dot{M}_{\rm OF, mol}$ & References \\ 
	 & [M$_{\odot}$/yr] & [M$_{\odot}$/yr] & [M$_{\odot}$/yr] & \\
	\hline
	\multicolumn{5}{c}{\textbf{MUSE sample}} \\
	\hline
 	IRAS 10257-4339 &  & 15.8 & 11.27 & ($\alpha$), (1)\\
 	IRAS 13120-5453 & 3.1 & 108 & 1134 & (a), ($\alpha$), (1) \\
  	IRAS 13229-2934 & 0.60 & & & (a), \\
  	Mrk 463 & 1.43 & & & (a) \\
 	IRAS 14378-3651 & 1.23 & 45.2 & 204 & (a), ($\alpha$), (1)  \\
 	IRAS 15115+0208 & 0.09 & 6.59 & 54.6 & (a), ($\alpha$), (1) \\
  	IRAS 17207-0014 & 0.13 & 82.9 & 176 & (a), (1) \\
  	IRAS 19542+1110 & 2.9 & 13.8 & & (a), ($\alpha$) \\
 	IRAS 20100-4156 & 130 & 61.1 & 1435 & (a), ($\alpha$), (1)  \\
 	IRAS 20551-4250 & 0.66 & 14.8 & 263 & (a), ($\alpha$), (1) \\
 	IRAS 21453-3511 & 0.95 & 4.22 & & (a), ($\alpha$) \\
 	IRAS 23060+0505 & 26 & & $\leq$504 & (a), (1) \\
 	IRAS 23128-5919 & 1.62 & & & (a),  \\
	\hline
	\multicolumn{5}{c}{\textbf{Extended sample}} \\
	\hline
	NGC 253 & 0.49 & & 3.34 & (b), (1) \\
	NGC 1433 & 0.05 & & 0.63 & (a), (1) \\
	NGC 1614 & 8.42 & 62.9 & 20.9 & (c), ($\beta$), (1) \\
	IRAS 05189-2524 & 0.16 & 40.4 & 216  & (d), ($\gamma$), (1) \\
	IRAS 07599+5506 & 0.004 & 32.3 & & (d), ($\gamma$) \\
	IRAS 08572+3915 & 0.07 & 11.3 & 365 & (e), ($\delta$), (1) \\
 	zC400528 & 32 & & 323 & (f), (2) \\
 	XID 2028 & 6.1 & & 110 & (g), (3) \\
	IRAS 10565+2448 & 0.07 & 37.7 & 92.1 & (e), ($\delta$), (1) \\
	ESO 320-G030 & 0.68 & 37.0 & 1.0 & (c), ($\beta$), (1) \\
	Mrk 231 & 0.017 & 38.8 & 323 & (e), ($\gamma$, $\delta$). (1) \\
	IRAS 13218+0552 & 0.35 & 3.85 & & (d), ($\gamma$) \\
	IRAS F13342+3932 & 0.71 & 4.86 & & (d), ($\gamma$) \\
	Mrk 273 & 0.22 & 9.5 & 187 & (e), ($\delta$), (1) \\
	HE 1353-1917 & 1.36 & & 29.2 & (h), (4) \\
	SDSS J1356+1026 & 0.26 & & 117 &  (i), (1) \\
 	3C 298 & 84 & & 636 & (j), (5) \\
	VV 705: NW & 0.07 & 27.8 & & (e), ($\delta$) \\
	Mrk 876 & 0.05 & & $\leq$609 & (d), (1) \\
	IC 5063 & 0.30 & & 21.2 &  (k), (1) \\
	\hline
\end{tabular}
\newpage
\begin{tablenotes}
\item Columns: (1): Galaxy name, (2): ionized mass outflow rate, (3): molecular mass outflow rate, (4): neutral outflow rate, (5) references: ionised outflow: (a): this work, (b): \cite{Westmoquette2011}, (c): \cite{Arribas2014}, (d): \cite{Rupke2017}, (e): \cite{Rupke2013}, (f): \cite{Herrera-Camus2019}, (g): \cite{Cresci2015a}, (h): \cite{Husemann2019}, (i): \cite{Greene2012}, (j): \cite{Vayner2017}, (k): \cite{Morganti2007}; neutral outflow: ($\alpha$): this work, ($\beta$): \cite{Cazzoli2016}, ($\gamma$): \cite{Rupke2017}, ($\delta$): \cite{Rupke2013}; molecular outflow: (1): \cite{Fluetsch2019} and references therein, (2): \cite{Herrera-Camus2019}, (3): \cite{Brusa2018}, (4): \cite{Husemann2019} (5): \cite{Vayner2017}
\end{tablenotes}
\end{table*}

\section{Kinematic maps}
\label{sec:map_outflows}

In this section, we show the maps of all galaxies in our sample with an outflow in at least one phase. In panel (a) we show the flux of H$\alpha$, [OIII], the gas velocity and gas velocity dispersion of the narrow (top) and broad (bottom) component. Some galaxies do have two components, but we still do not classify the broad component as an outflow, because the components are not kinematically distinct and/or the broad component is not strong enough to be larger than 3 $\sigma$ in the integrated spectrum (panel (d)). In panel (b), the sodium absorption profile as well as the He~I emission line is shown. Panel (c) depicts the velocity ($v_{\rm 50}$) and the gas velocity dispersion of the different components. Finally, panel (d) shows the integrated spectrum of the galaxy covering all main emission lines.
\onecolumn


\begin{figure}%
\centering
\subfigure[Maps of the ionized gas]{
\label{fig:sfig1}
\includegraphics[width=\textwidth]{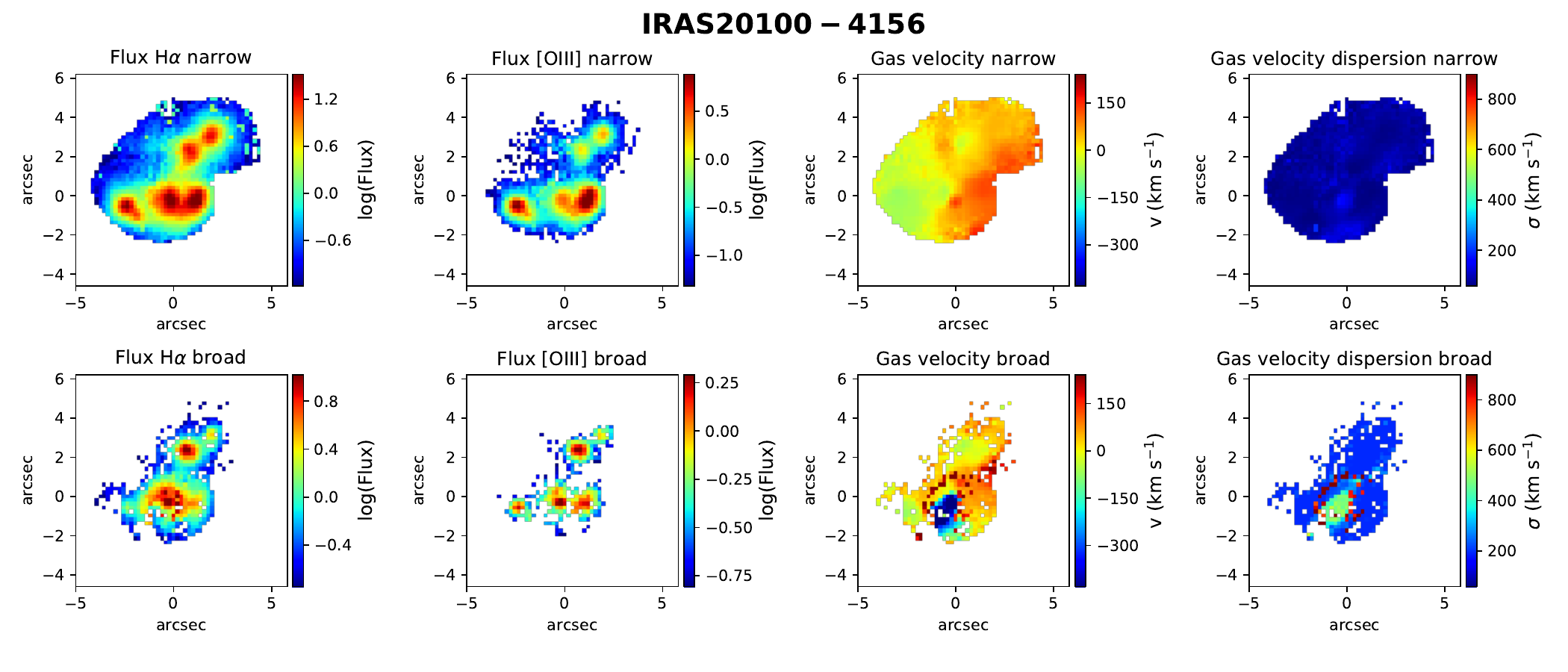}}%
\qquad
\subfigure[Integrated spectrum of the He~I emission and the sodium absorption feature (blue) with fit (red)]{
\label{fig:sfig2}
\includegraphics[width=.495\textwidth]{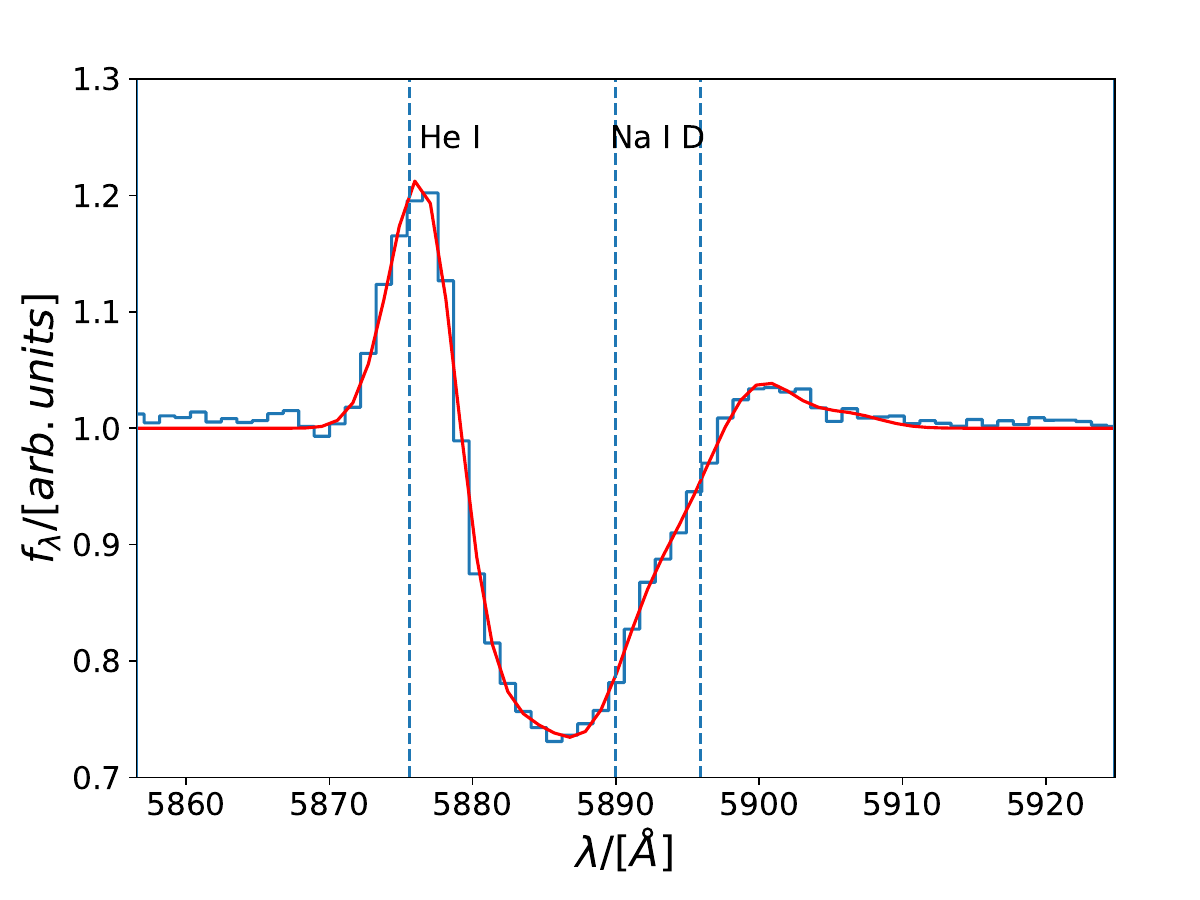}}%
\qquad\hspace{-0.2em}
\subfigure[Maps of the neutral atomic gas]{
\label{fig:sfig3}
\includegraphics[trim={11cm 0 0 0},clip,width=.45\textwidth]{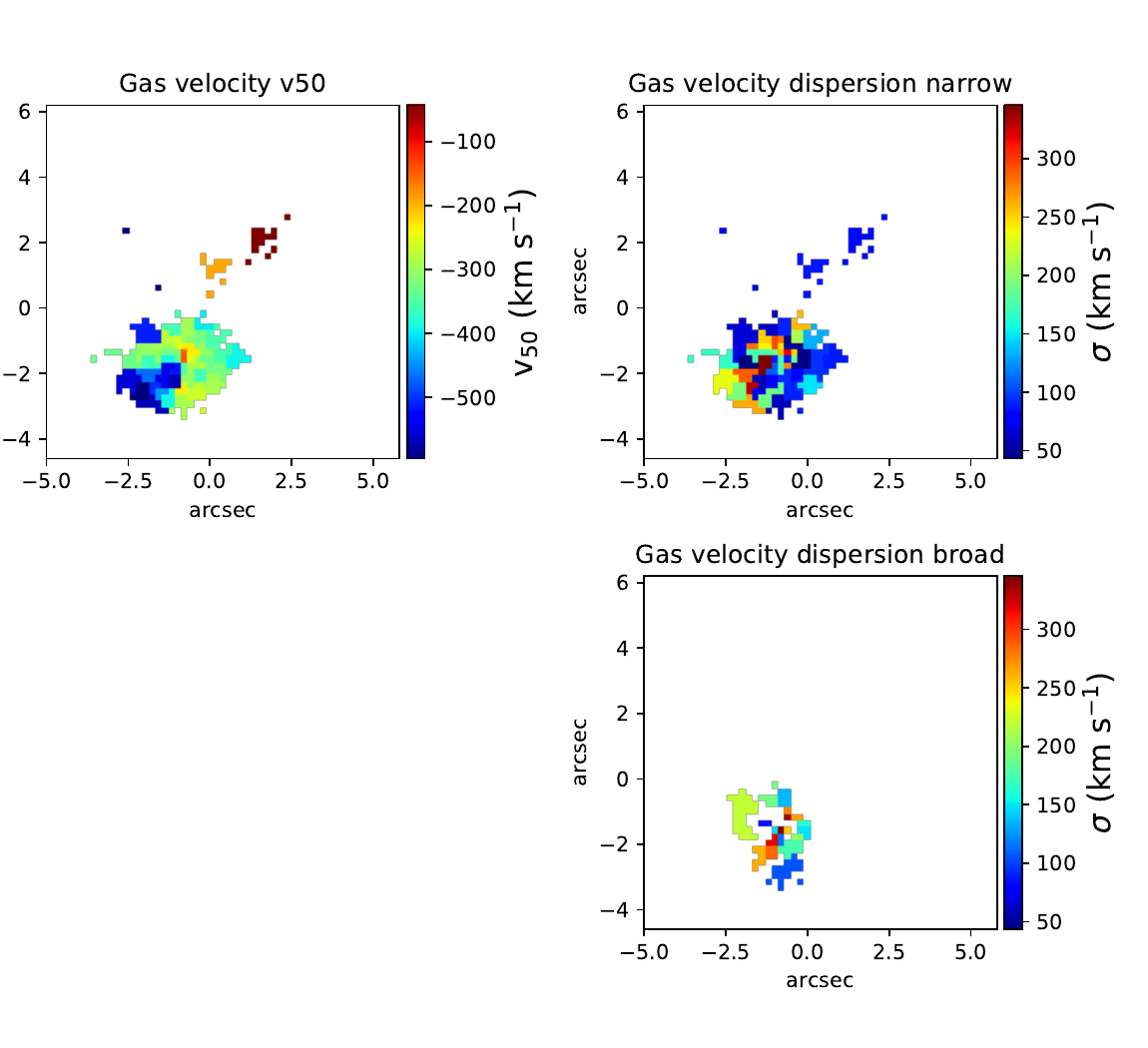}}%
\qquad
\subfigure[Integrated spectrum: The spectrum is shown in light blue, the two-component fit in green, the one component fit in dark blue and for the broad components of H$\alpha$ and H$\beta$ in cyan.]{
\label{fig:sfig4}
\includegraphics[width=.8\textwidth]{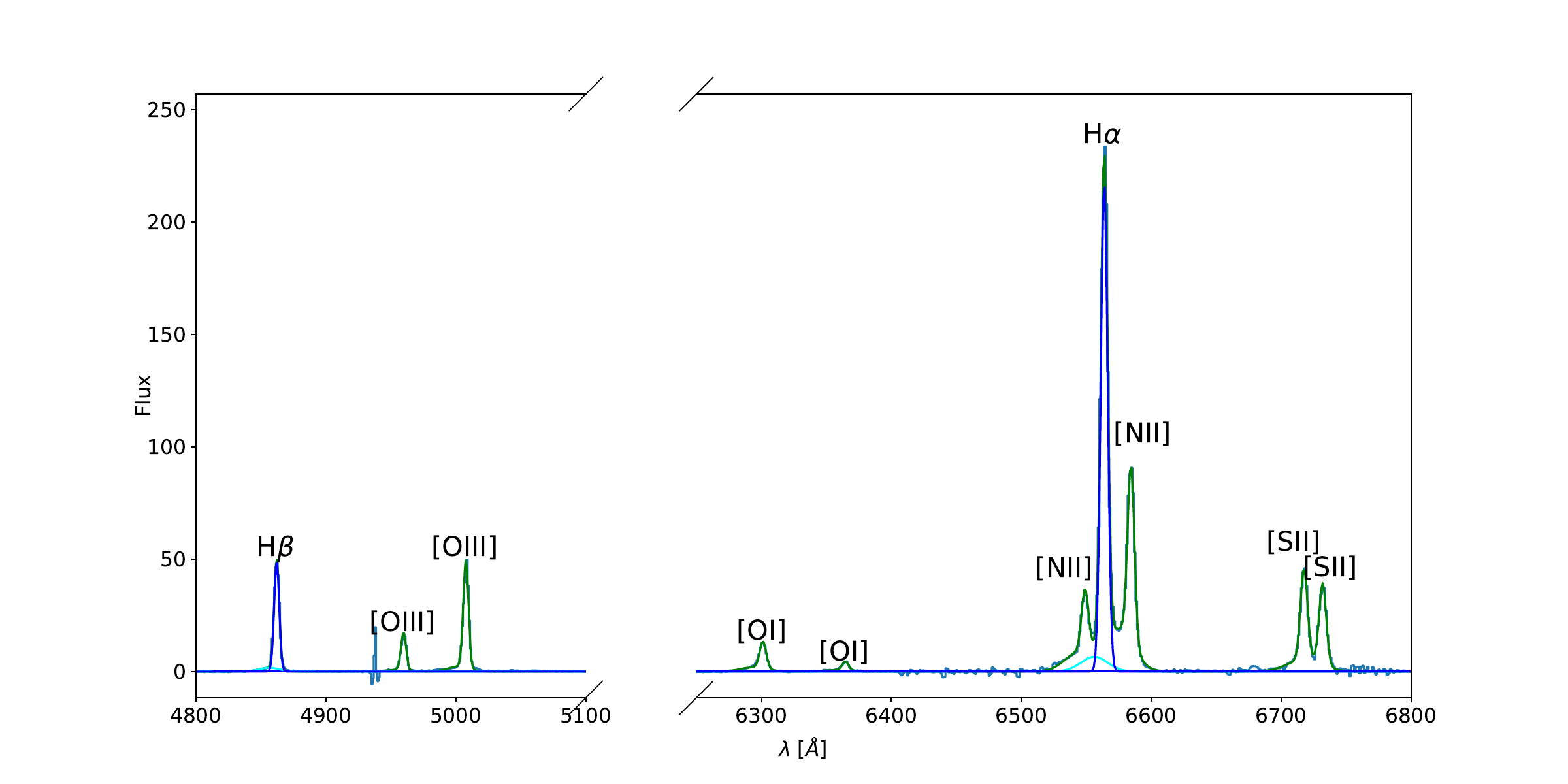}}%
\caption{IRAS 20100-4156}
\end{figure}

\begin{figure}%
\centering
\subfigure[Maps of the ionized gas]{
\label{fig:sfig1}
\includegraphics[width=\textwidth]{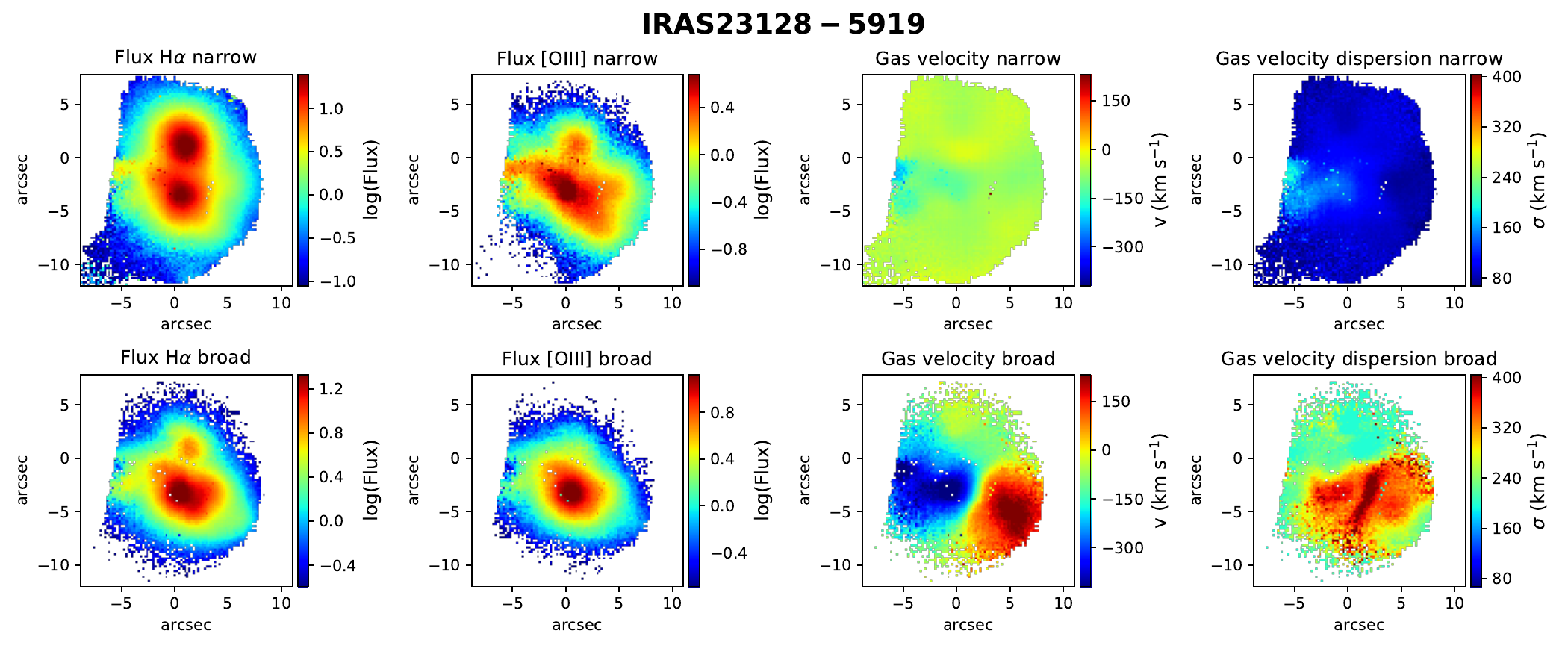}}%
\qquad
\subfigure[Integrated spectrum of the He~I emission and the sodium absorption feature (blue) with fit (red)]{
\label{fig:sfig2}
\includegraphics[width=.495\textwidth]{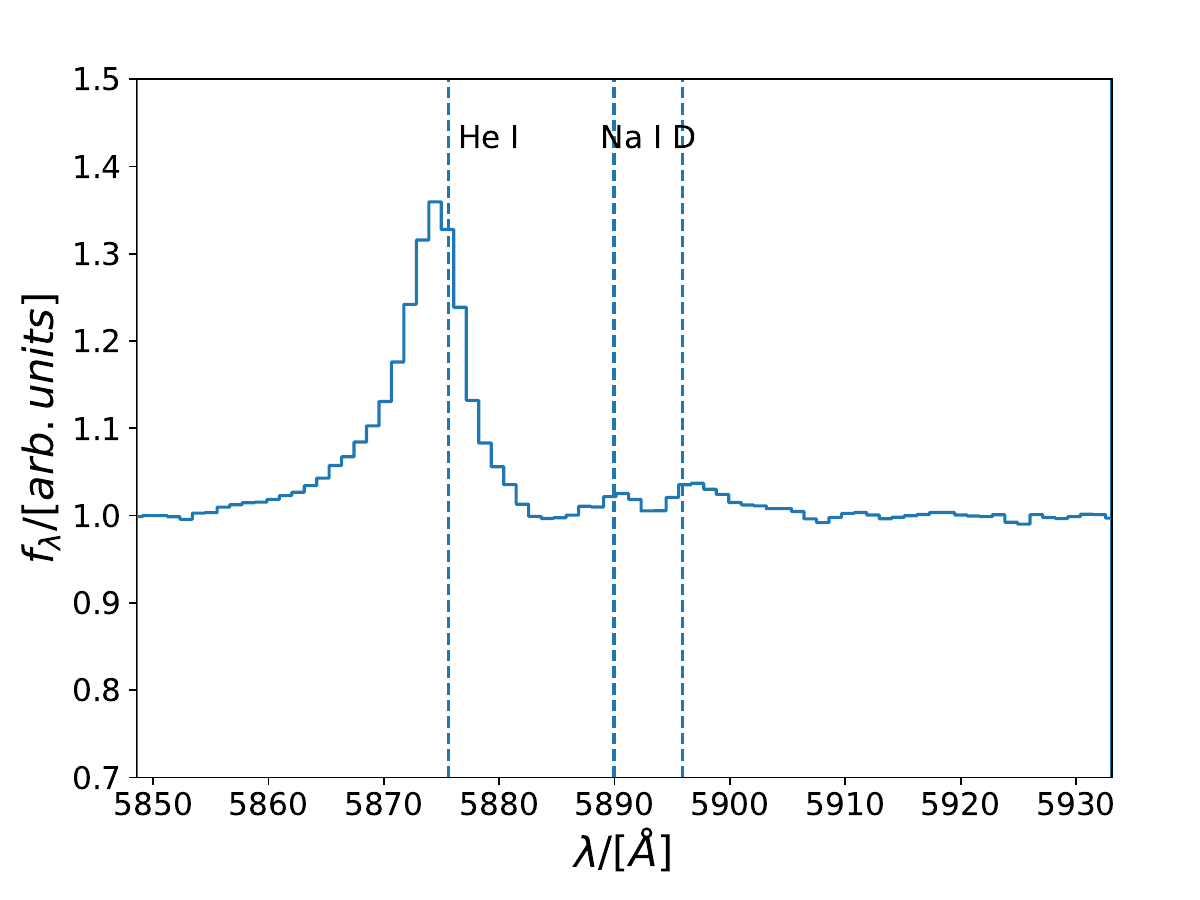}}%
\qquad\hspace{0em}
\subfigure[Maps of the neutral atomic gas]{
\label{fig:sfig3}
\includegraphics[trim={11cm 0 0 0},clip,width=.45\textwidth]{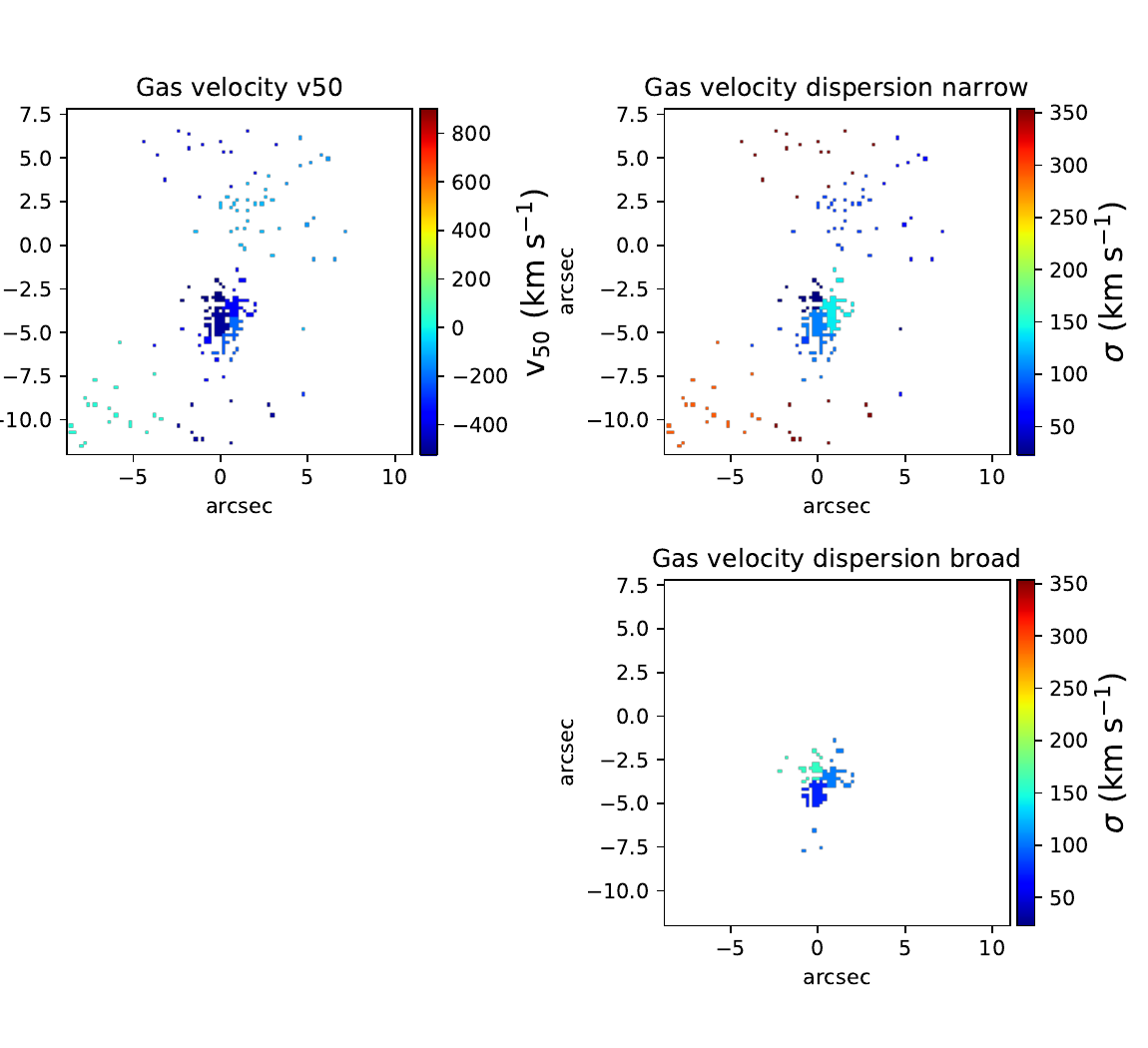}}%
\qquad
\subfigure[Integrated spectrum: The spectrum is shown in light blue, the two-component fit in green, the one component fit in dark blue and for the broad components of H$\alpha$ and H$\beta$ in cyan.]{
\label{fig:sfig4}
\includegraphics[width=.8\textwidth]{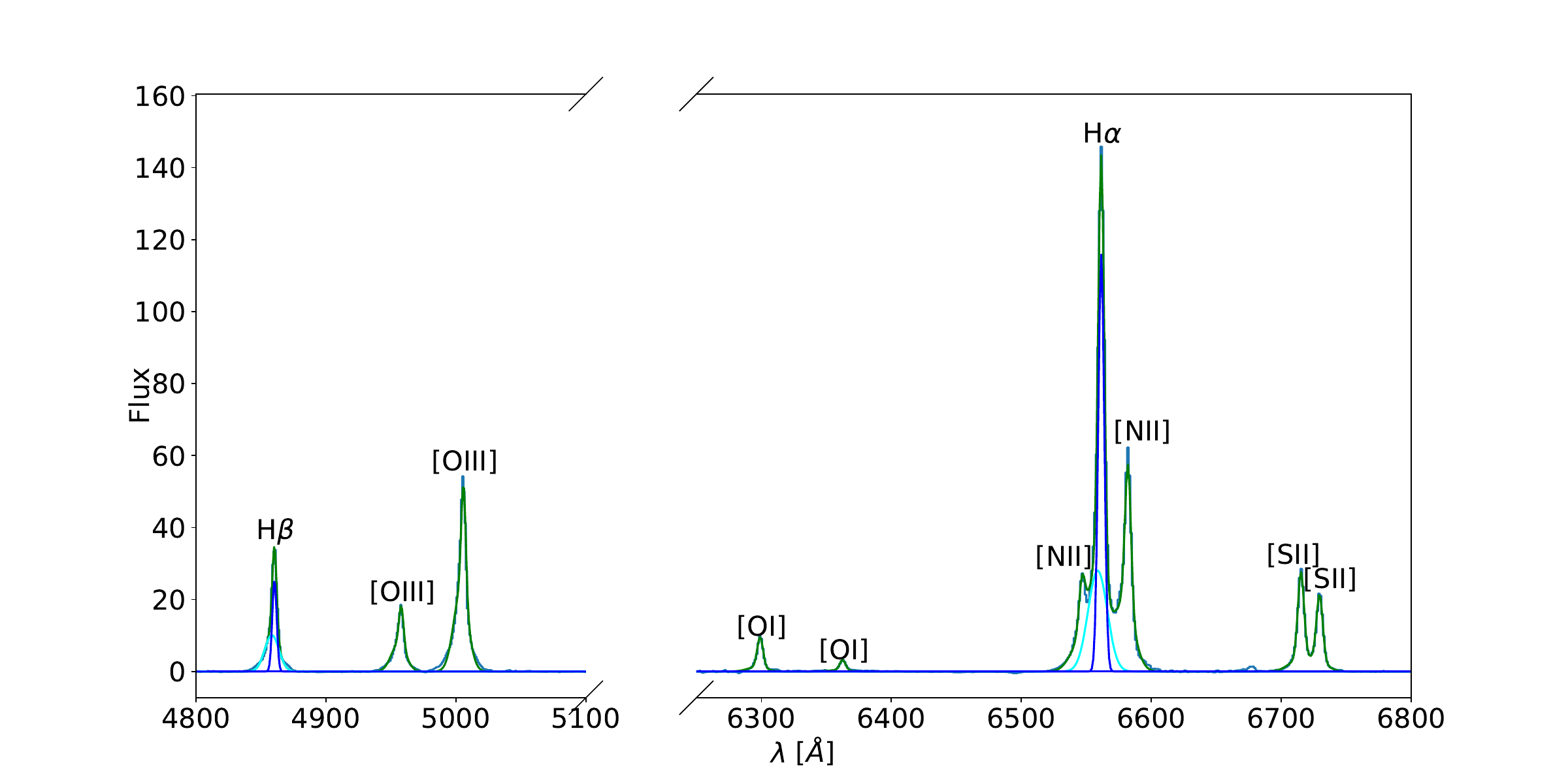}}%
\caption{IRAS 23128-5919}
\end{figure}

\begin{figure}%
\centering
\subfigure[Maps of the ionized gas]{
\label{fig:sfig1}
\includegraphics[width=\textwidth]{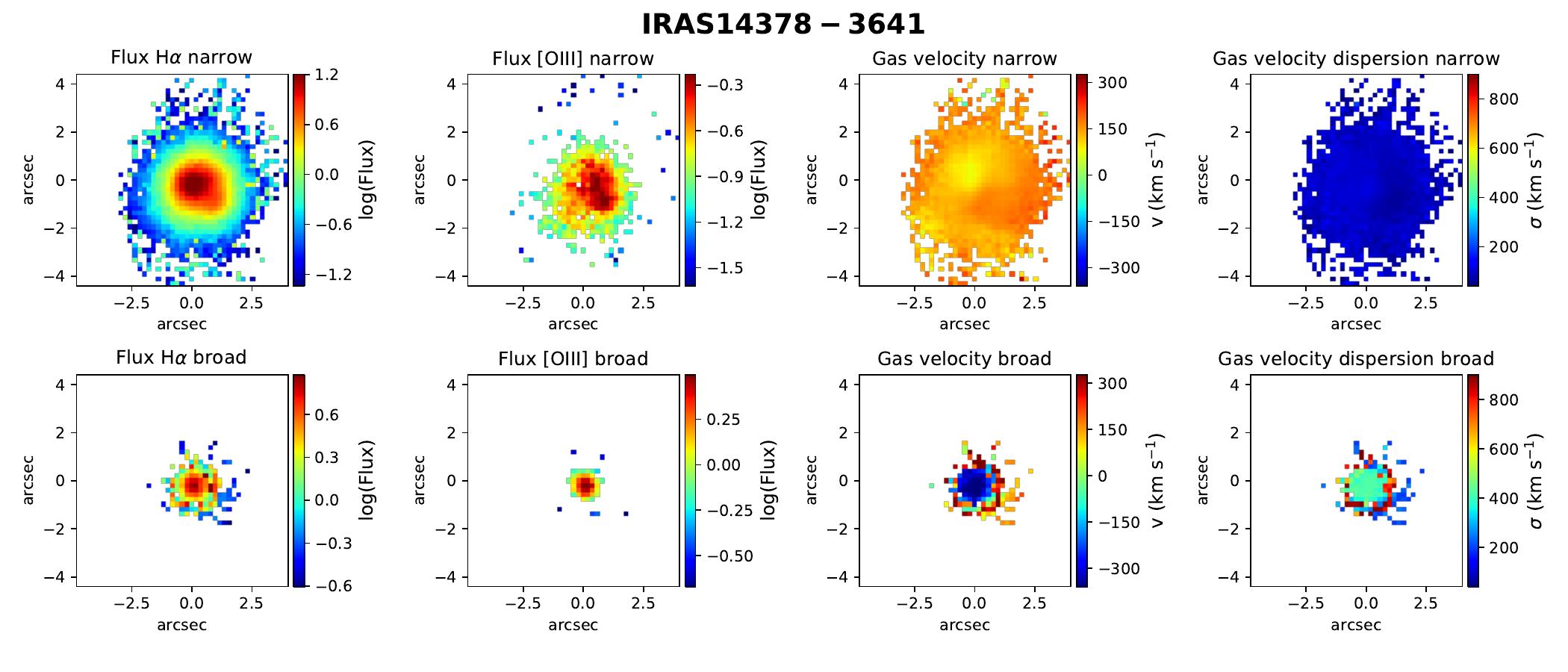}}%
\qquad
\subfigure[Integrated spectrum of the He~I emission and the sodium absorption feature (blue) with fit (red)]{
\label{fig:sfig2}
\includegraphics[width=.495\textwidth]{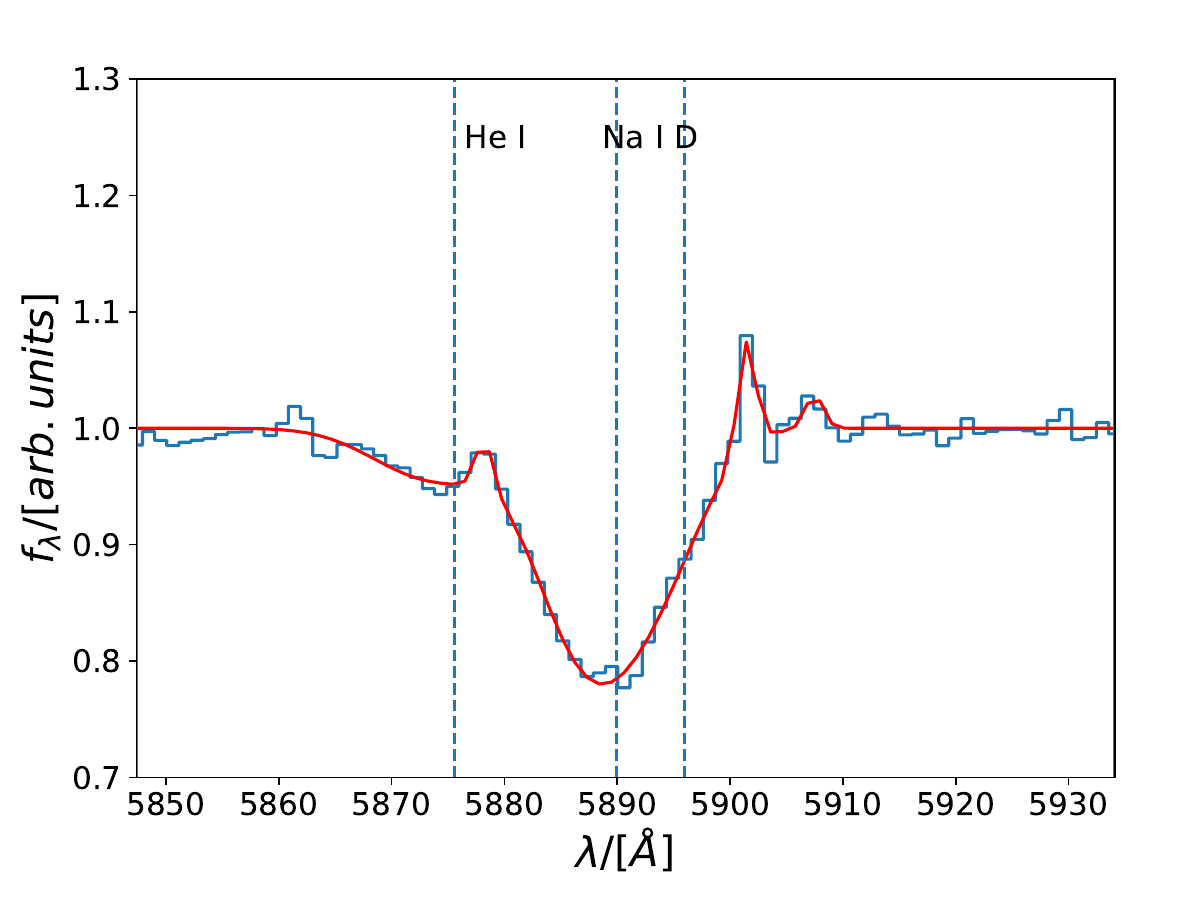}}%
\qquad\hspace{0em}
\subfigure[Maps of the neutral atomic gas]{
\label{fig:sfig3}
\includegraphics[trim={11cm 0 0 0},clip,width=.45\textwidth]{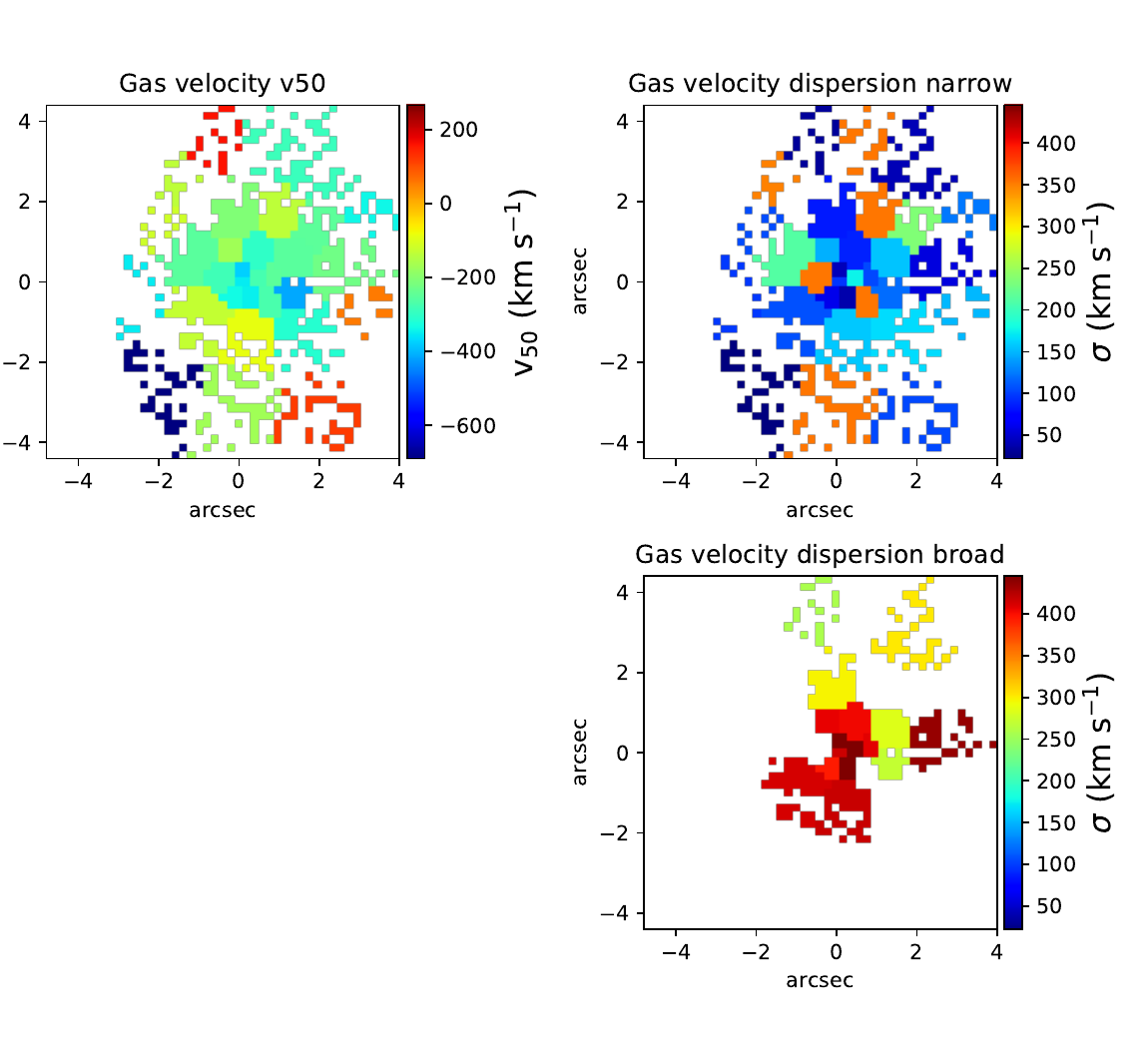}}%
\qquad
\subfigure[Integrated spectrum: The spectrum is shown in light blue, the two-component fit in green, the one component fit in dark blue and for the broad components of H$\alpha$ and H$\beta$ in cyan.]{
\label{fig:sfig4}
\includegraphics[width=.8\textwidth]{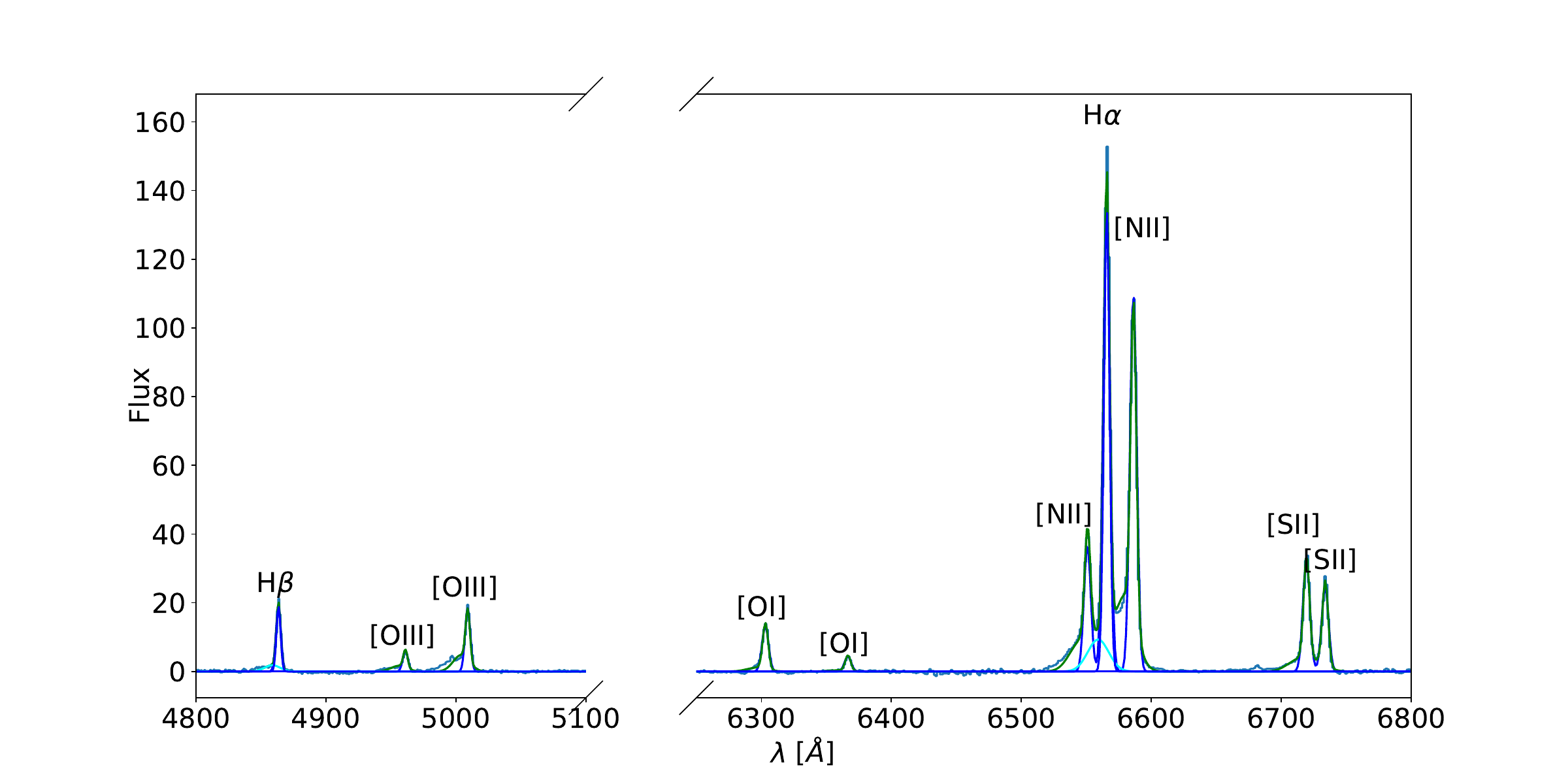}}%
\caption{IRAS 14378-3651}
\label{fig:IRAS14378}
\end{figure}

\begin{figure}%
\centering
\subfigure[Maps of the ionized gas]{
\label{fig:sfig1}
\includegraphics[width=\textwidth]{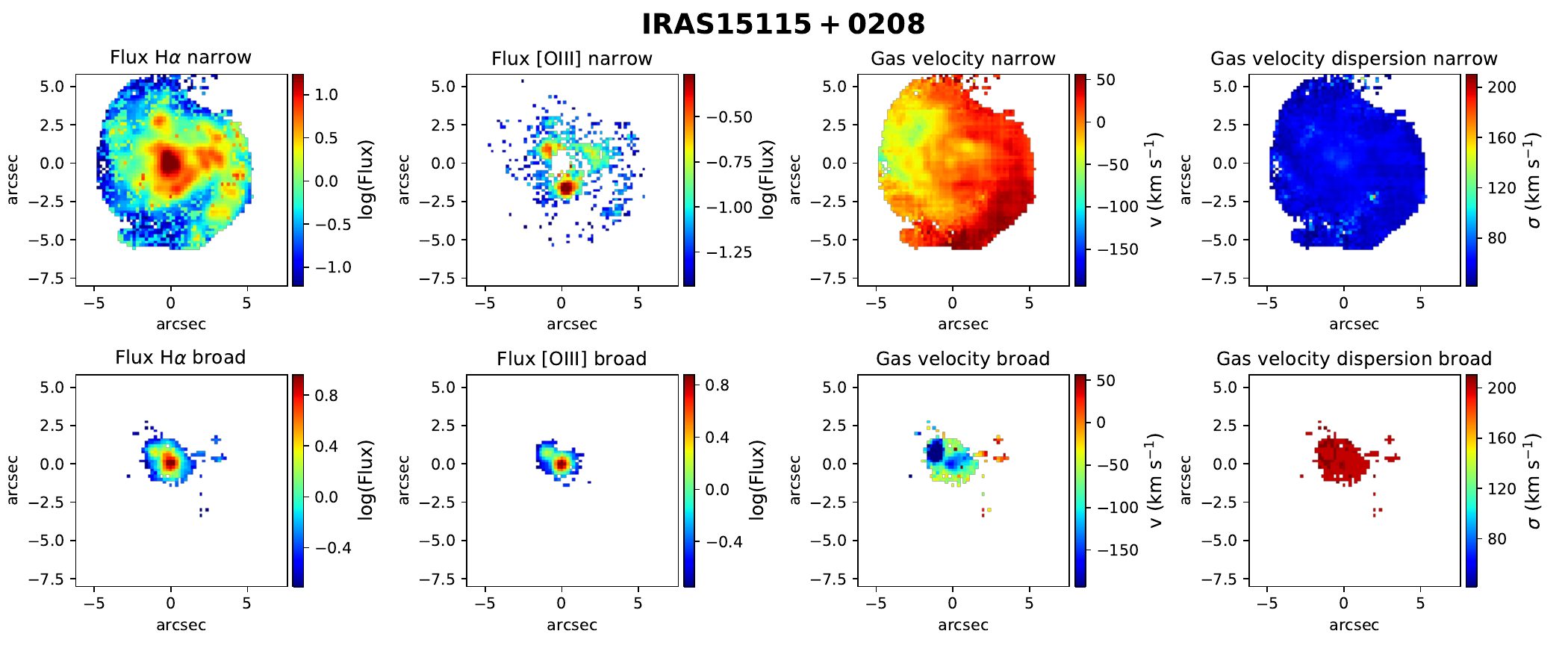}}%
\qquad
\subfigure[Integrated spectrum of the He~I emission and the sodium absorption feature (blue) with fit (red)]{
\label{fig:sfig2}
\includegraphics[width=.495\textwidth]{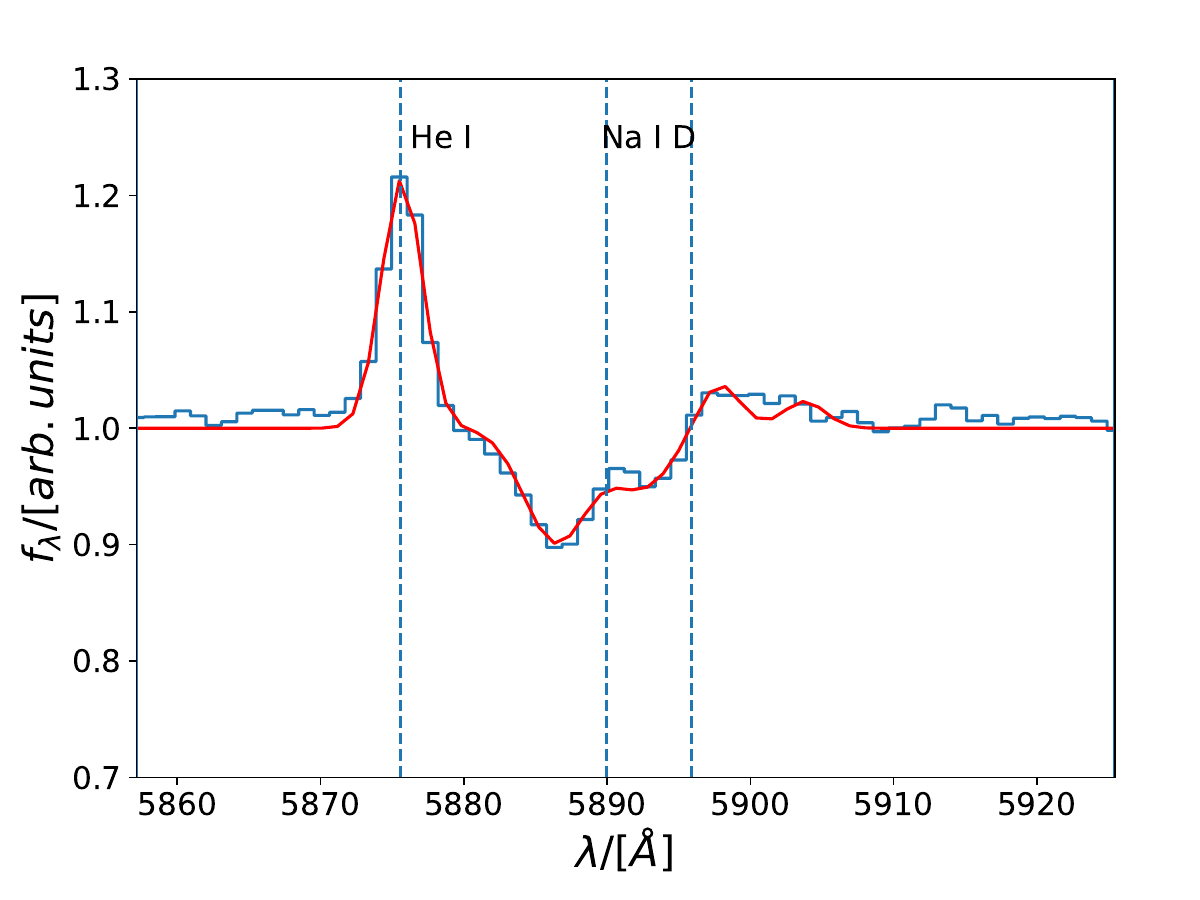}}%
\qquad\hspace{0em}
\subfigure[Maps of the neutral atomic gas]{
\label{fig:sfig3}
\includegraphics[trim={11cm 0 0 0},clip,width=.45\textwidth]{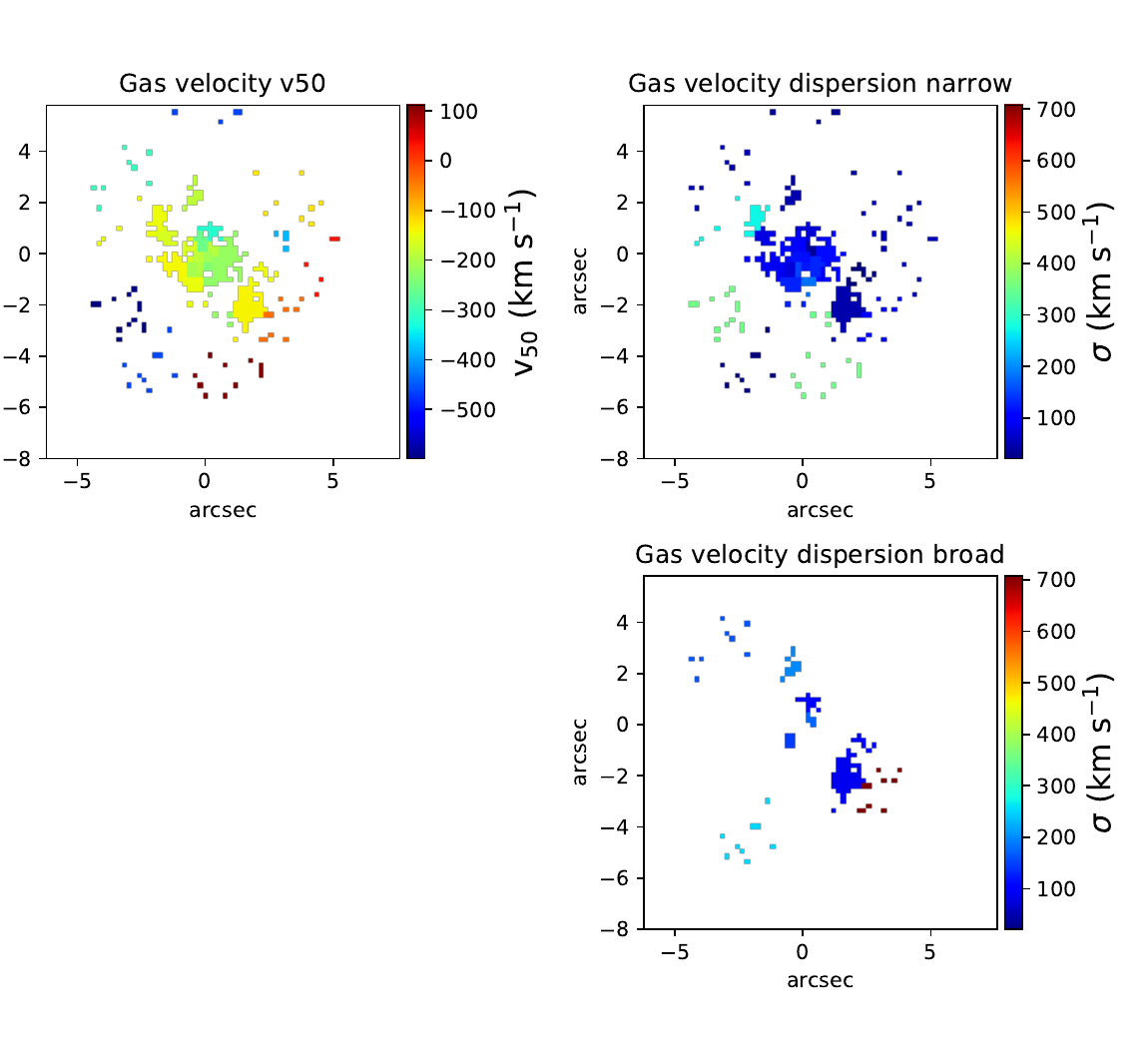}}%
\qquad
\subfigure[Integrated spectrum: The spectrum is shown in light blue, the two-component fit in green, the one component fit in dark blue and for the broad components of H$\alpha$ and H$\beta$ in cyan.]{
\label{fig:sfig4}
\includegraphics[width=.8\textwidth]{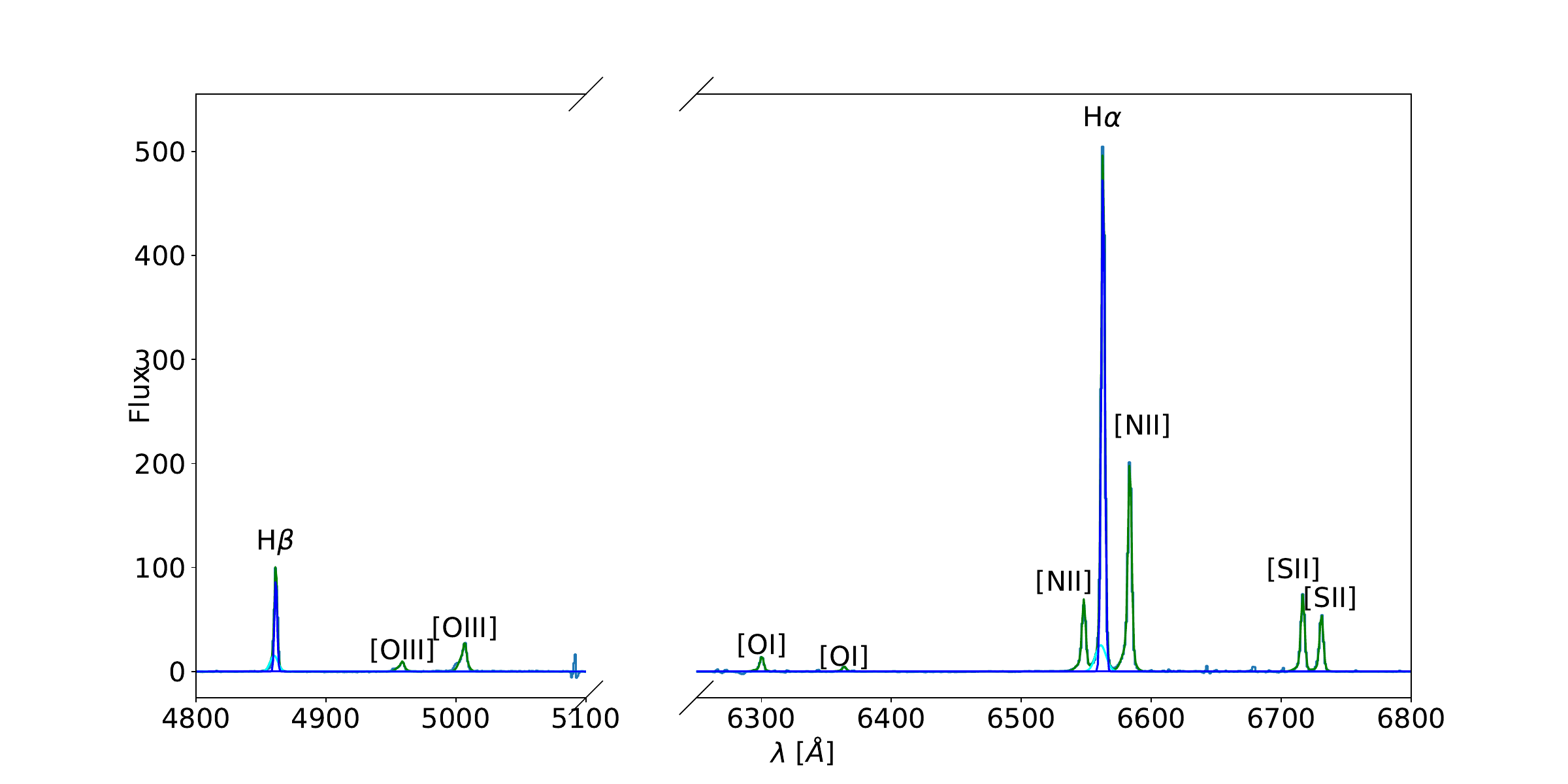}}%
\caption{IRAS 15115+0208}
\end{figure}

\begin{figure}%
\centering
\subfigure[Maps of the ionized gas]{
\label{fig:sfig1}
\includegraphics[width=\textwidth]{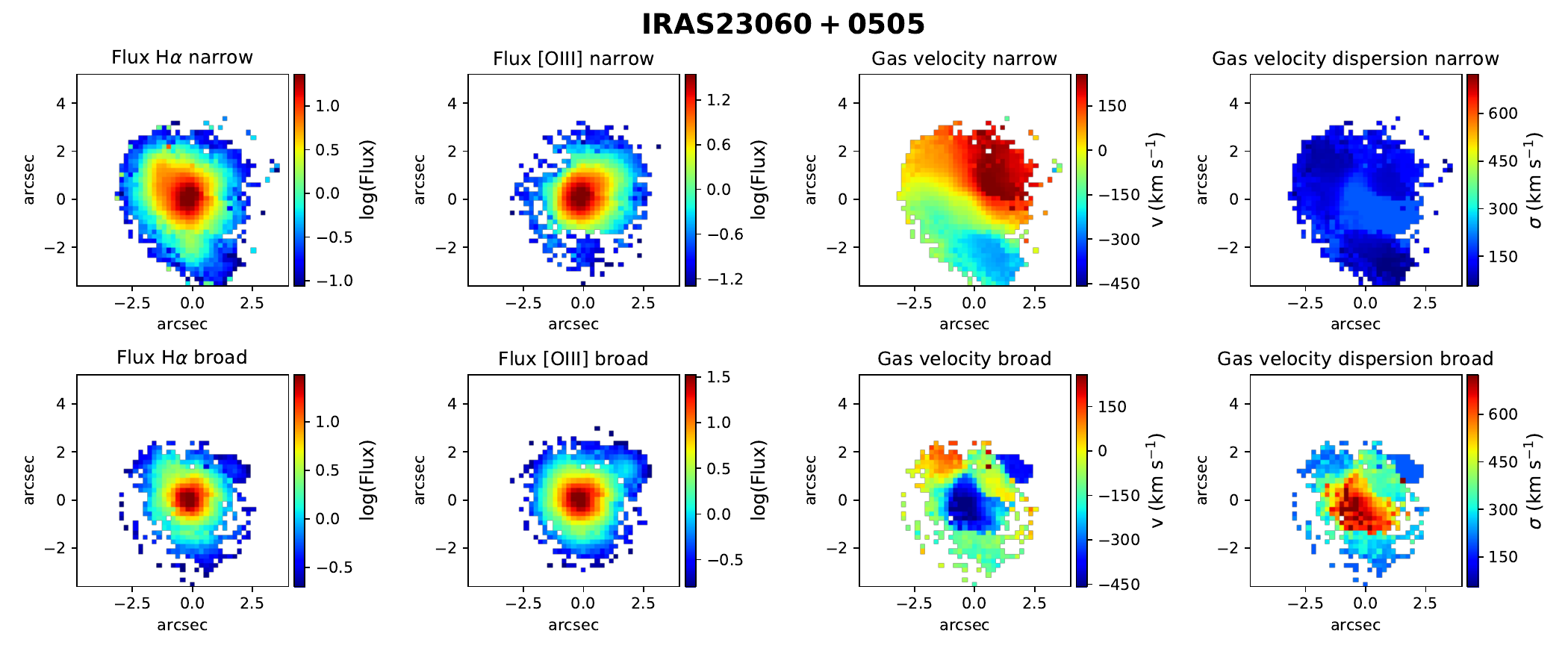}}%
\qquad
\subfigure[Integrated spectrum of the He~I emission and the sodium absorption feature (blue) with fit (red)]{
\label{fig:sfig2}
\includegraphics[width=.495\textwidth]{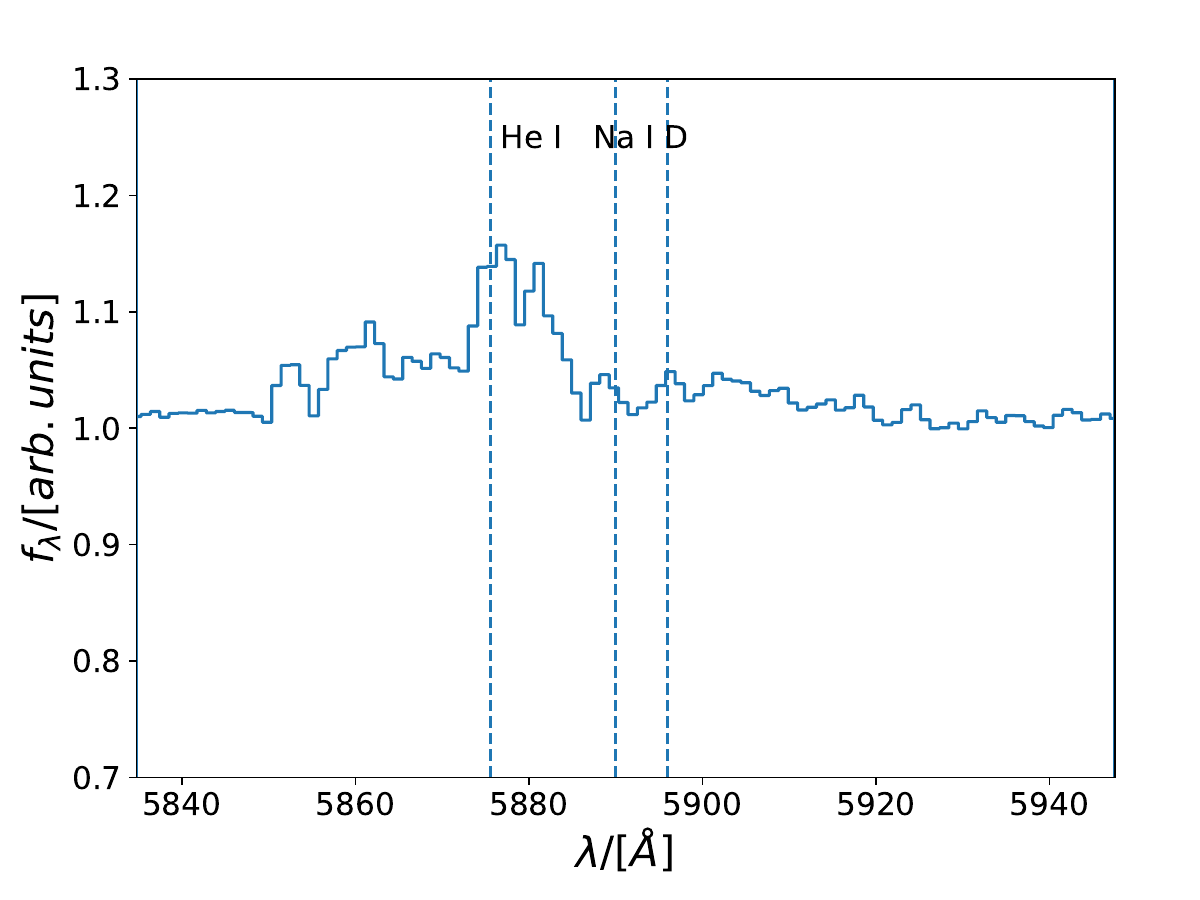}}%
\qquad\hspace{0em}
\subfigure[Integrated spectrum: The spectrum is shown in light blue, the three component fit in red, the narrow, broad and very broad components of H$\alpha$ and H$\beta$ in blue,  cyan and green, respectively.]{
\label{fig:sfig4}
\includegraphics[width=.8\textwidth]{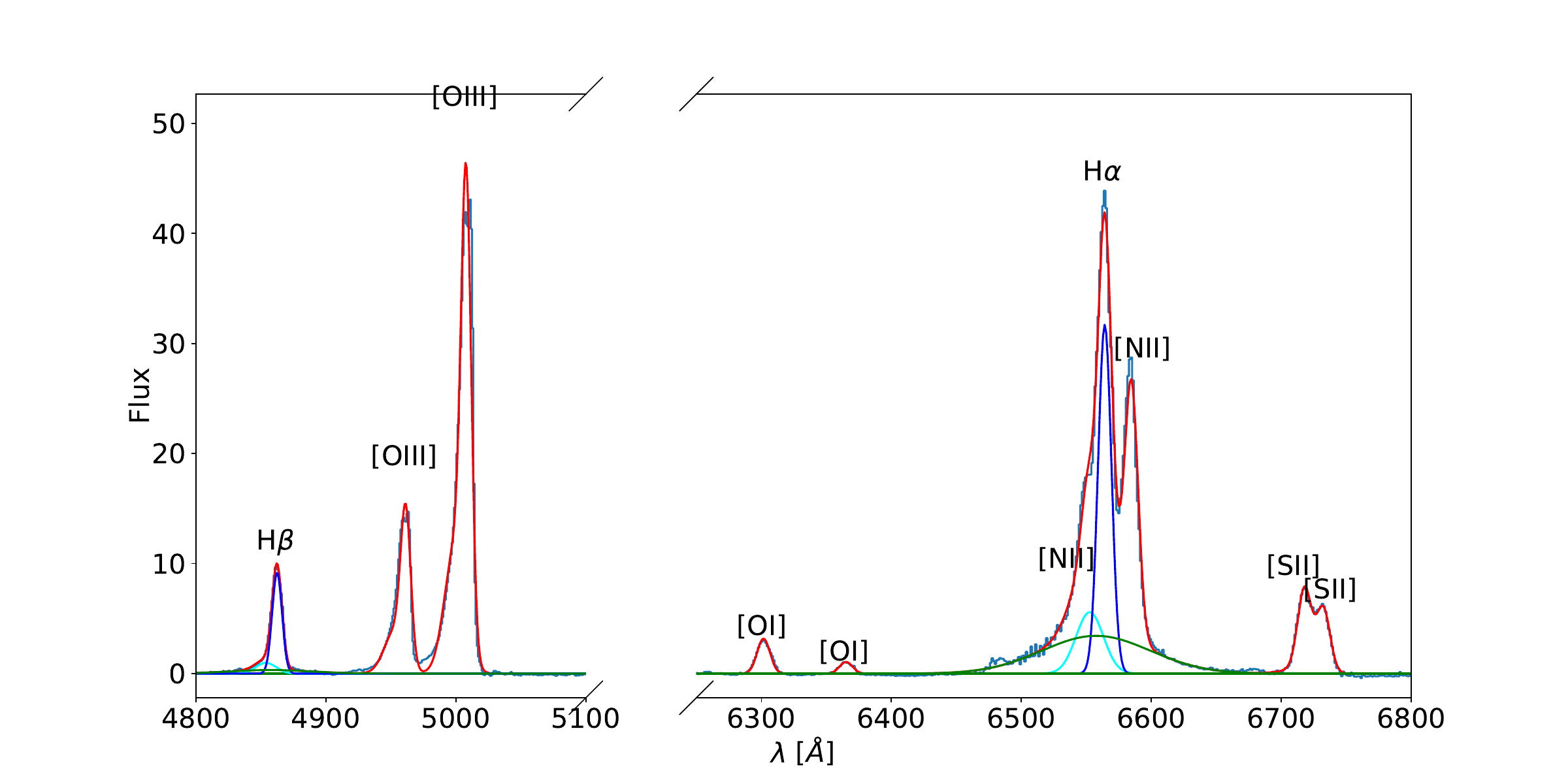}}%
\caption{IRAS 23060+0505}
\end{figure}

\begin{figure}%
\centering
\subfigure[Maps of the ionized gas]{
\label{fig:sfig1}
\includegraphics[width=\textwidth]{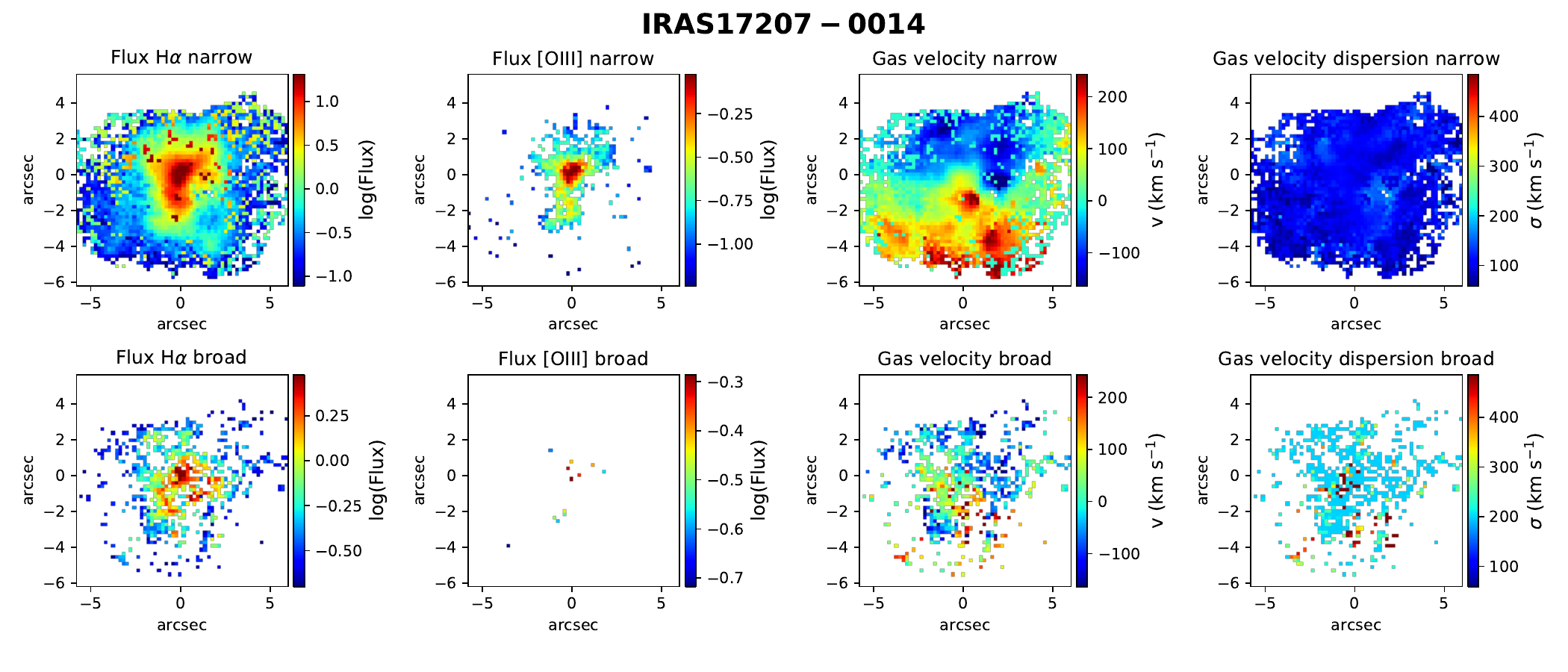}}%
\qquad
\subfigure[Integrated spectrum of the He~I emission and the sodium absorption feature (blue) with fit (red)]{
\label{fig:sfig2}
\includegraphics[width=.495\textwidth]{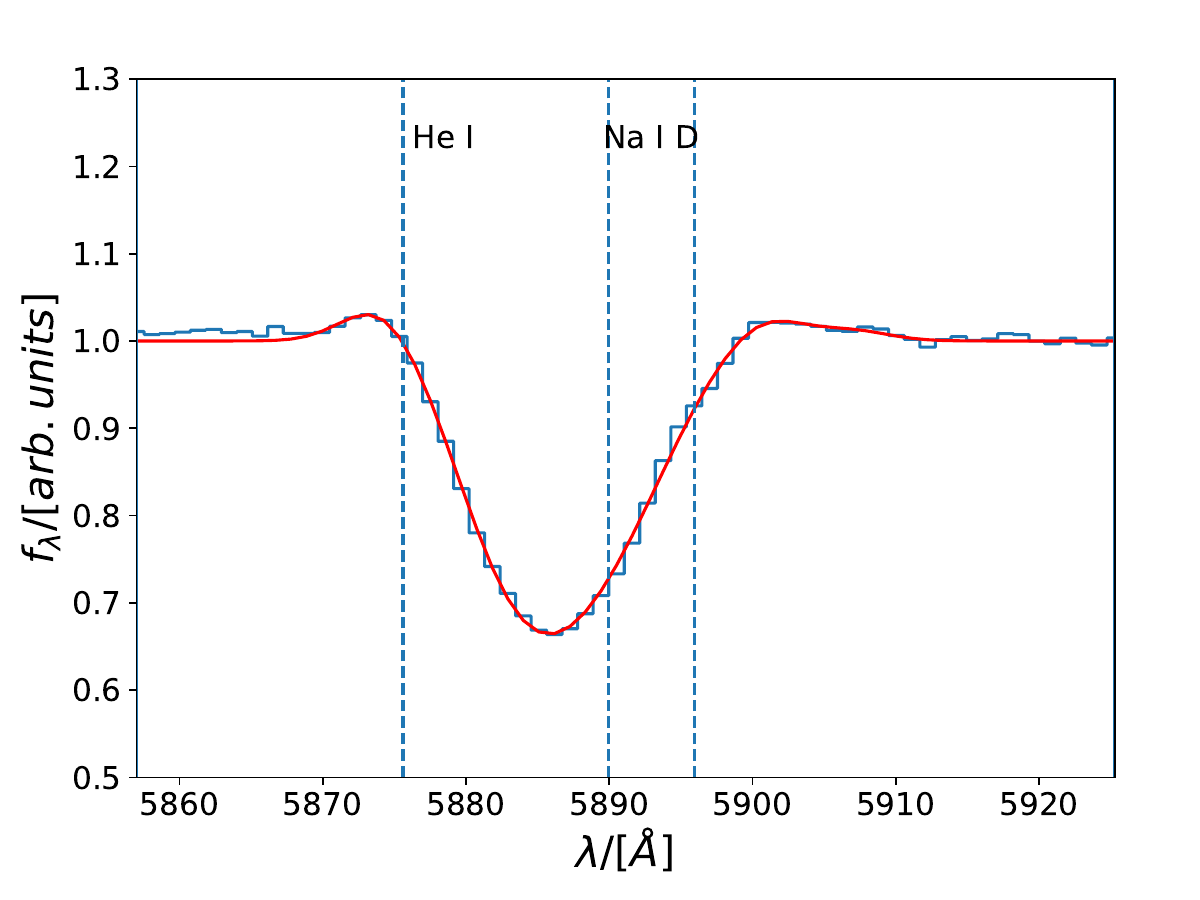}}%
\qquad\hspace{0em}
\subfigure[Maps of the neutral atomic gas]{
\label{fig:sfig3}
\includegraphics[trim={11cm 0 0 0},clip,width=.45\textwidth]{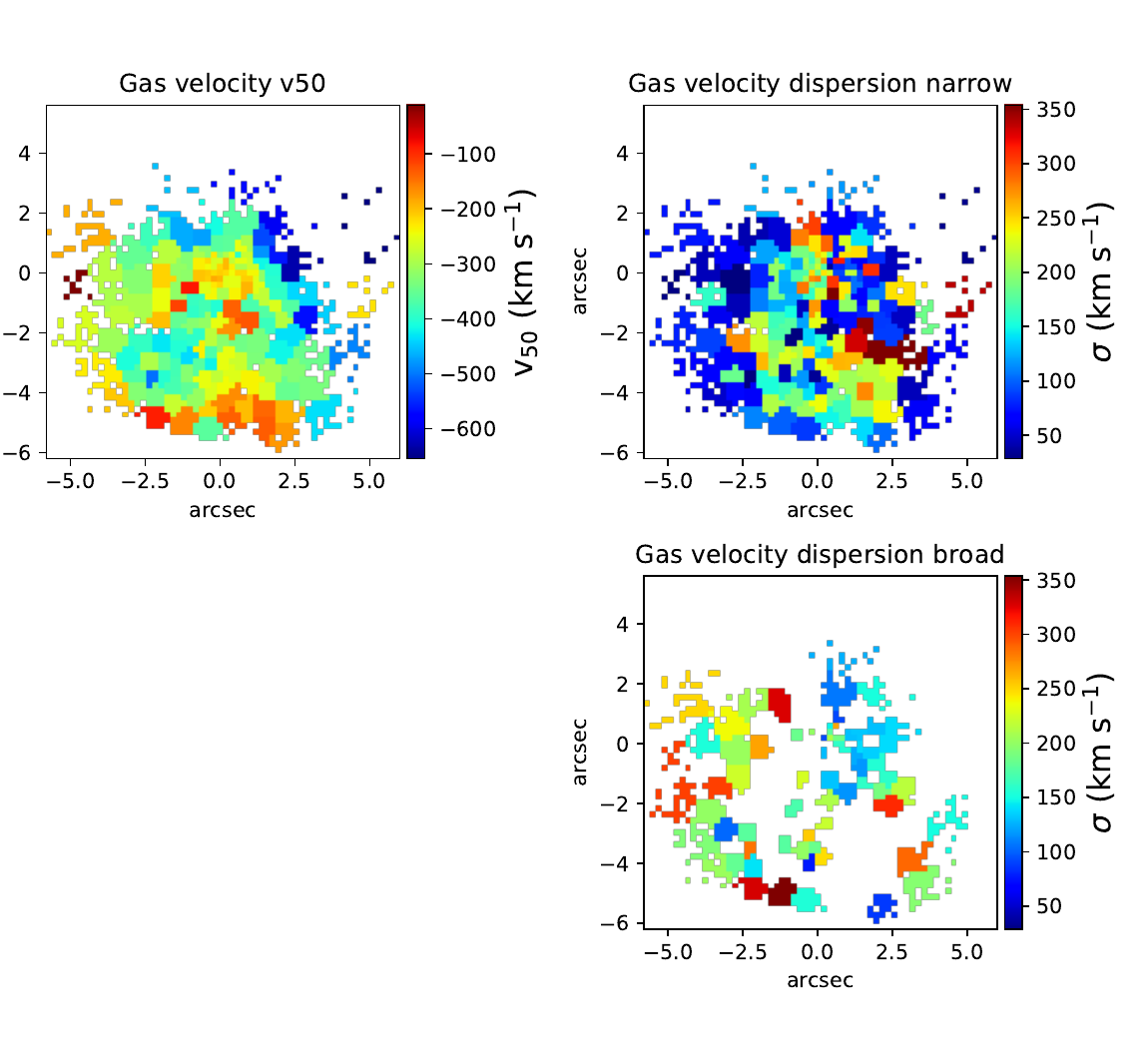}}%
\qquad
\subfigure[Integrated spectrum: The spectrum is shown in light blue, the two-component fit in green, the one component fit in dark blue and for the broad components of H$\alpha$ and H$\beta$ in cyan.]{
\label{fig:sfig4}
\includegraphics[width=.8\textwidth]{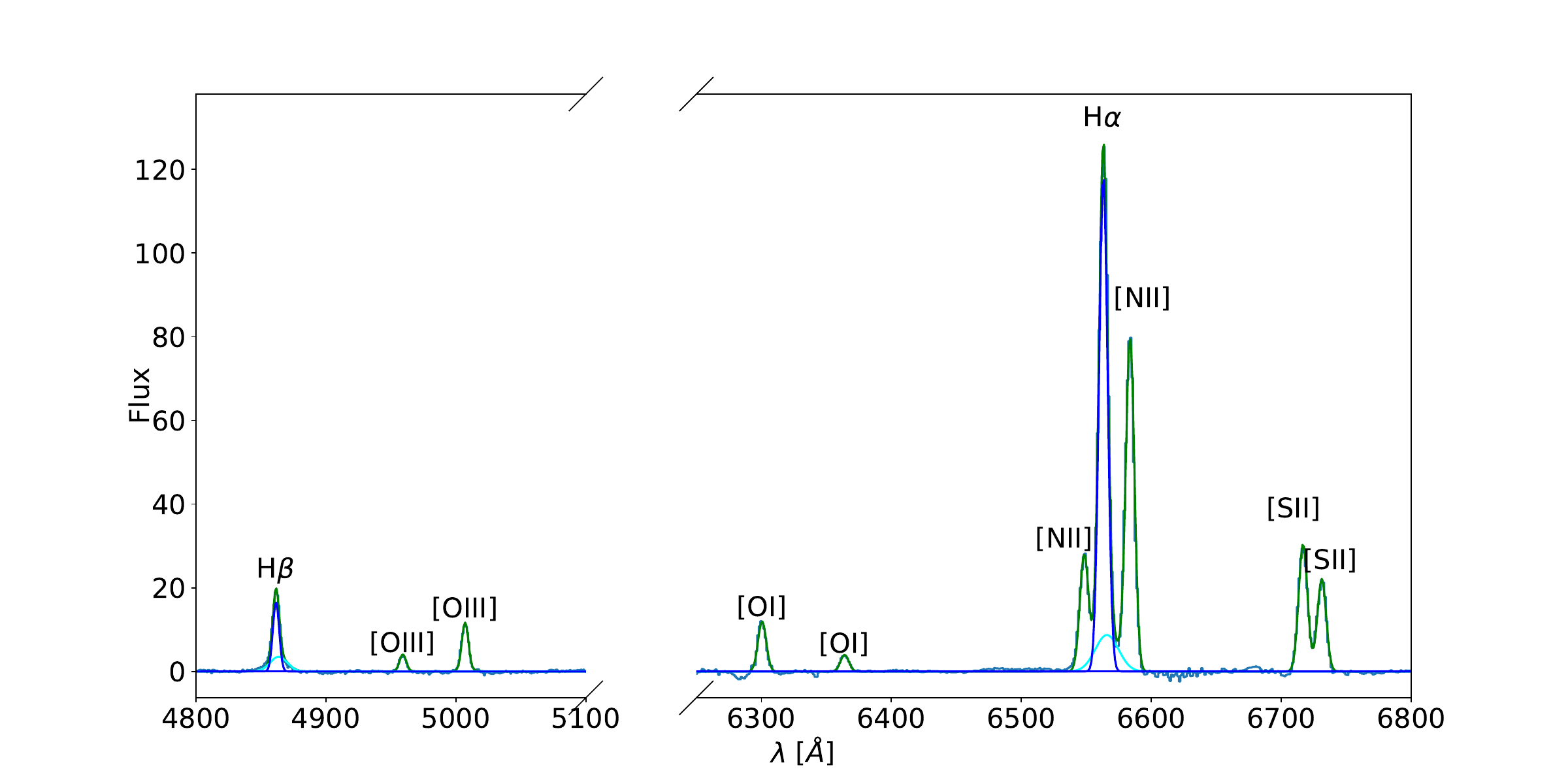}}%
\caption{IRAS 17207-0014}
\end{figure}

\begin{figure}%
\centering
\subfigure[Maps of the ionized gas]{
\label{fig:sfig1}
\includegraphics[width=\textwidth]{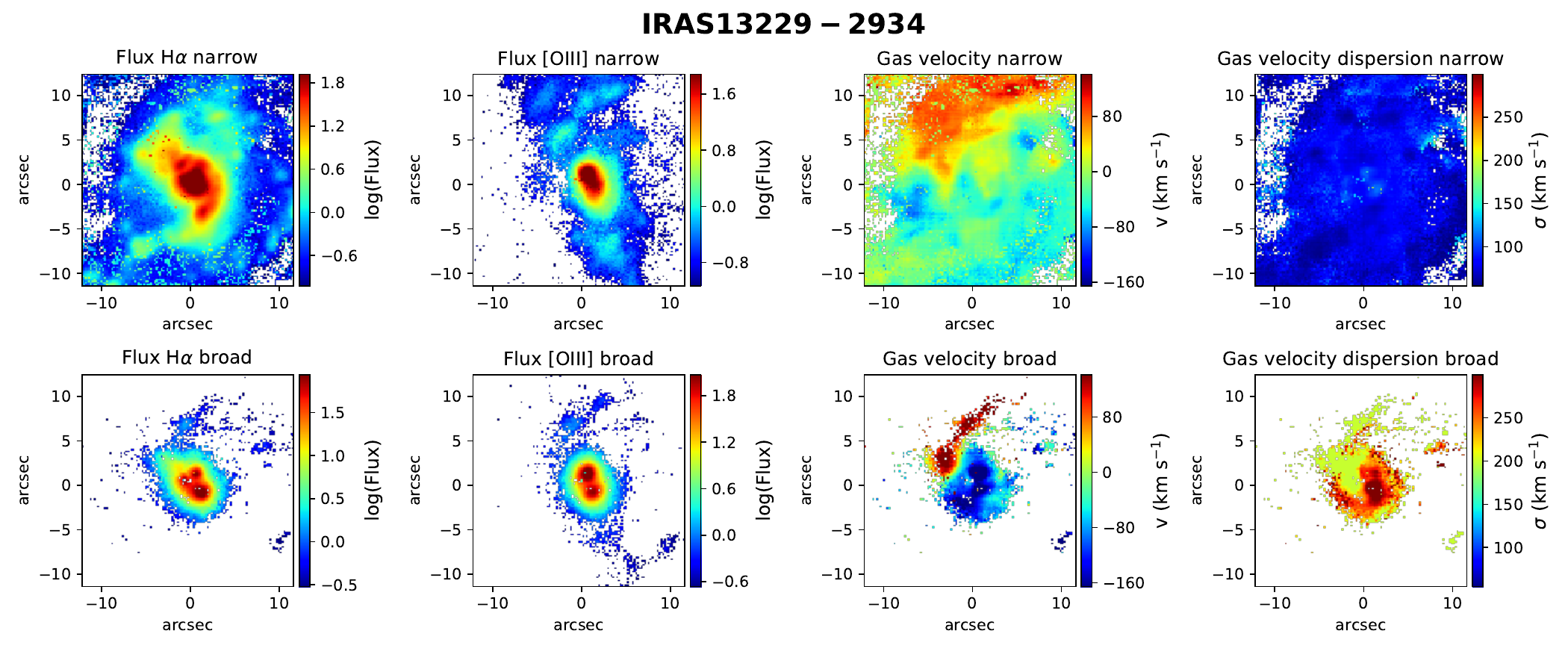}}%
\qquad
\subfigure[Integrated spectrum of the He~I emission and the sodium absorption feature (blue) with fit (red)]{
\label{fig:sfig2}
\includegraphics[width=.495\textwidth]{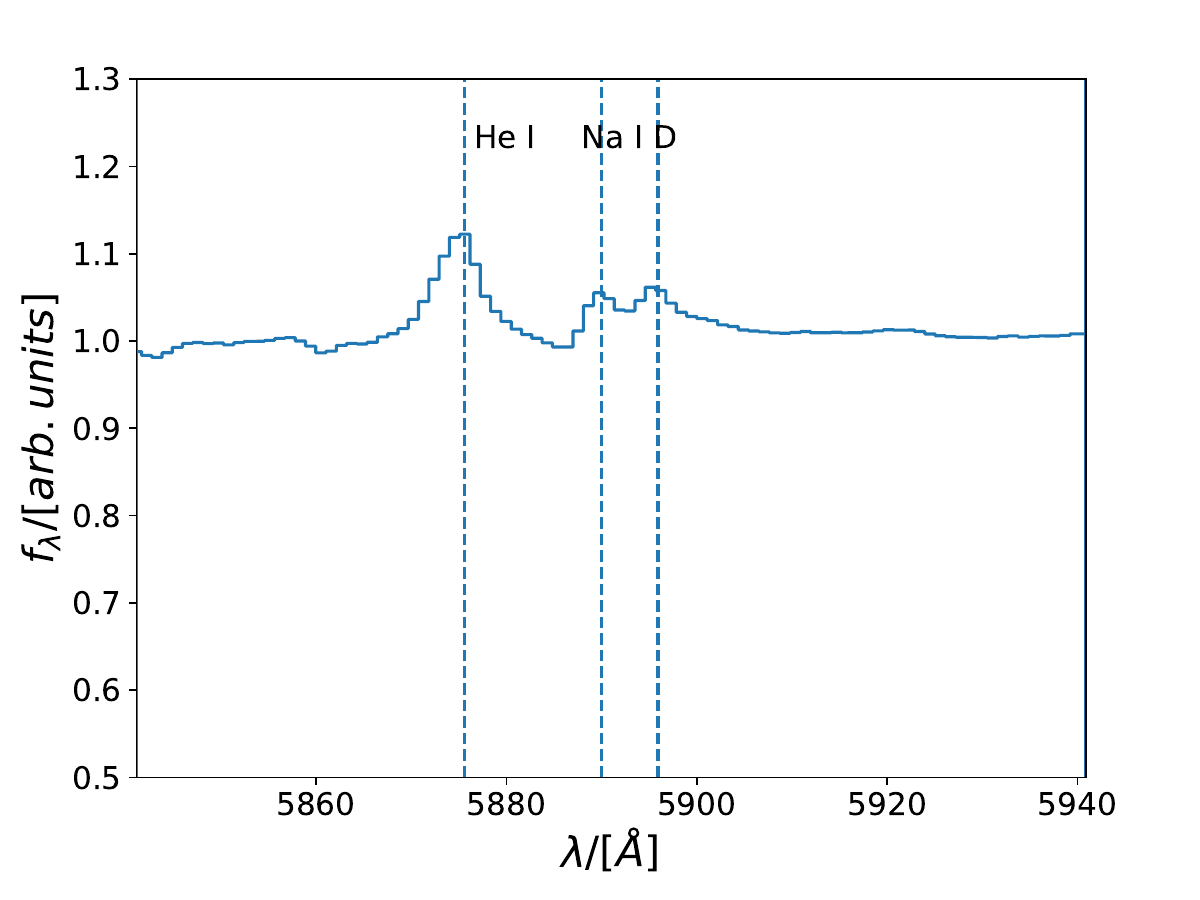}}%
\qquad\hspace{0em}
\subfigure[Maps of the neutral atomic gas]{
\label{fig:sfig3}
\includegraphics[trim={11cm 0 0 0},clip,width=.45\textwidth]{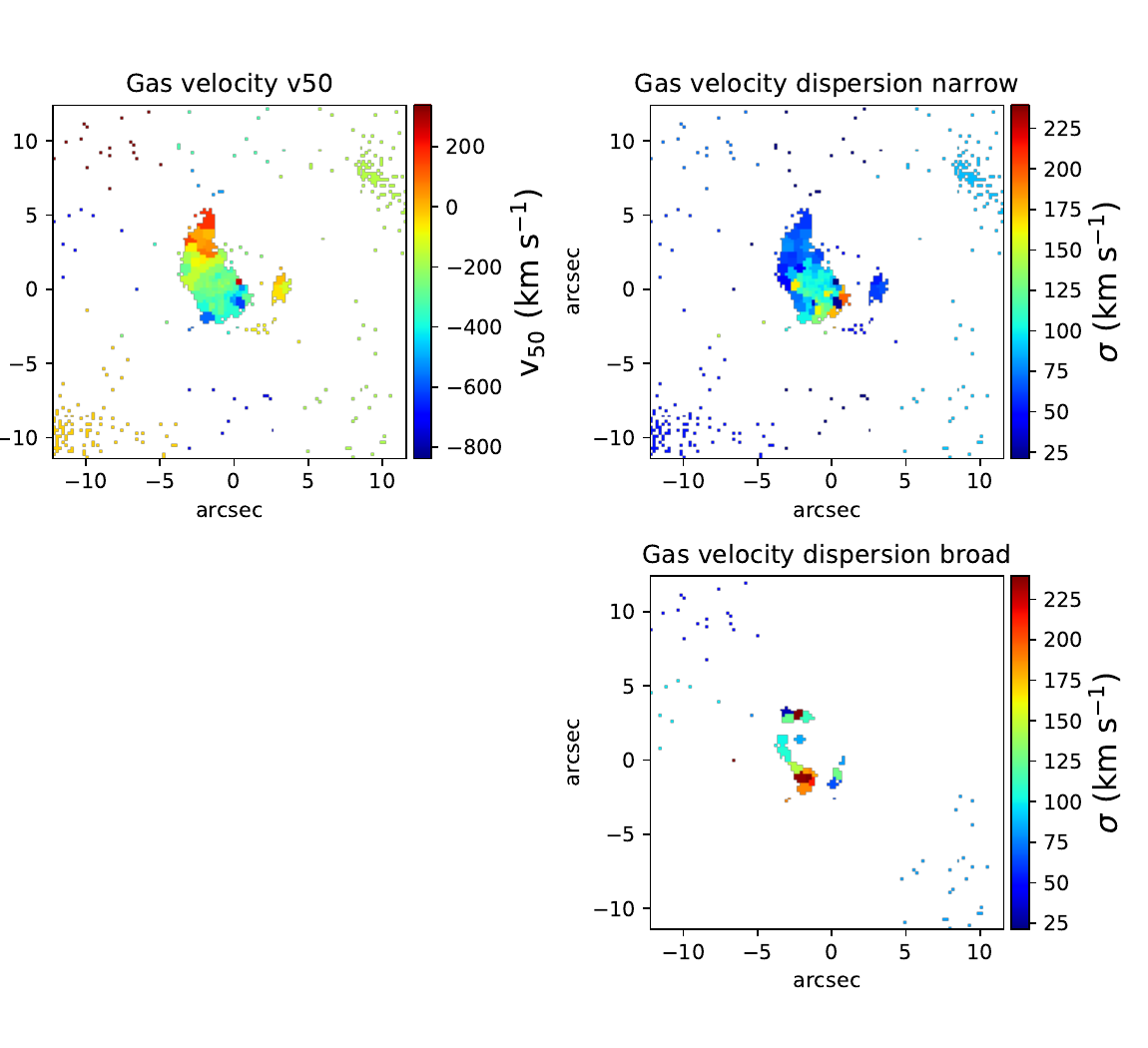}}%
\qquad
\subfigure[Integrated spectrum: The spectrum is shown in light blue, the two-component fit in green, the one component fit in dark blue and for the broad components of H$\alpha$ and H$\beta$ in cyan.]{
\label{fig:sfig4}
\includegraphics[width=.8\textwidth]{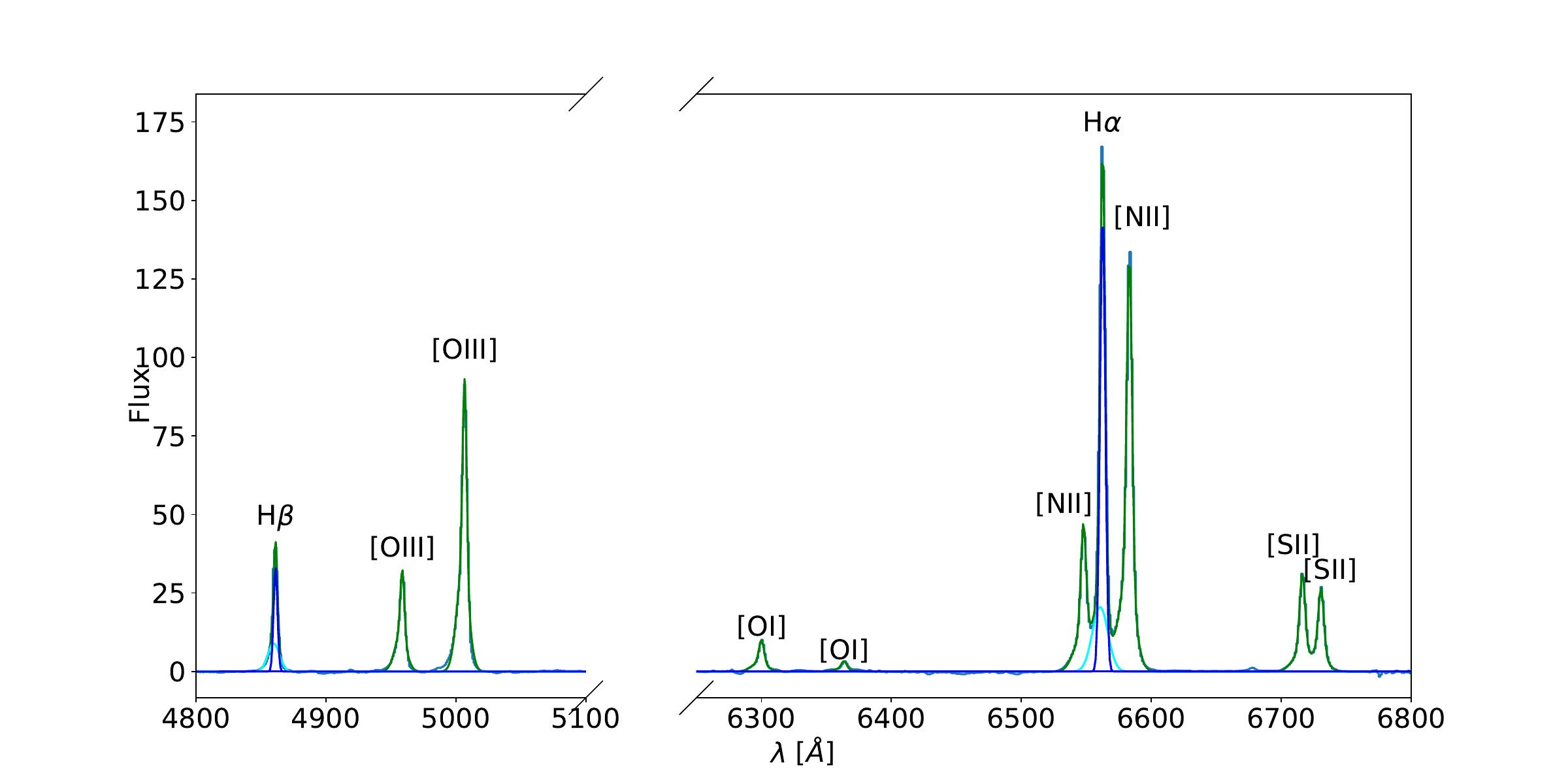}}%
\caption{IRAS 13229-2934}
\end{figure}

\begin{figure}%
\centering
\subfigure[Maps of the ionized gas]{
\label{fig:sfig1}
\includegraphics[width=\textwidth]{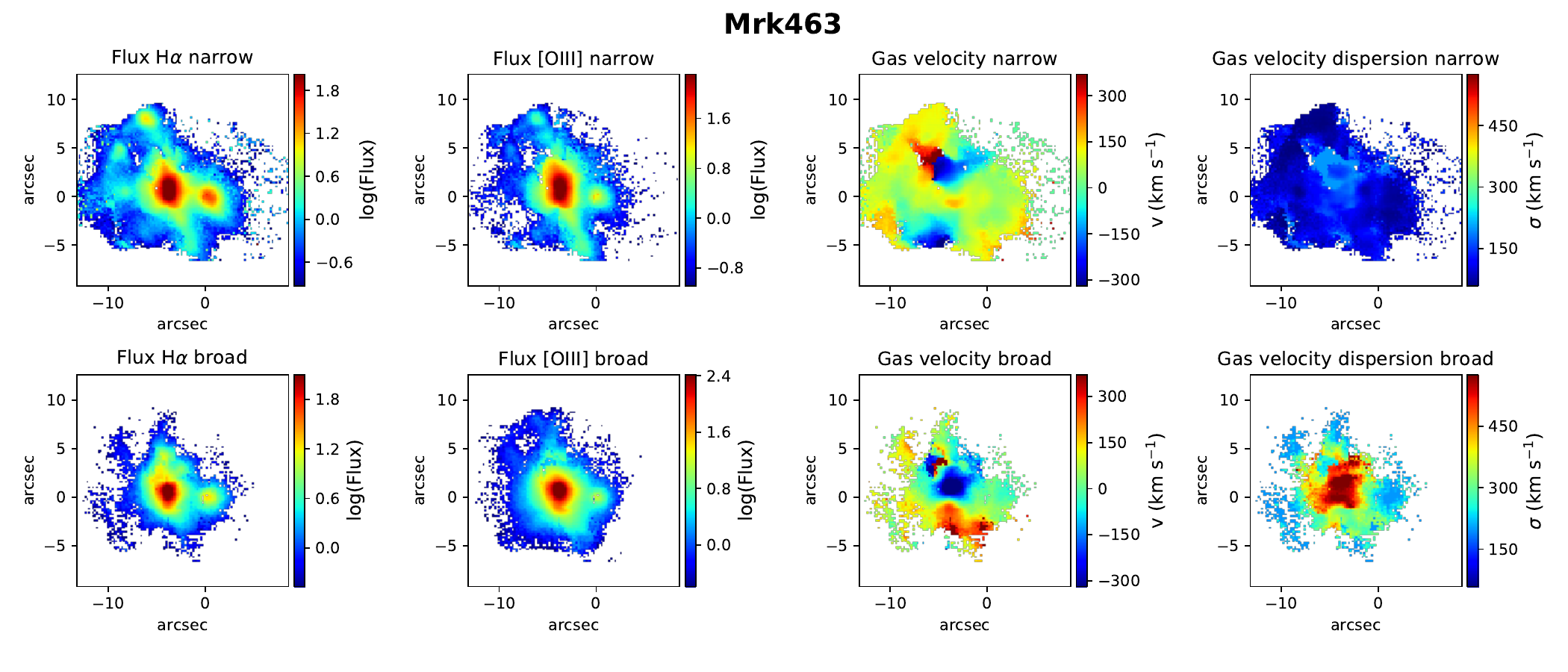}}%
\qquad
\subfigure[Integrated spectrum of the He~I emission and the sodium absorption feature (blue) with fit (red)]{
\label{fig:sfig2}
\includegraphics[width=.495\textwidth]{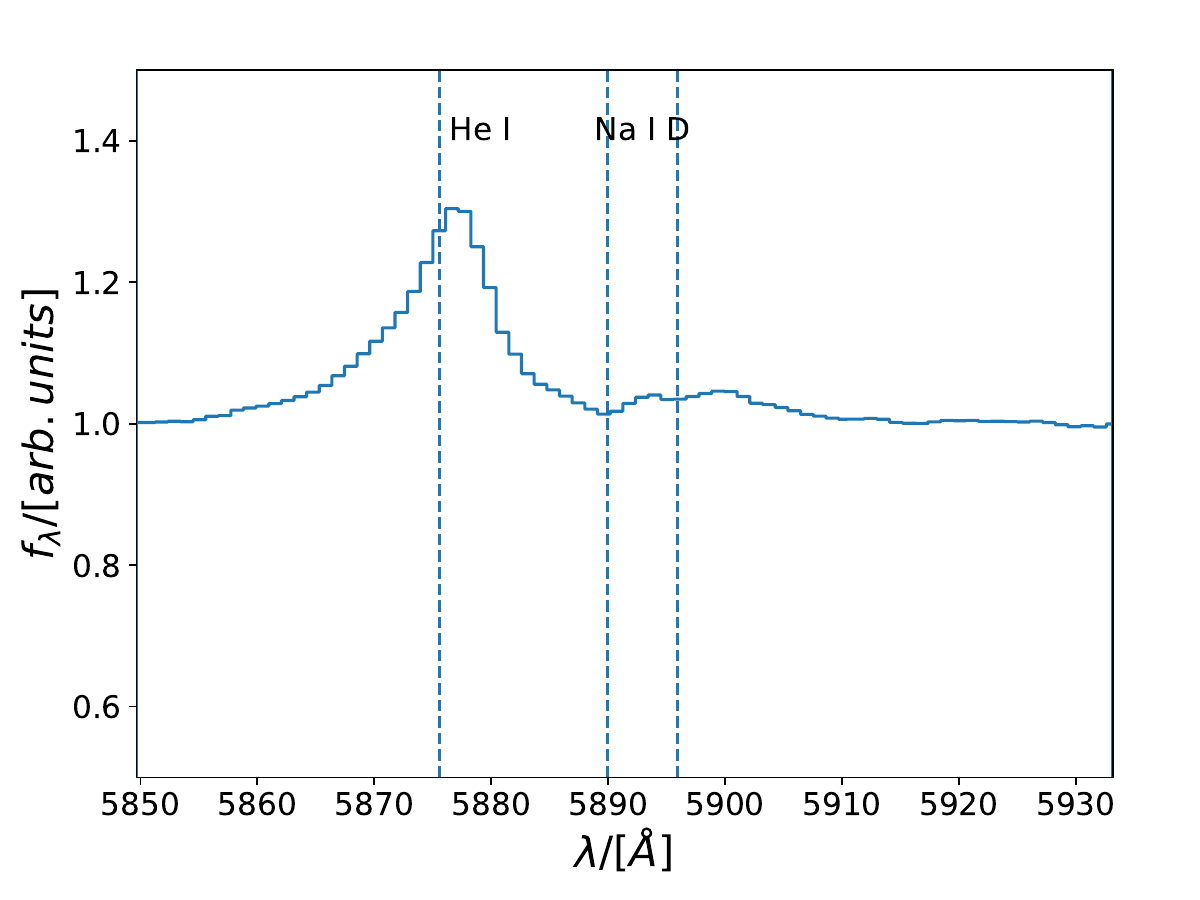}}%
\qquad\hspace{0em}
\subfigure[Integrated spectrum: The spectrum is shown in light blue, the two-component fit in green, the one component fit in dark blue and for the broad components of H$\alpha$ and H$\beta$ in cyan.]{
\label{fig:sfig4}
\includegraphics[width=.8\textwidth]{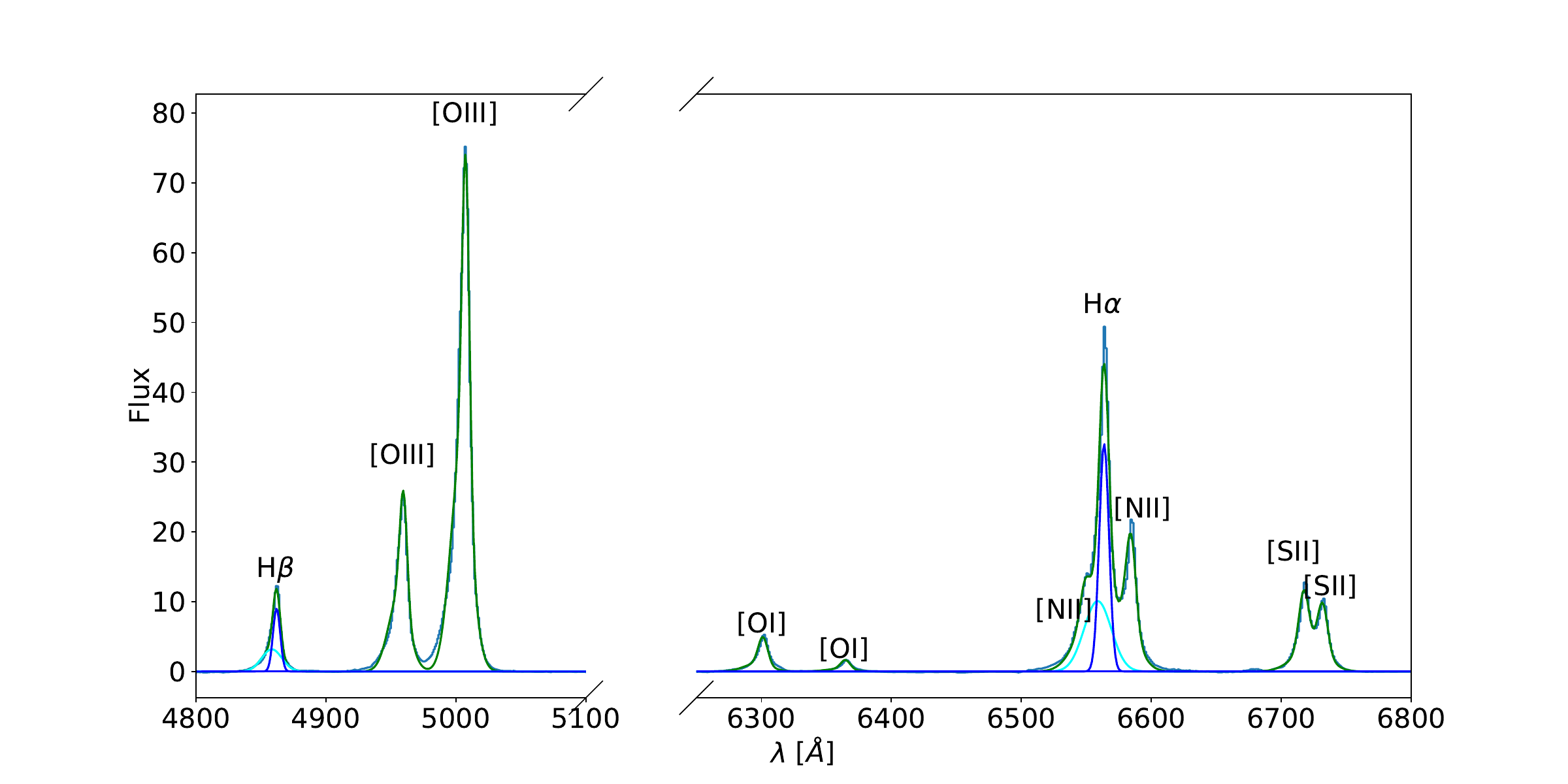}}%
\caption{Mrk 463}
\end{figure}

\begin{figure}%
\centering
\subfigure[Maps of the ionized gas]{
\label{fig:sfig1}
\includegraphics[width=\textwidth]{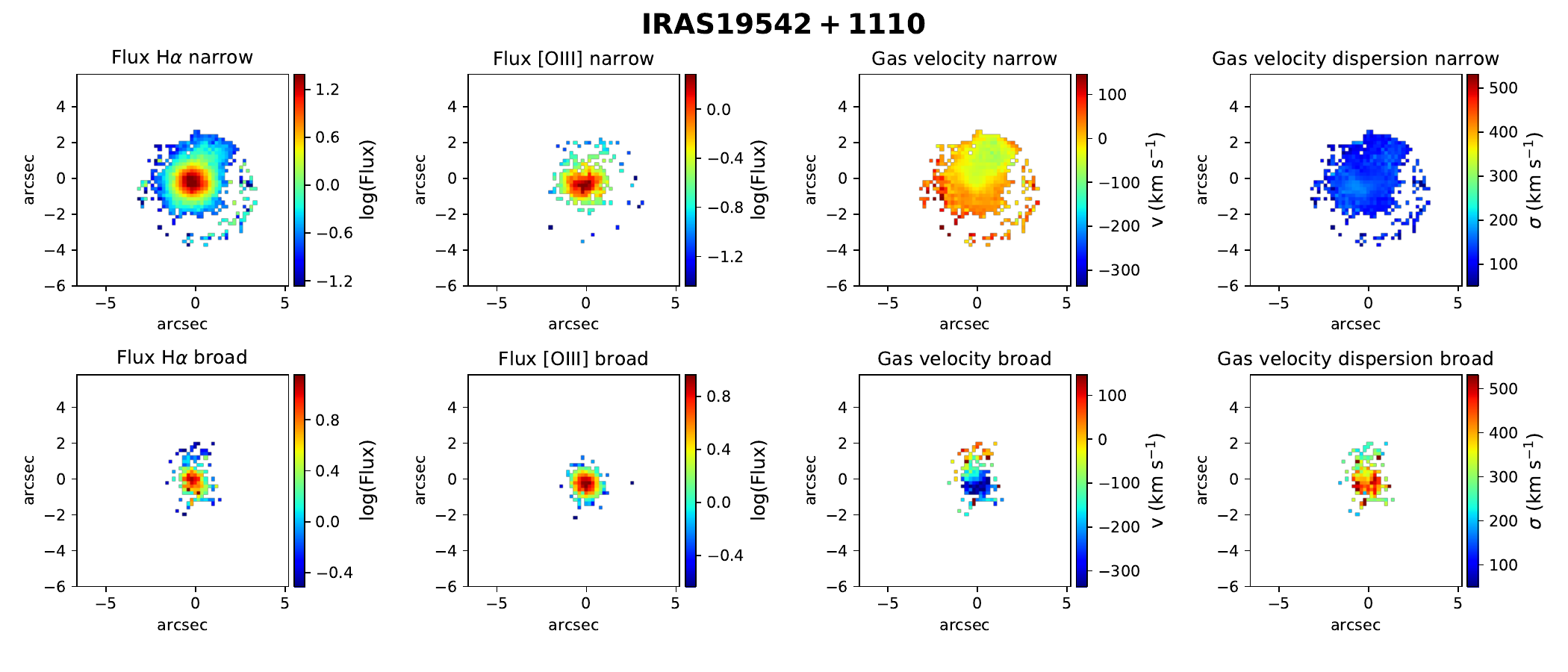}}%
\qquad
\subfigure[Integrated spectrum of the He~I emission and the sodium absorption feature (blue) with fit (red)]{
\label{fig:sfig2}
\includegraphics[width=.495\textwidth]{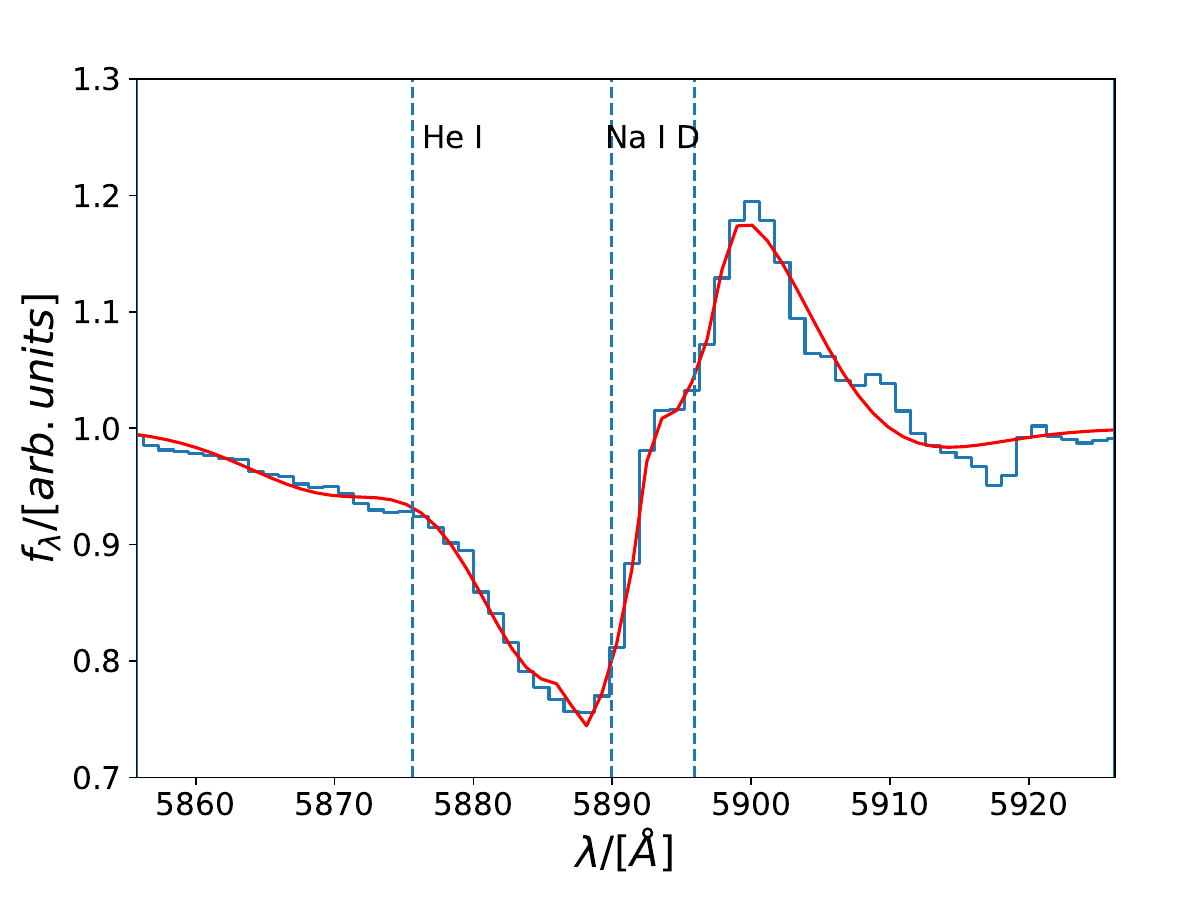}}%
\qquad\hspace{0em}
\subfigure[Maps of the neutral atomic gas]{
\label{fig:sfig3}
\includegraphics[trim={11cm 0 0 0},clip,width=.45\textwidth]{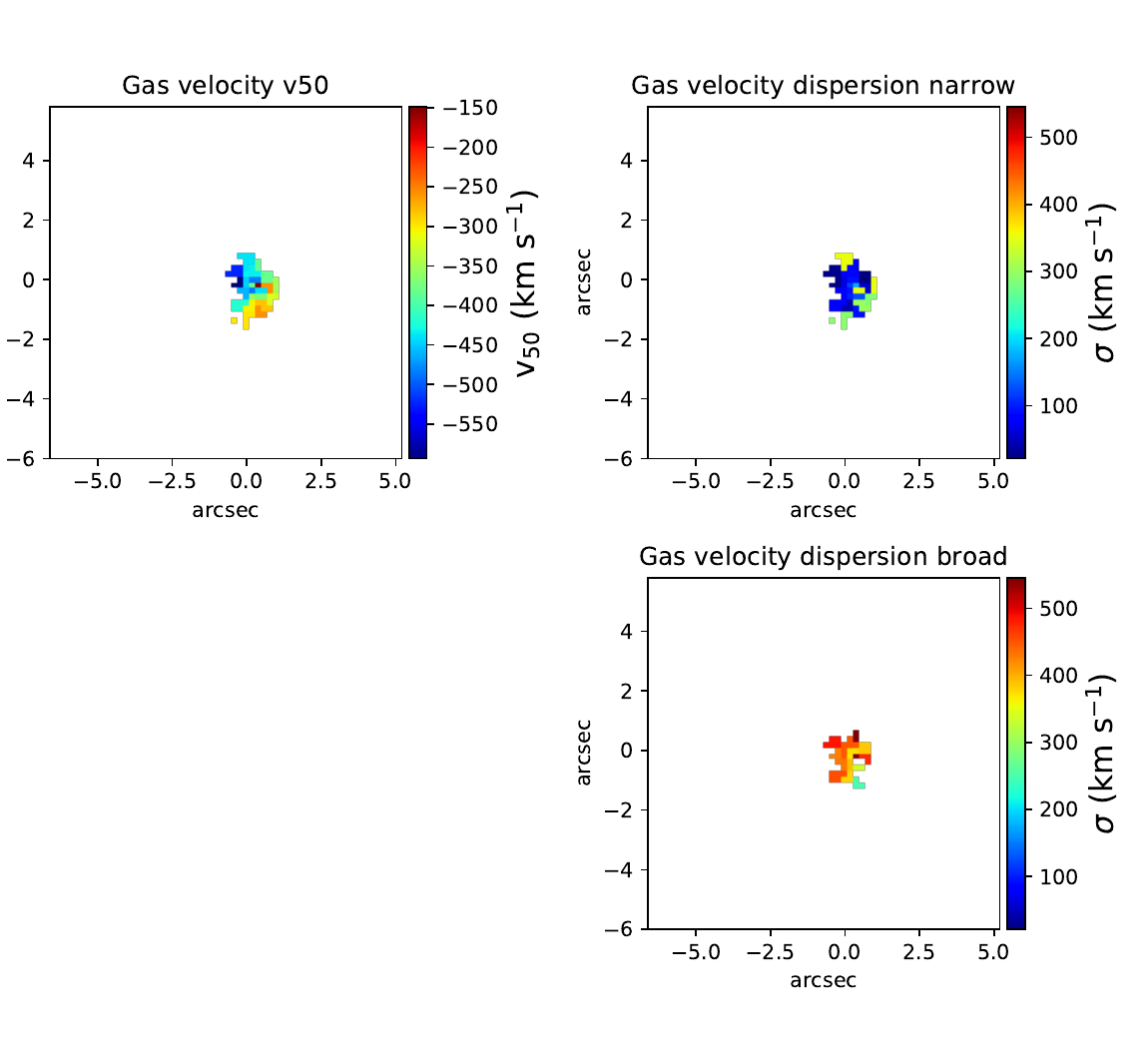}}%
\qquad
\subfigure[Integrated spectrum: The spectrum is shown in light blue, the two-component fit in green, the one component fit in dark blue and for the broad components of H$\alpha$ and H$\beta$ in cyan.]{
\label{fig:sfig4}
\includegraphics[width=.8\textwidth]{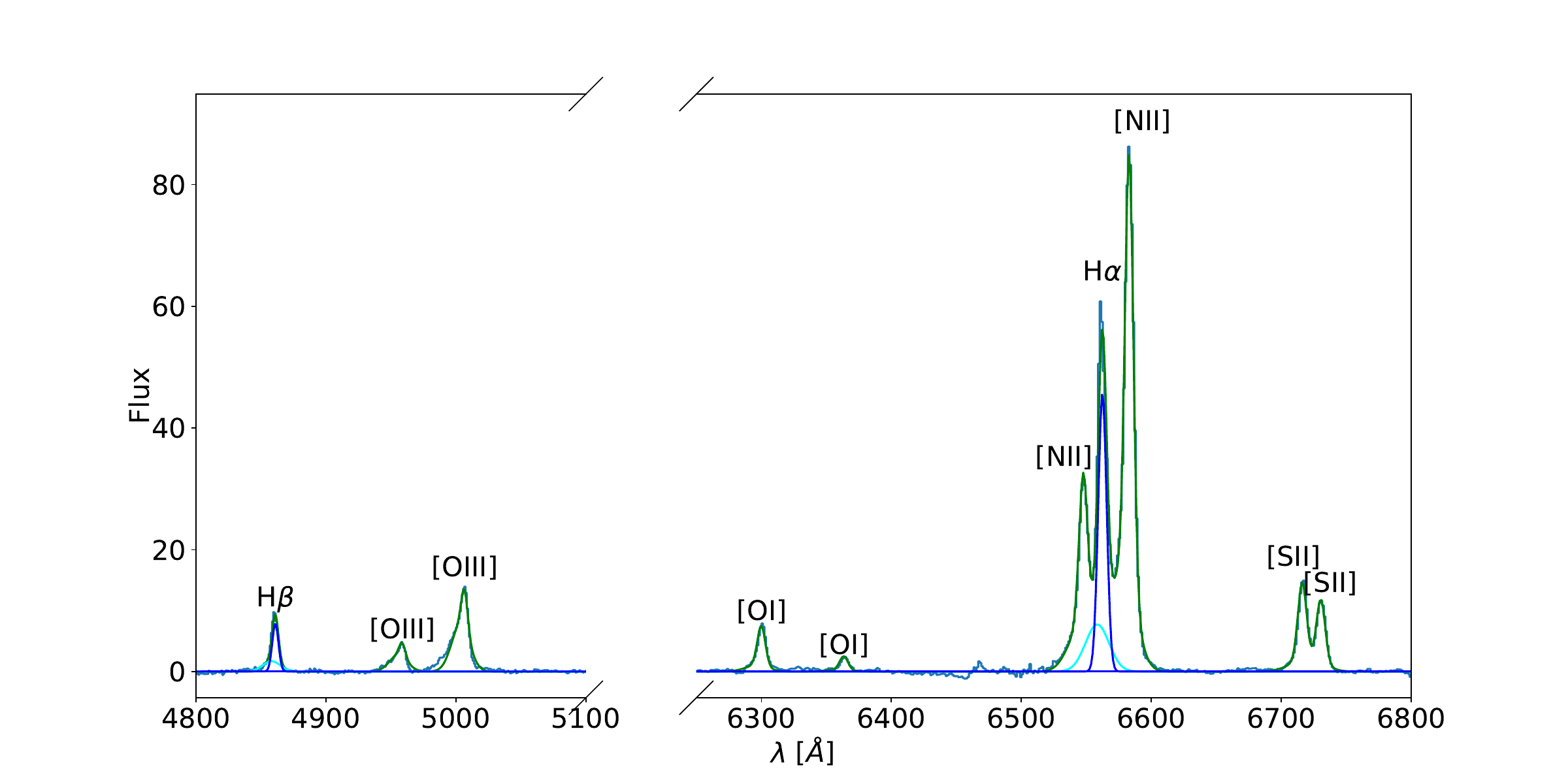}}%
\caption{IRAS 19542+1110}
\end{figure}

\begin{figure}%
\centering
\subfigure[Maps of the ionized gas]{
\label{fig:sfig1}
\includegraphics[width=\textwidth]{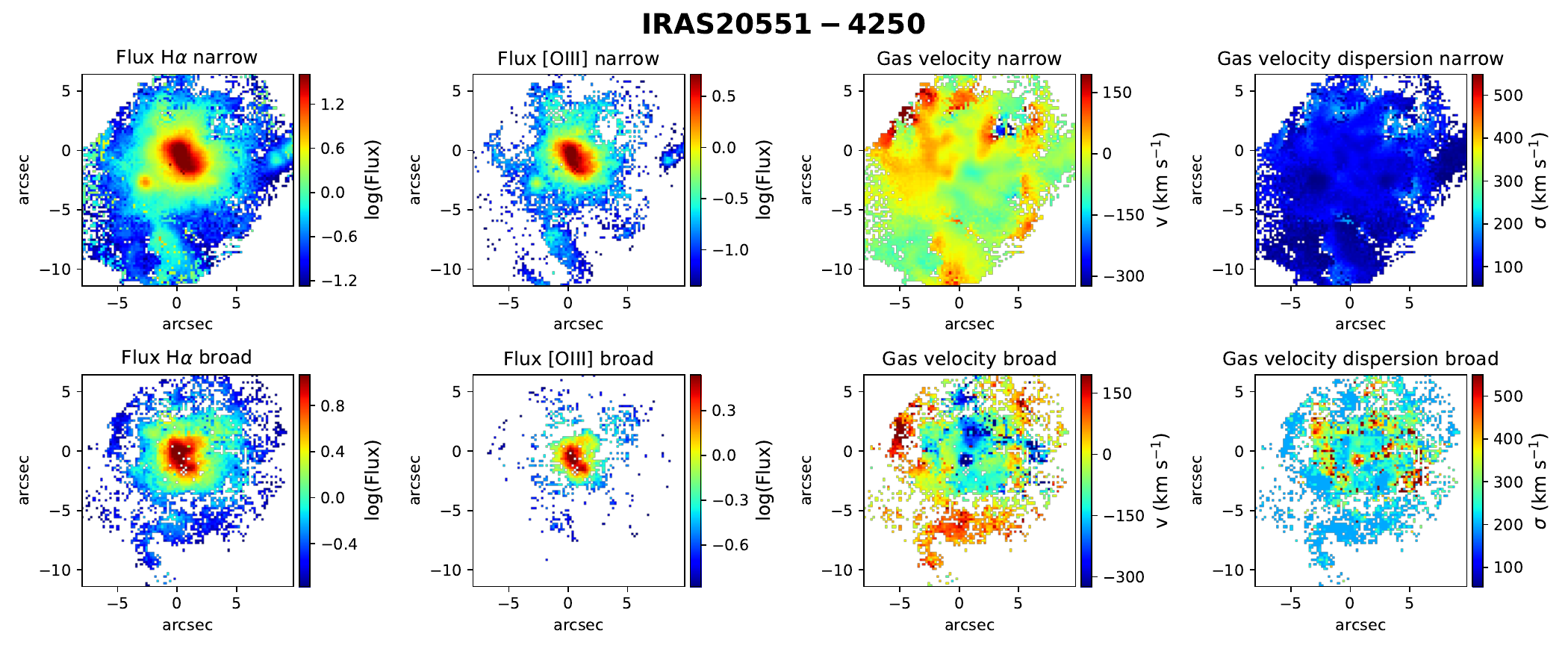}}%
\qquad
\subfigure[Integrated spectrum of the He~I emission and the sodium absorption feature (blue) with fit (red)]{
\label{fig:sfig2}
\includegraphics[width=.495\textwidth]{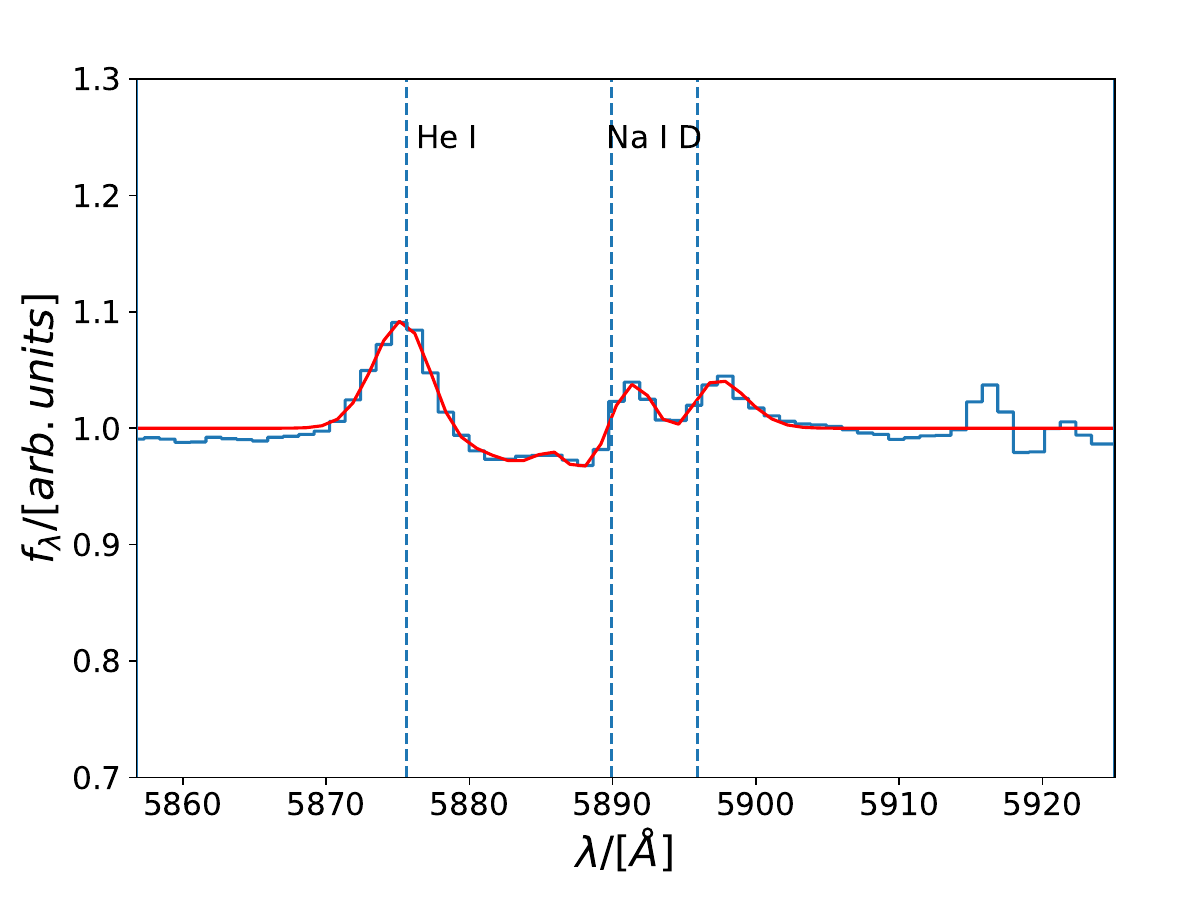}}%
\qquad\hspace{0em}
\subfigure[Maps of the neutral atomic gas]{
\label{fig:sfig3}
\includegraphics[trim={11cm 0 0 0},clip,width=.45\textwidth]{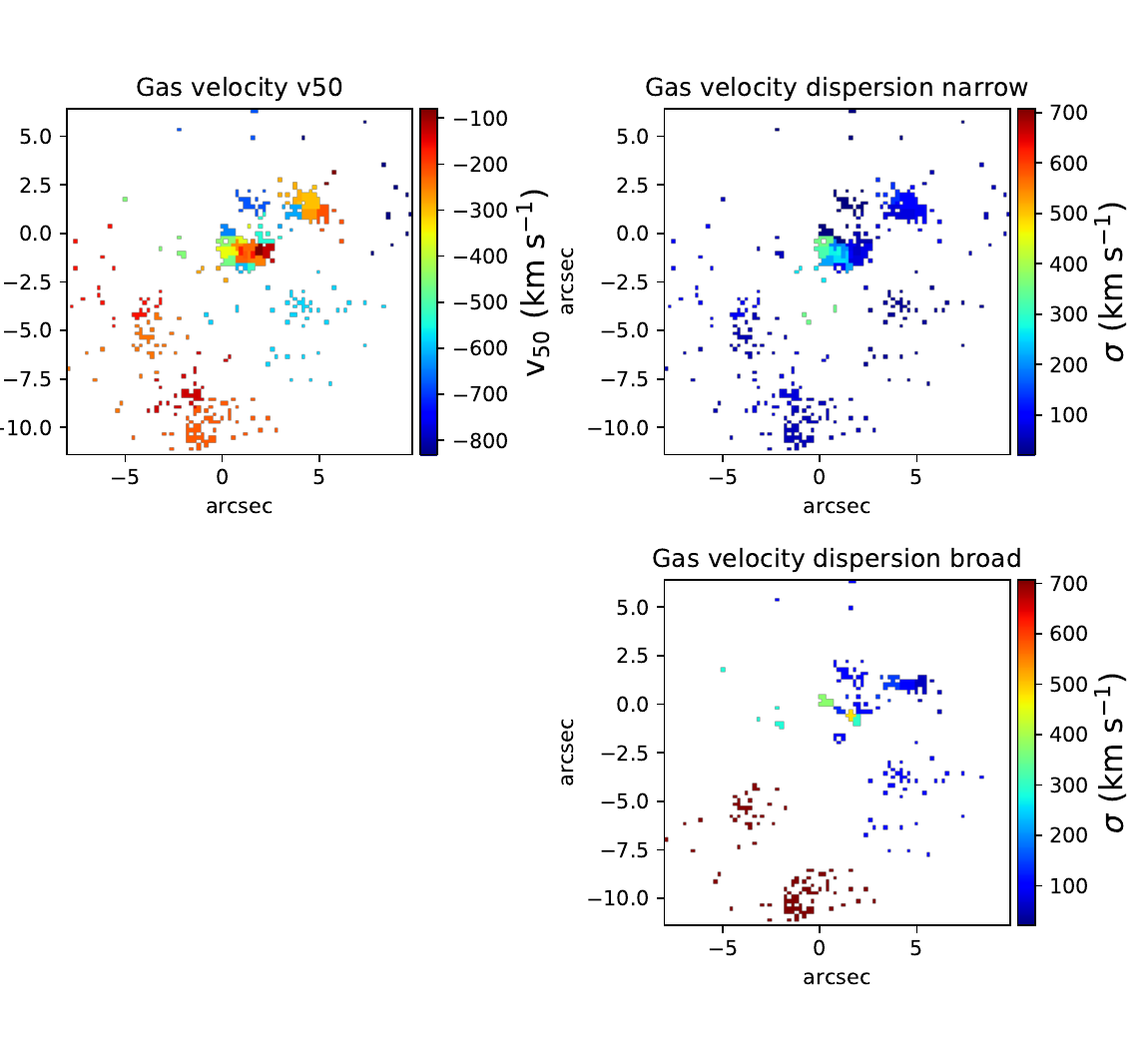}}%
\qquad
\subfigure[Integrated spectrum: The spectrum is shown in light blue, the two-component fit in green, the one component fit in dark blue and for the broad components of H$\alpha$ and H$\beta$ in cyan.]{
\label{fig:sfig4}
\includegraphics[width=.8\textwidth]{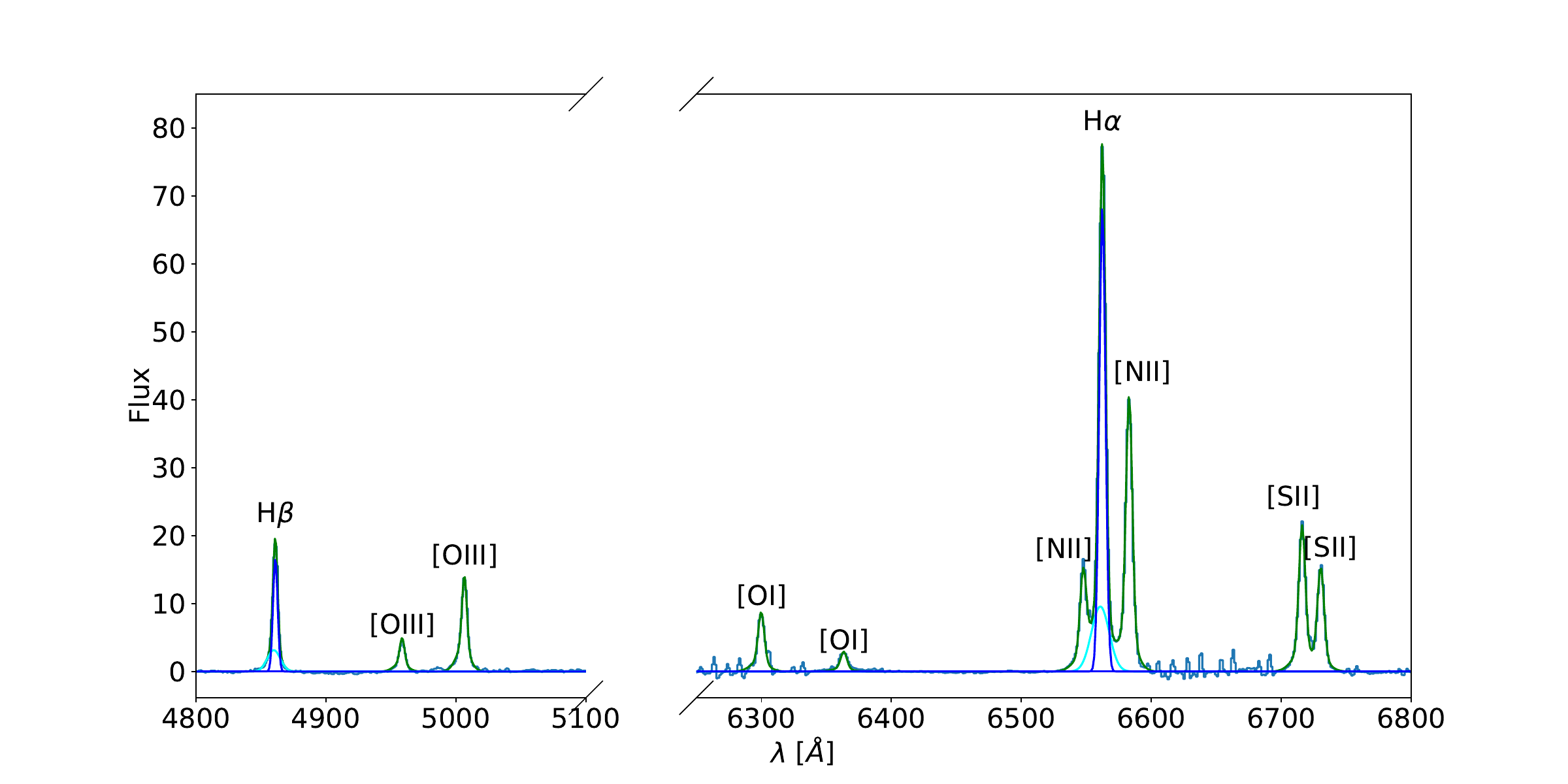}}%
\caption{IRAS 20551-4250}
\end{figure}

\begin{figure}%
\centering
\subfigure[Maps of the ionized gas. Although two components are shown, they are not kinematically distinct and we therefore do not classify the broad component as an outflow]{
\label{fig:sfig1}
\includegraphics[width=\textwidth]{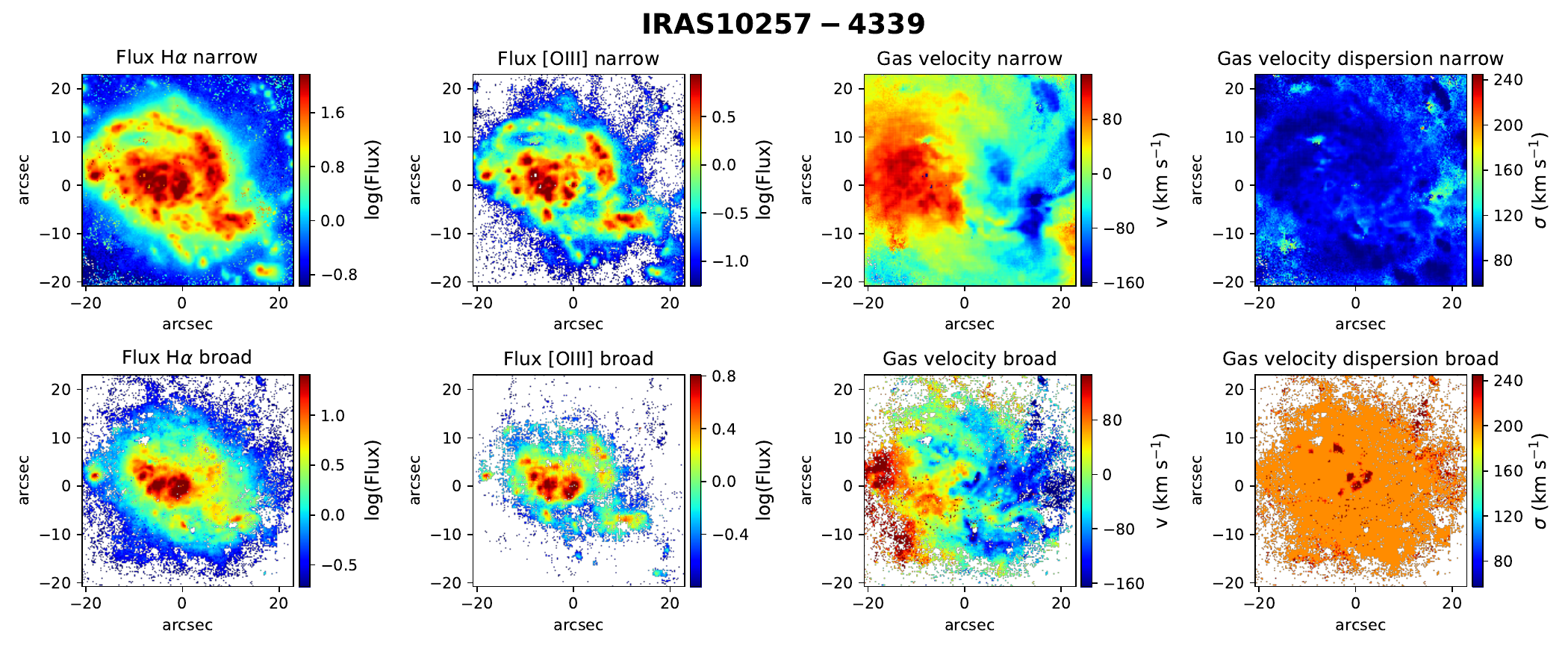}}%
\qquad
\subfigure[Integrated spectrum of the He~I emission and the sodium absorption feature (blue) with fit (red)]{
\label{fig:sfig2}
\includegraphics[width=.495\textwidth]{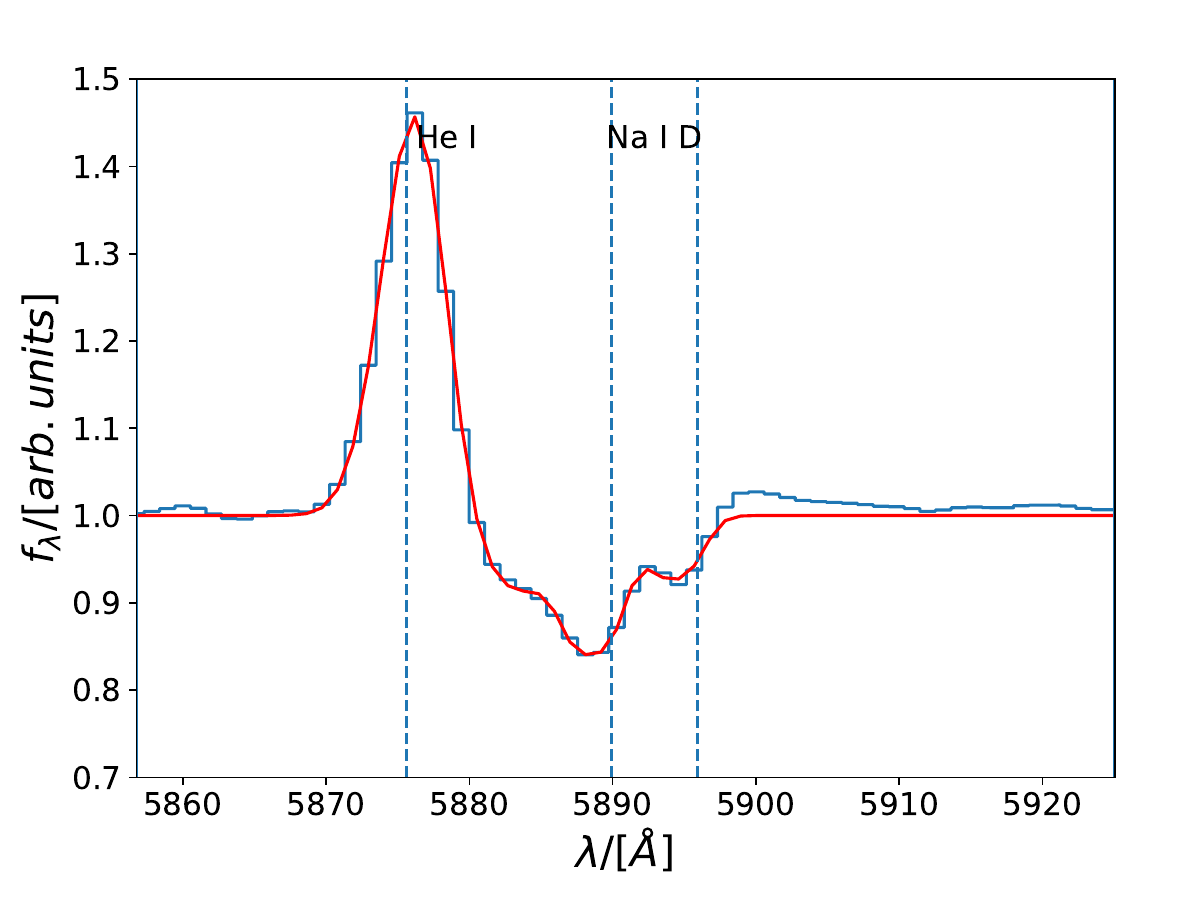}}%
\qquad\hspace{0em}
\subfigure[Maps of the neutral atomic gas]{
\label{fig:sfig3}
\includegraphics[trim={11cm 0 0 0},clip,width=.45\textwidth]{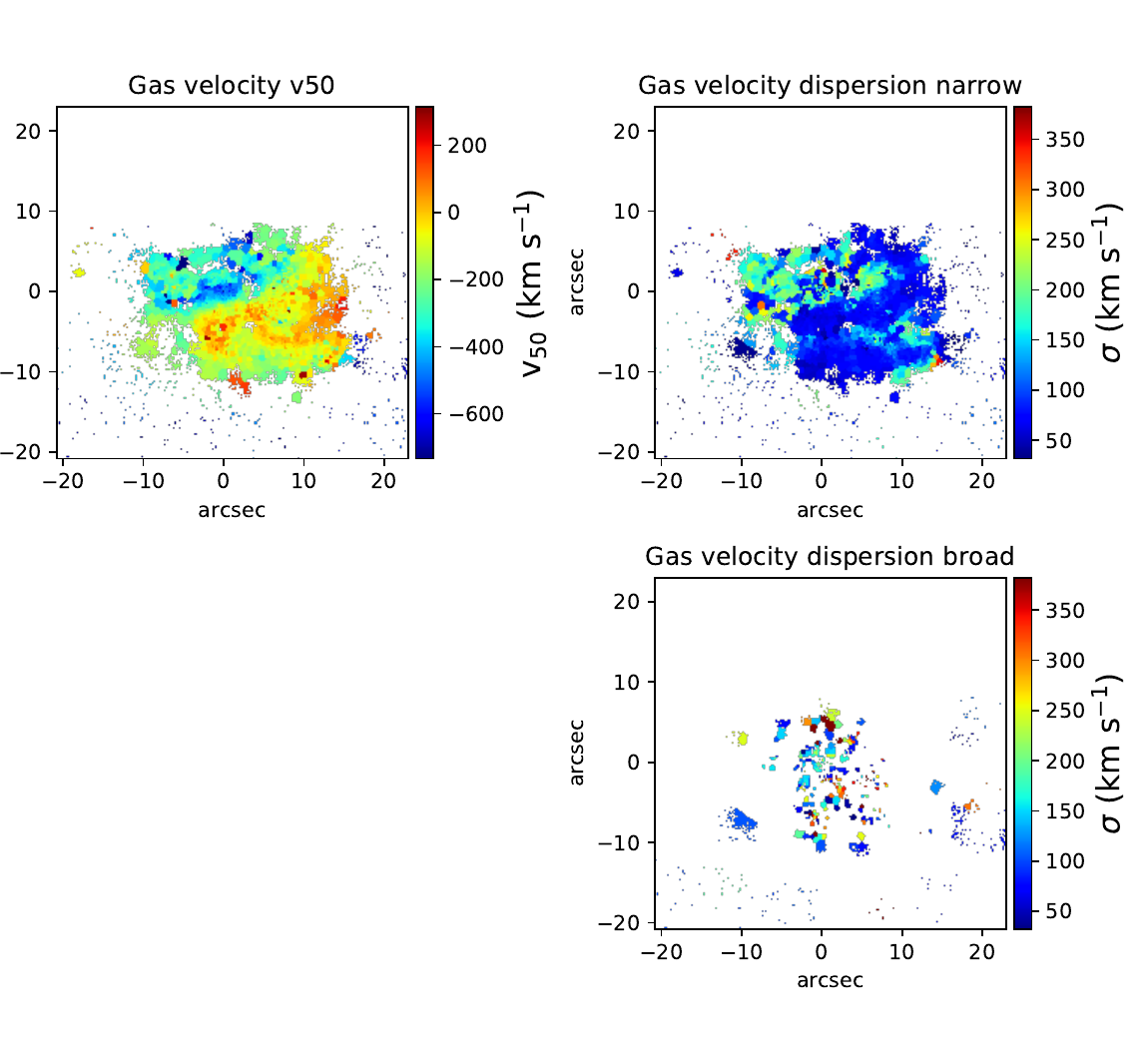}}%
\qquad
\subfigure[Integrated spectrum: The spectrum is shown in light blue and the one component fit in dark blue.]{
\label{fig:sfig4}
\includegraphics[width=.8\textwidth]{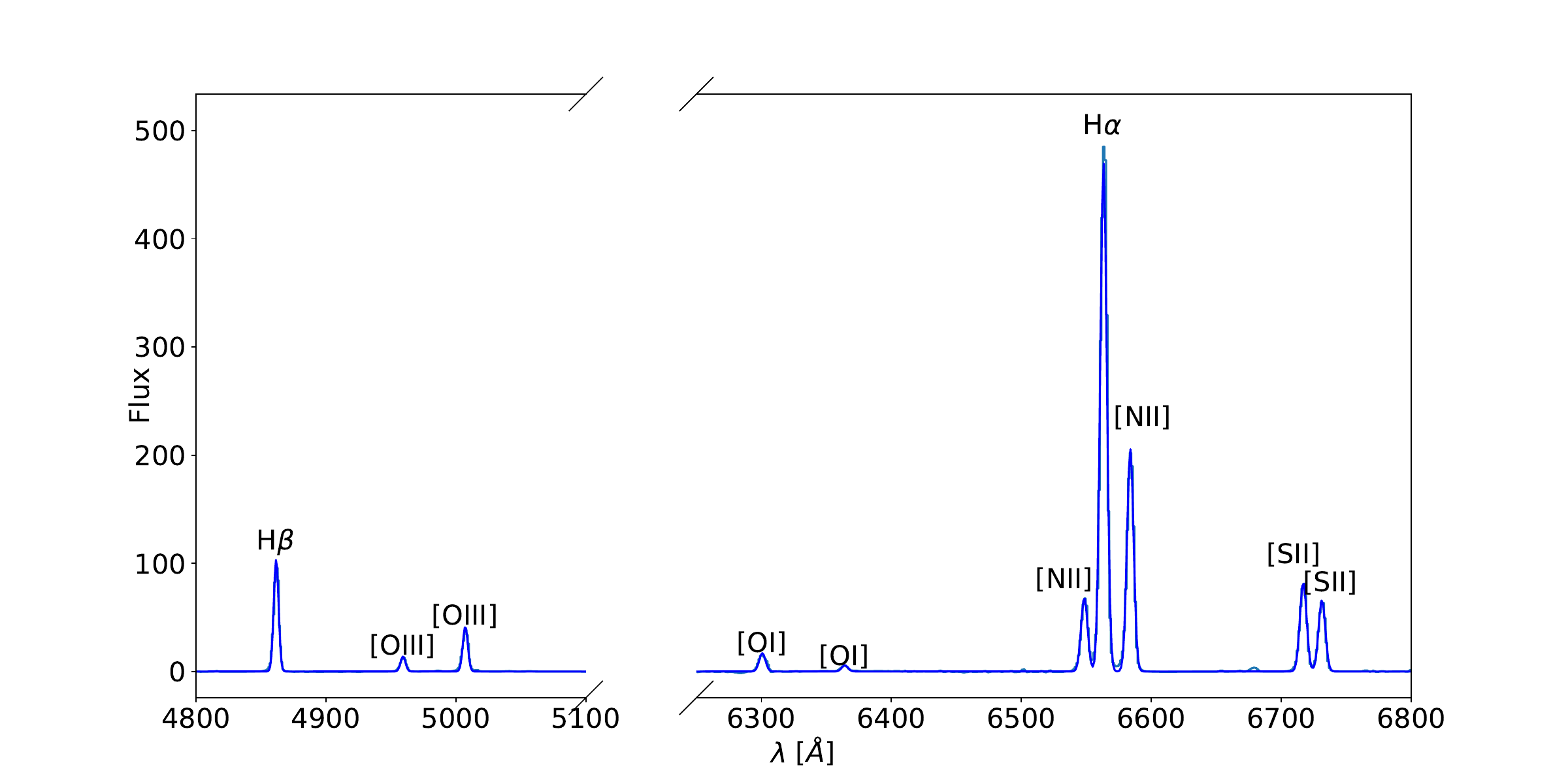}}%
\caption{IRAS 10257-4339}
\end{figure}

\begin{figure}%
\centering
\subfigure[Maps of the ionized gas]{
\label{fig:sfig1}
\includegraphics[width=\textwidth]{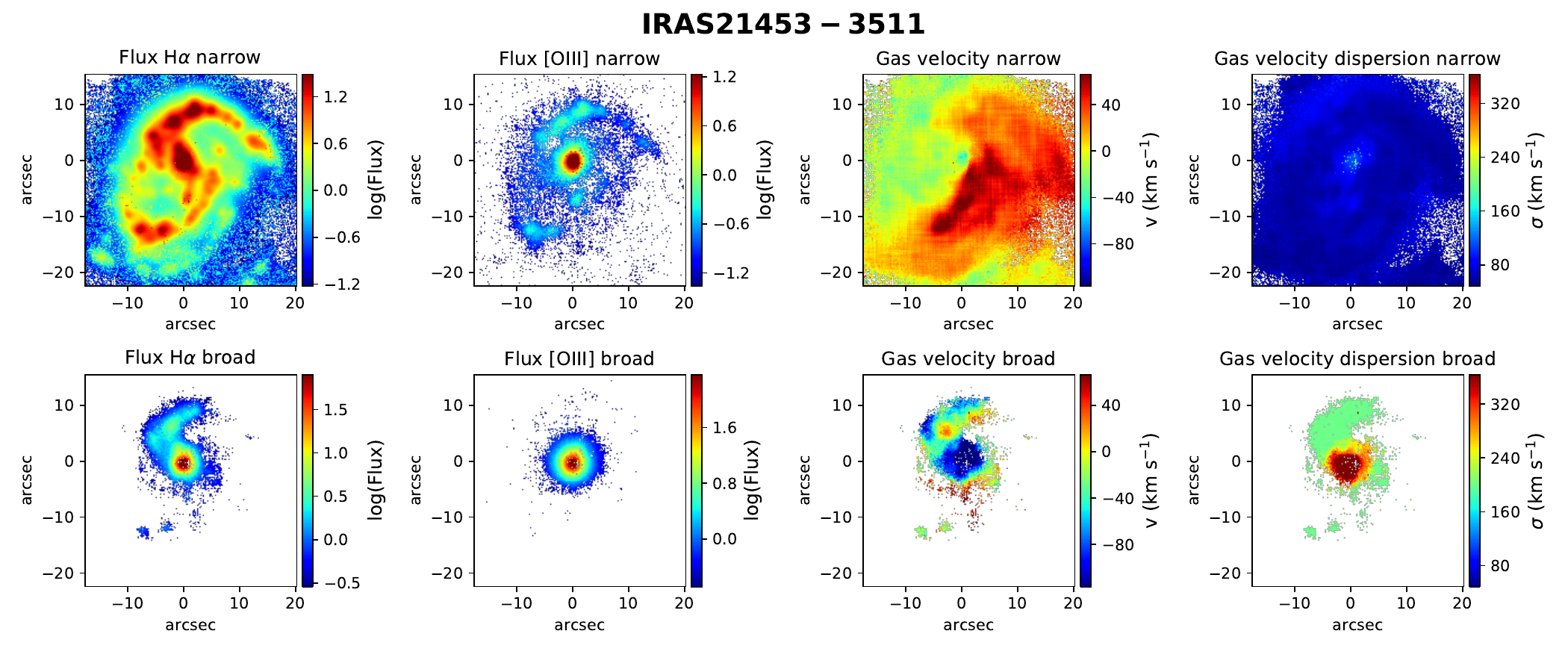}}%
\qquad
\subfigure[Integrated spectrum of the He~I emission and the sodium absorption feature (blue) with fit (red)]{
\label{fig:sfig2}
\includegraphics[width=.495\textwidth]{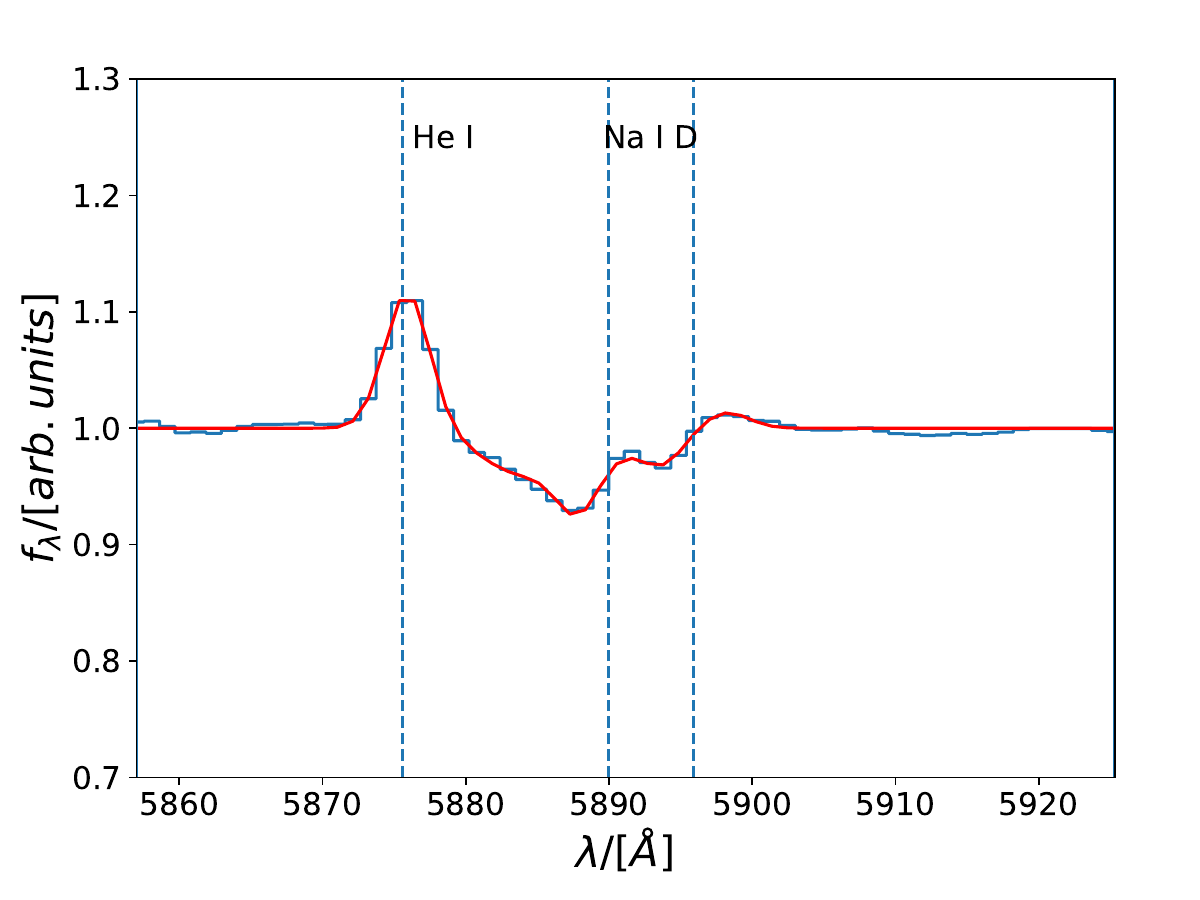}}%
\qquad\hspace{0em}
\subfigure[Maps of the neutral atomic gas]{
\label{fig:sfig3}
\includegraphics[trim={11cm 0 0 0},clip,width=.45\textwidth]{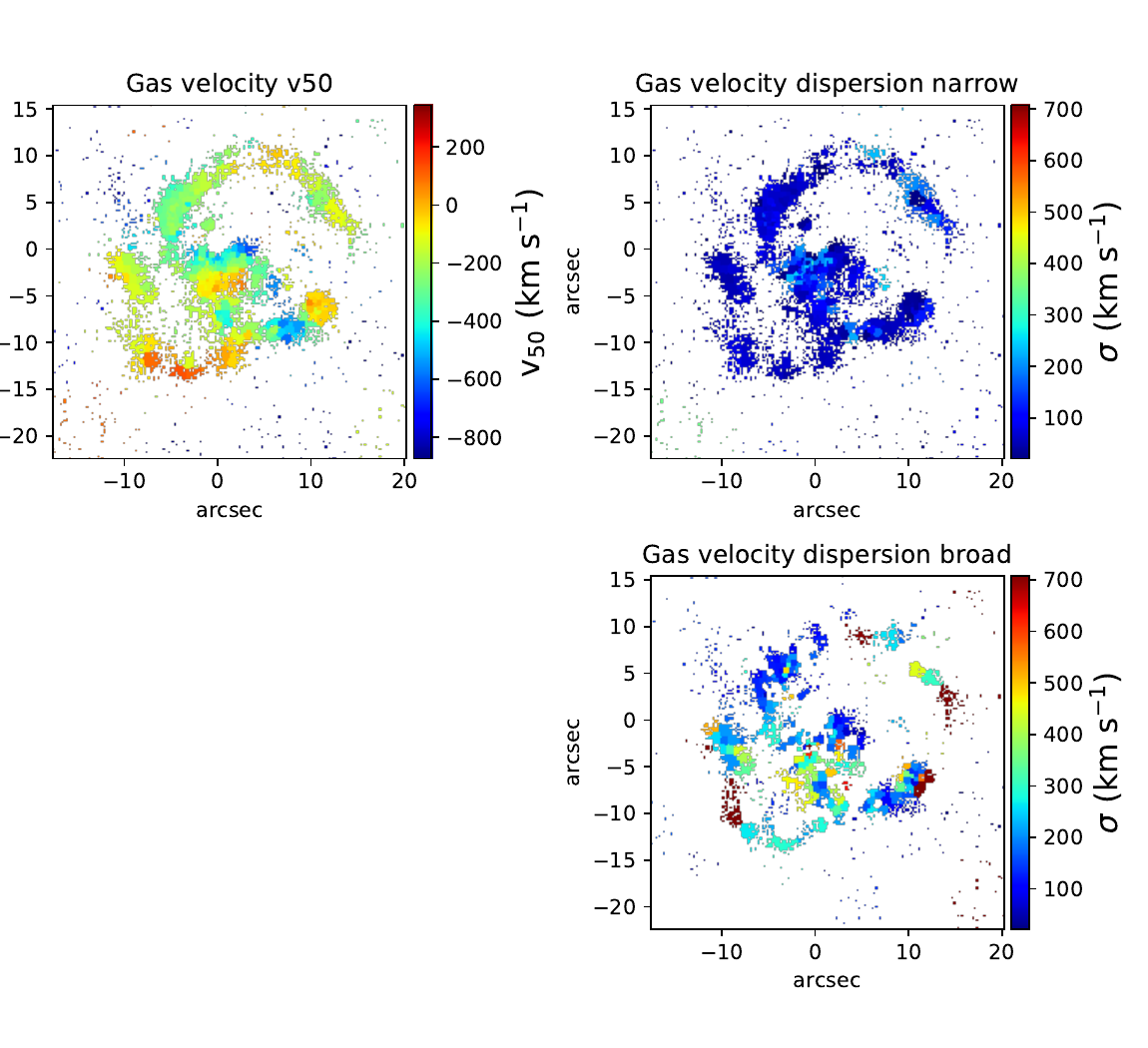}}%
\qquad
\subfigure[Integrated spectrum: The spectrum is shown in light blue, the two-component fit in green, the one component fit in dark blue and for the broad components of H$\alpha$ and H$\beta$ in cyan.]{
\label{fig:sfig4}
\includegraphics[width=.8\textwidth]{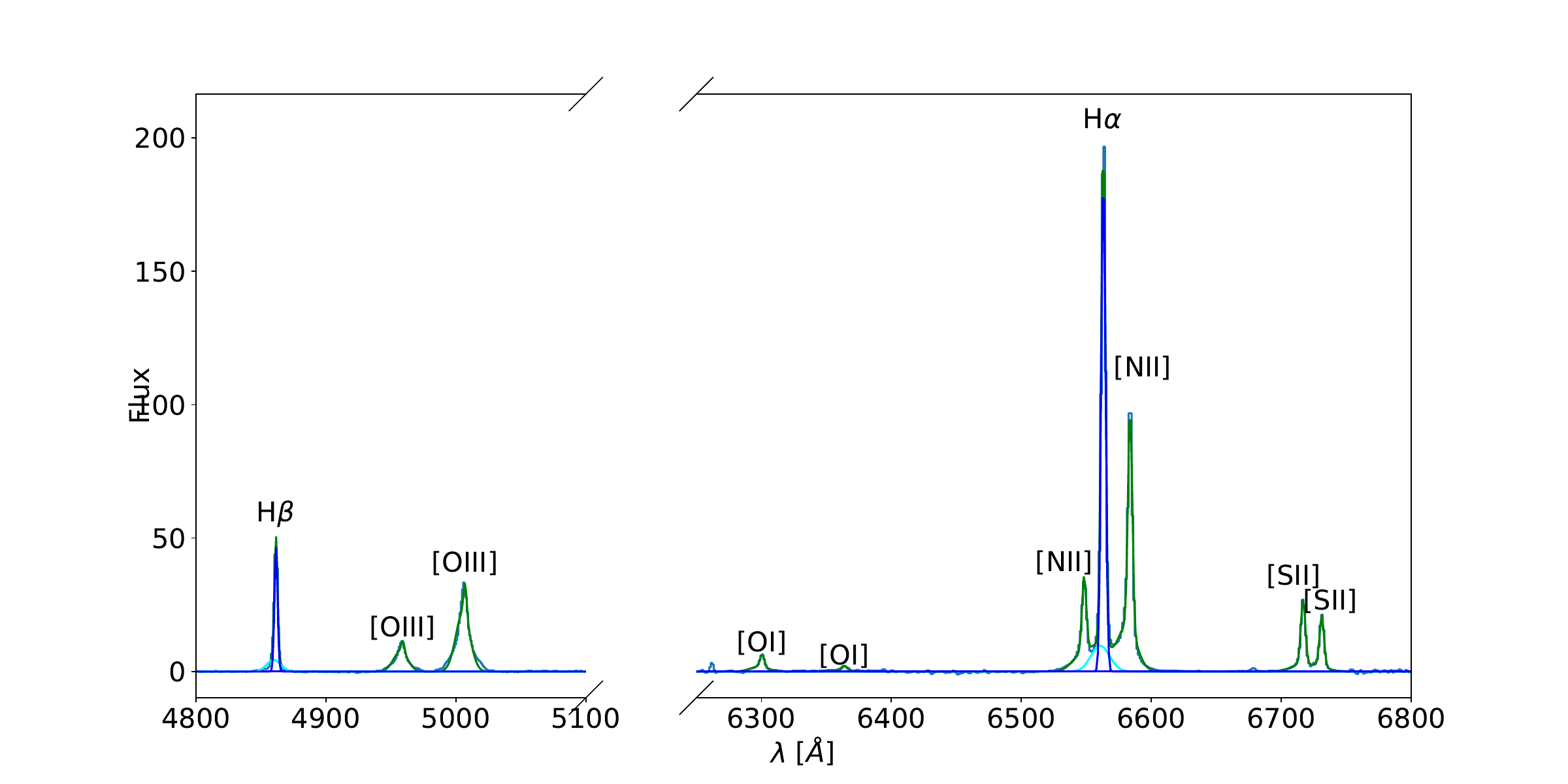}}%
\caption{IRAS 21453-3511}
\end{figure}

\begin{figure}%
\centering
\subfigure[Maps of the ionized gas]{
\label{fig:sfig1}
\includegraphics[width=\textwidth]{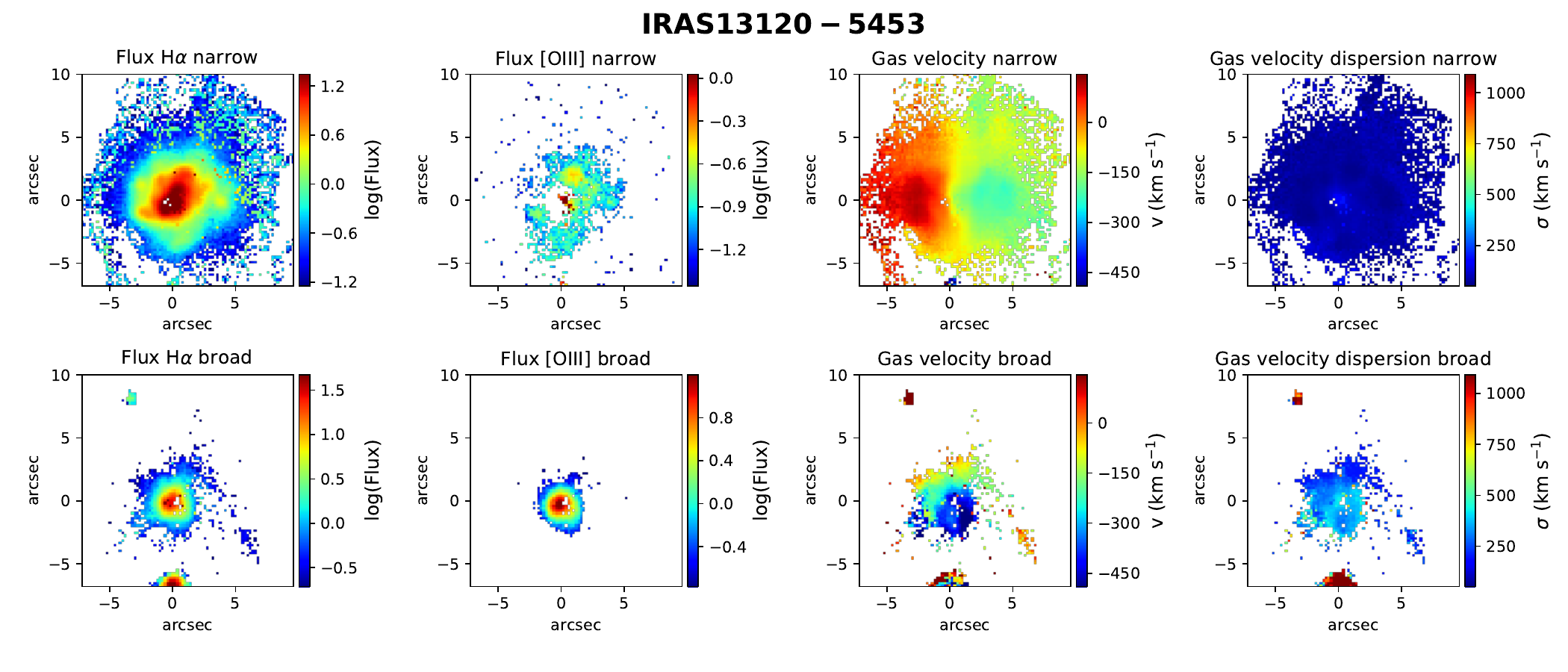}}%
\qquad
\subfigure[Integrated spectrum of the He~I emission and the sodium absorption feature (blue) with fit (red)]{
\label{fig:sfig2}
\includegraphics[width=.495\textwidth]{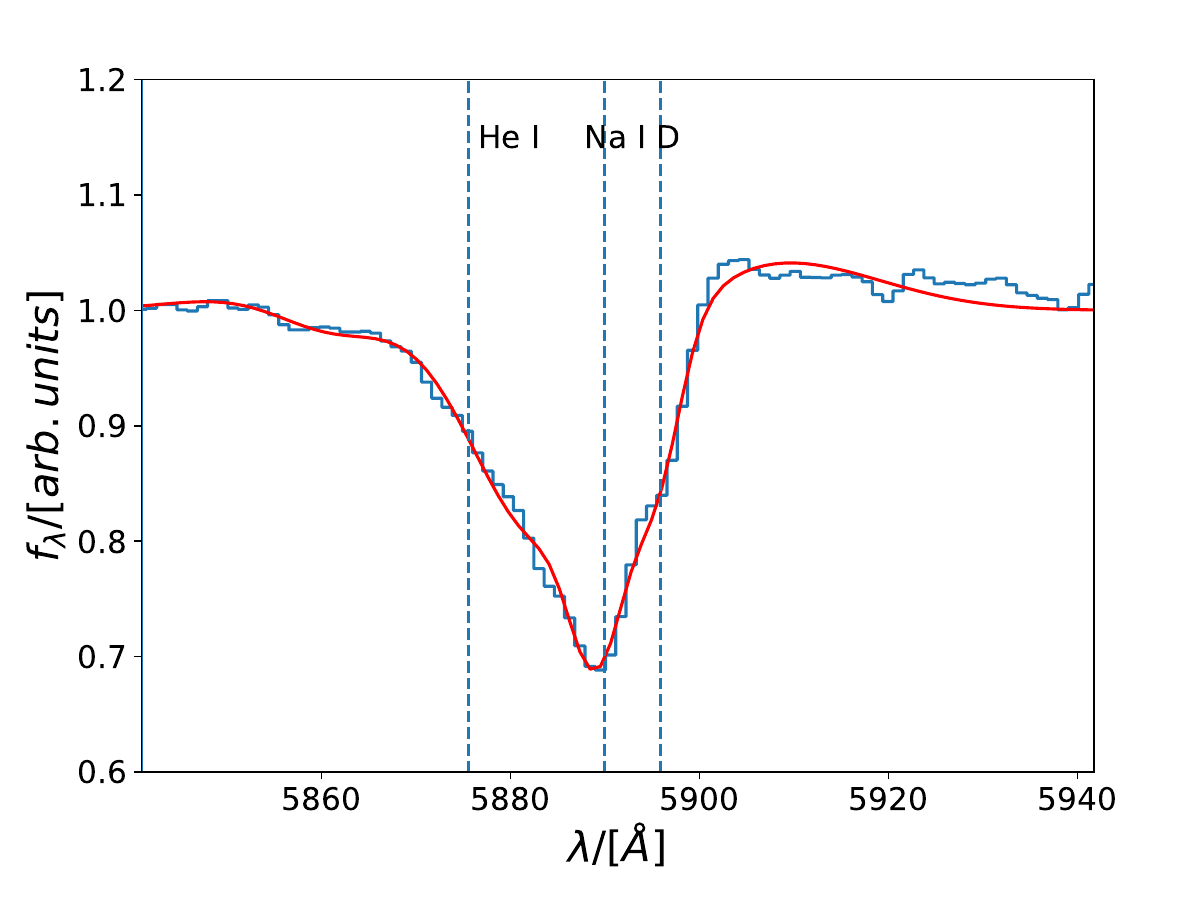}}%
\qquad\hspace{0em}
\subfigure[Maps of the neutral atomic gas]{
\label{fig:sfig3}
\includegraphics[trim={11cm 0 0 0},clip,width=.45\textwidth]{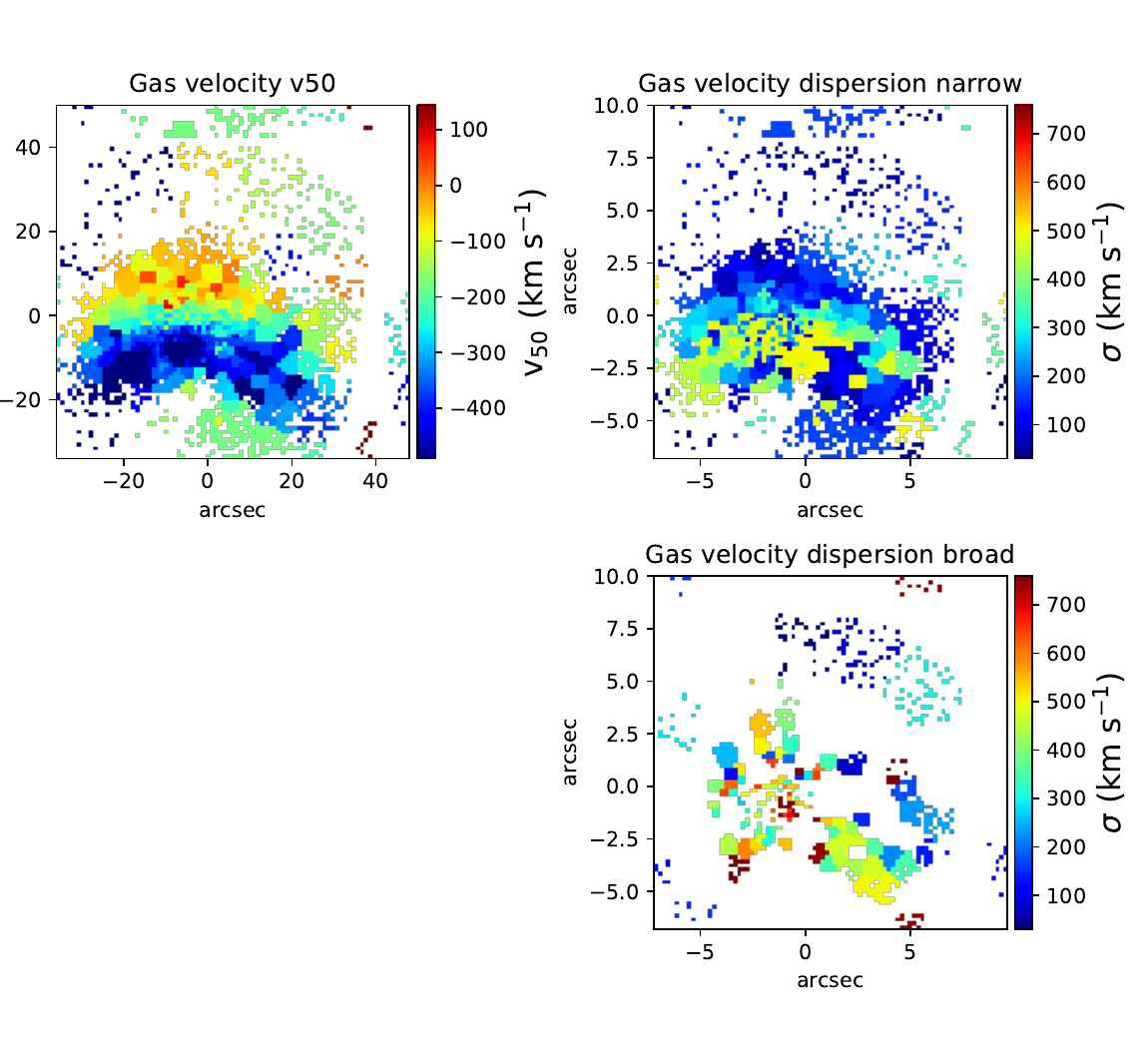}}%
\qquad
\subfigure[Integrated spectrum: The spectrum is shown in light blue, the two-component fit in green, the one component fit in dark blue and for the broad components of H$\alpha$ and H$\beta$ in cyan.]{
\label{fig:sfig4}
\includegraphics[width=.8\textwidth]{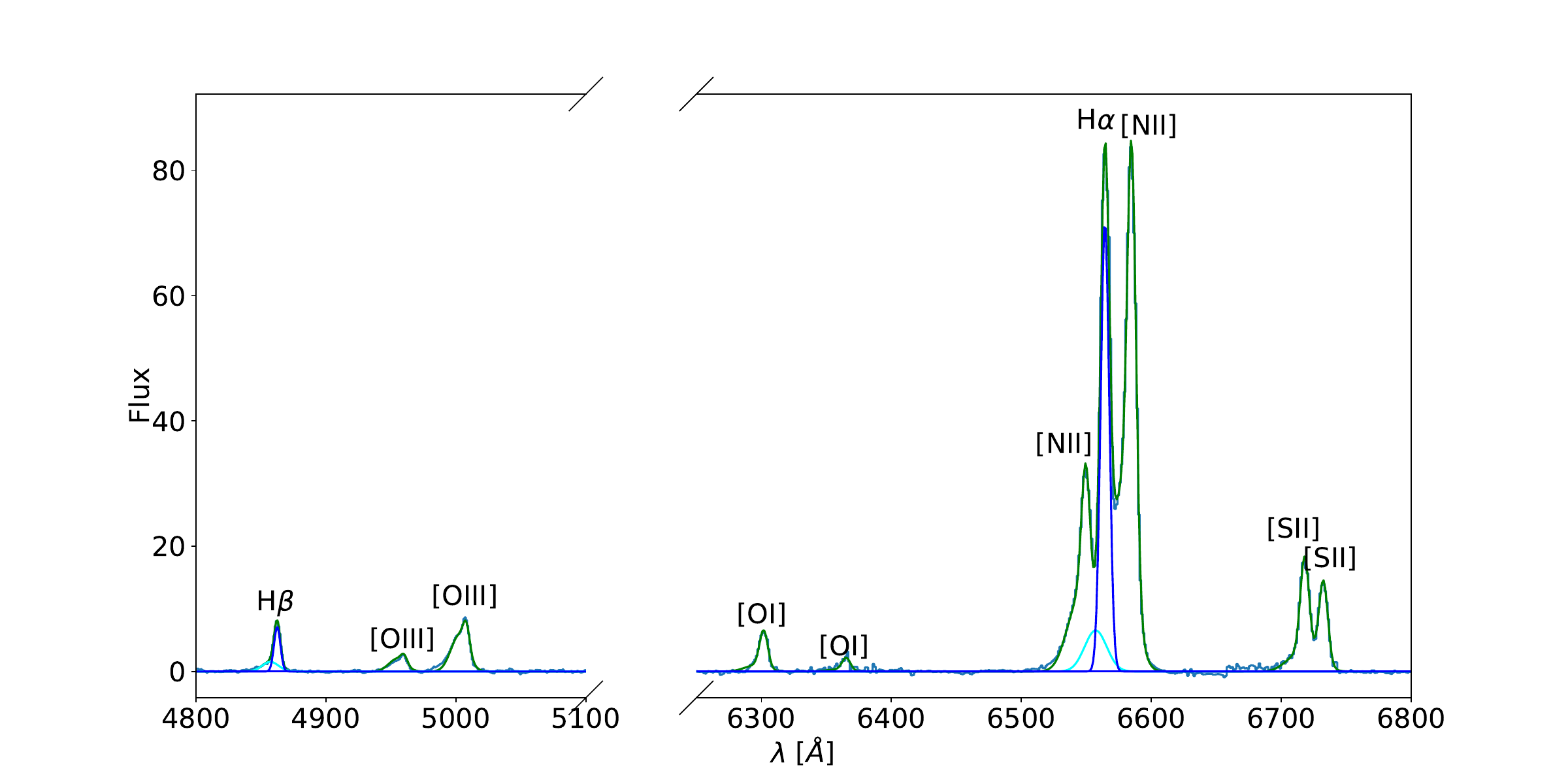}}%
\caption{IRAS 13120-5453}
\end{figure}

\begin{figure}%
\centering
\subfigure[Maps of the ionized gas]{
\label{fig:sfig1}
\includegraphics[width=\textwidth]{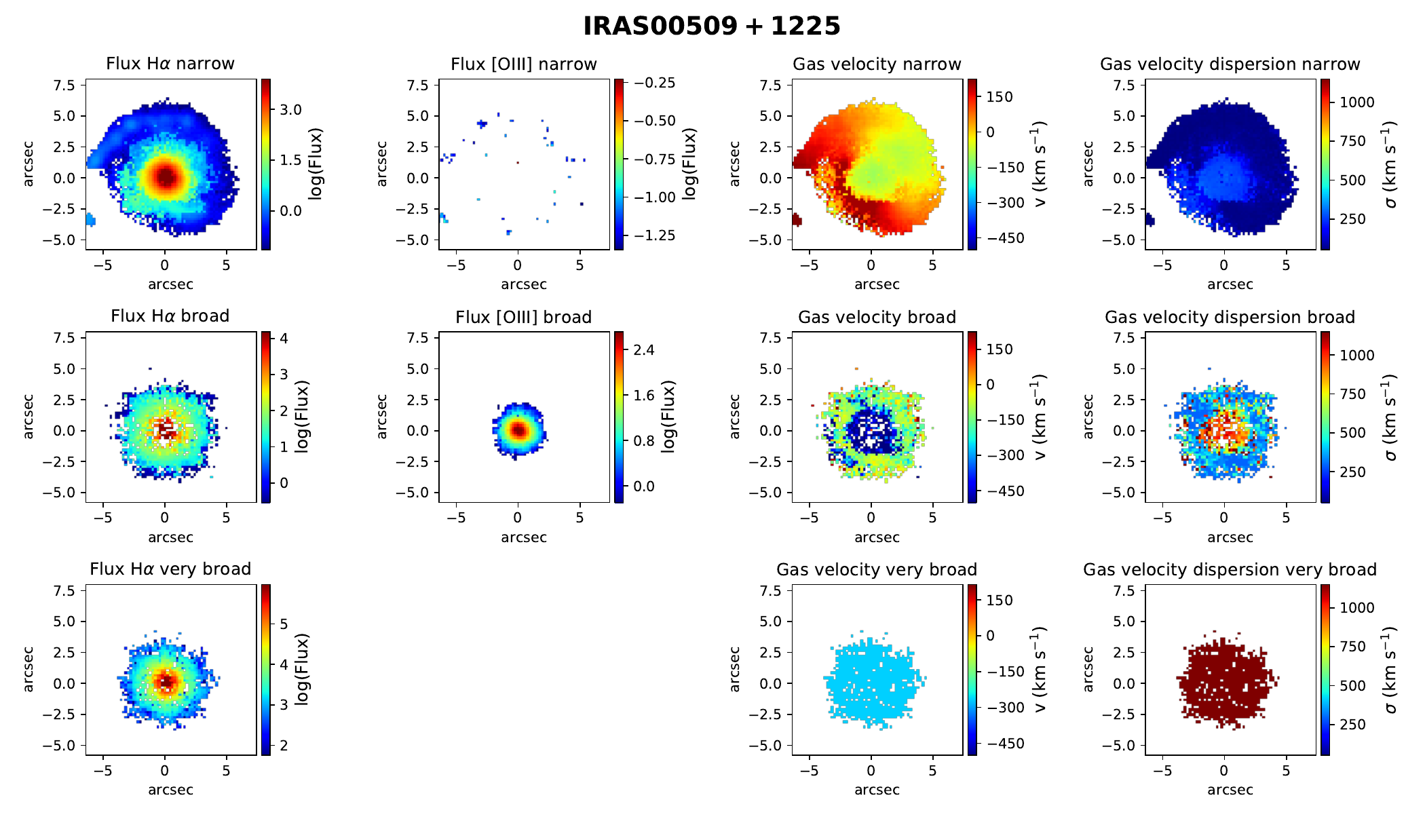}}%
\qquad
\subfigure[Integrated spectrum: The spectrum is shown in light blue, the three component fit in red, the narrow, broad and very broad components of H$\alpha$ and H$\beta$ in blue,  cyan and green, respectively.]{
\label{fig:sfig4}
\includegraphics[width=.8\textwidth]{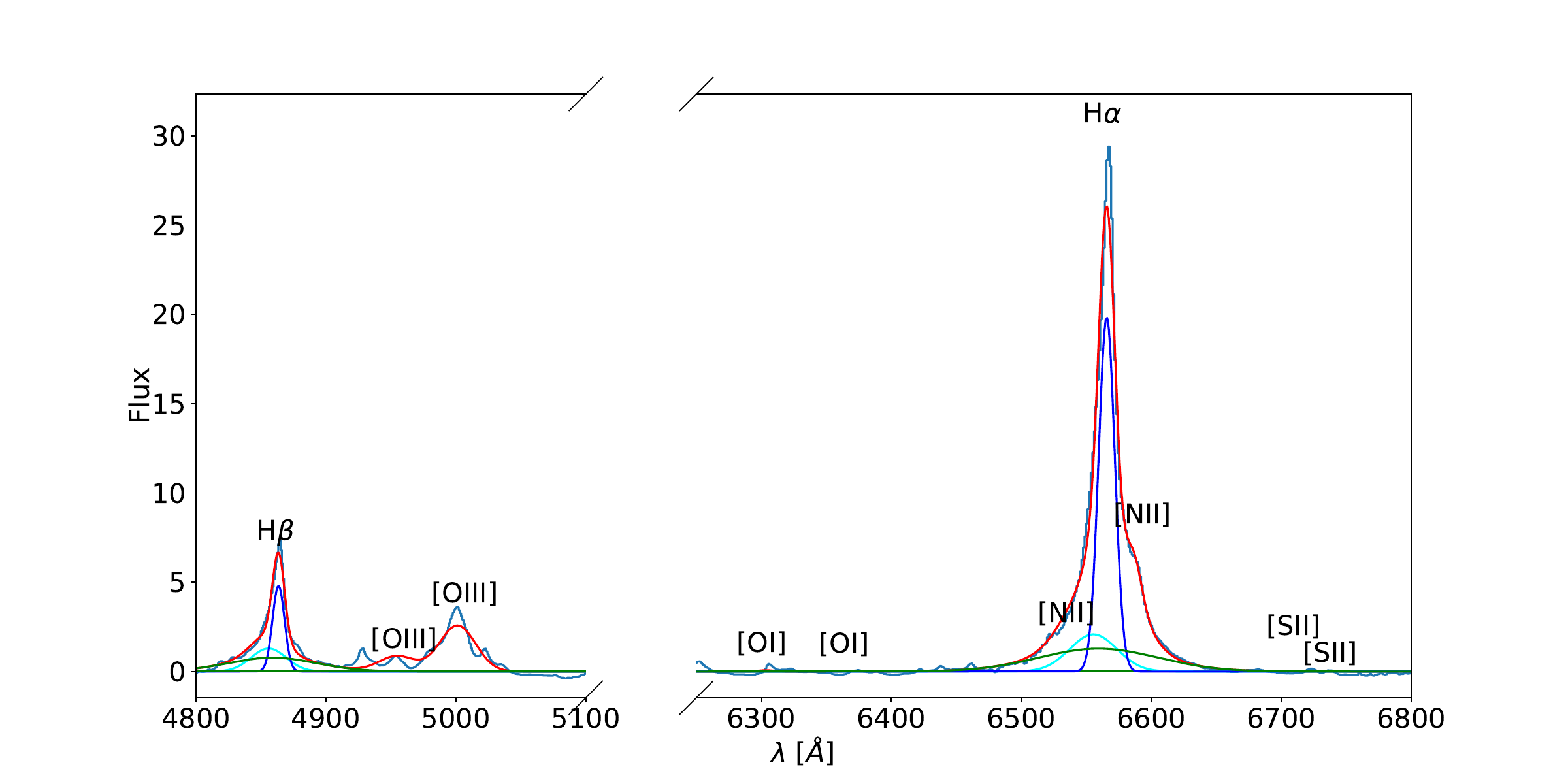}}%
\caption{IRAS 00509+1225}
\end{figure}

\bsp	
\label{lastpage}
\end{document}